\documentclass[journal=jacsat,manuscript=article]{achemso}

\usepackage[version=3]{mhchem} 
\usepackage[T1]{fontenc}       

\usepackage[labelfont=bf, font=small]{caption}

\usepackage{changepage}

\usepackage{amsmath,amssymb}

\SectionNumbersOn 

\usepackage{units, xfrac}

\newcommand\globalScalePlots{1} 

\usepackage{xparse}

\usepackage[normalem]{ulem}

\newcommand{\captionStroke}[1]{\textbf{#1}}
\newcommand{\captionSymbol}[1]{\boldsymbol{#1}} 

\DeclareDocumentCommand \eref{oooo} {\IfNoValueTF{#2}{Eq~(\ref{#1})}{\IfNoValueTF{#3}{Eqs~(\ref{#1}) and (\ref{#2})}{\IfNoValueTF{#4}{Eqs~(\ref{#1})-(\ref{#3})}{Eqs~(\ref{#1})-(\ref{#4})}}}}

\DeclareDocumentCommand \fref{ooo} {\IfNoValueTF{#2}{Fig~\ref{#1}}{\IfNoValueTF{#3}{Figs~\ref{#1} and \ref{#2}}{Figs~\ref{#1}-\ref{#3}}}}

\DeclareDocumentCommand \tref{ooo} {\IfNoValueTF{#2}{Table \ref{#1}}{\IfNoValueTF{#3}{Tables \ref{#1} and \ref{#2}}{Tables \ref{#1}-\ref{#3}}}}

\newcommand{\letter}[1]{#1} 
\newcommand{\letterParen}[1]{(#1)} 

\newcommand{\hyphen}{\hbox{\sout{ }}} 

\newcommand{\KOWT}{K_\text{O}}
\newcommand{\KCWT}{K_\text{C}}
\newcommand{\KOMut}{K_{\text{O}}^{*}}
\newcommand{\KCMut}{K_{\text{C}}^{*}}
\newcommand{\nWT}{n} 
\newcommand{\KOeff}{\tilde{K}_\text{O}} 
\newcommand{\m}{m} 
\newcommand{\kB}{k_\text{B}} 

\usepackage{nameref}
\usepackage{hyperref}

\author{Tal Einav}
\affiliation{Department of Physics, California Institute of Technology, Pasadena, California 91125, United States}
\author{Rob Phillips}
\affiliation{Department of Applied Physics and Division of Biology and Biological Engineering, California Institute of Technology, Pasadena, California 91125, United States}
\email{phillips@pboc.caltech.edu}

\title[Monod-Wyman-Changeux Analysis of Ligand-Gated Ion Channel Mutants]
{Monod-Wyman-Changeux Analysis of Ligand-Gated Ion Channel Mutants}

\begin{document}
	\singlespacing
		
		\begin{abstract}
			We present a framework for computing the gating properties of ligand-gated ion
			channel mutants using the Monod-Wyman-Changeux (MWC) model of allostery. We
			derive simple analytic formulas for key functional properties such as the
			leakiness, dynamic range, half-maximal effective concentration ($[EC_{50}]$),
			and effective Hill coefficient, and explore the full spectrum of phenotypes that
			are accessible through mutations. Specifically, we consider mutations in the
			channel pore of nicotinic acetylcholine receptor (nAChR) and the ligand binding
			domain of a cyclic nucleotide-gated (CNG) ion channel, demonstrating how each
			mutation can be characterized as only affecting a subset of the biophysical
			parameters. In addition, we show how the unifying perspective offered by the MWC
			model allows us, perhaps surprisingly, to collapse the plethora of dose-response
			data from different classes of ion channels into a universal family of curves.
		\end{abstract}
\renewcommand{\thepage}{S\arabic{page}}
\renewcommand{\thefigure}{S\arabic{figure}}
\renewcommand{\thetable}{S\arabic{table}}
\renewcommand{\theequation}{S\arabic{equation}}

\title[Monod-Wyman-Changeux Analysis of Ligand-Gated Ion Channel Mutants Supporting Information]
{Monod-Wyman-Changeux Analysis of Ligand-Gated Ion Channel Mutants Supporting Information}

\begin{document}
	\singlespacing 				

\section{Introduction}

Ion channels are signaling proteins responsible for a huge variety of
physiological functions ranging from responding to membrane voltage, tension,
and temperature to serving as the primary players in the signal transduction we
experience as vision \cite{Hille2001}. Broadly speaking, these channels are
classified based upon the driving forces that gate them. In this work, we
explore one such classification for ligand-gated ion channel mutants based on
the Monod-Wyman-Changeux (MWC) model of allostery. In particular, we focus on
mutants in two of the arguably best studied ligand-gated ion channels: the
nicotinic acetylcholine receptor (nAChR) and the cyclic nucleotide-gated (CNG)
ion channel shown schematically in \fref[figReceptorHighLevelImage]
\cite{Changeux2014, Kaupp2002}.

\begin{figure}[h]
	\centering
	\includegraphics[scale=\globalScalePlots]{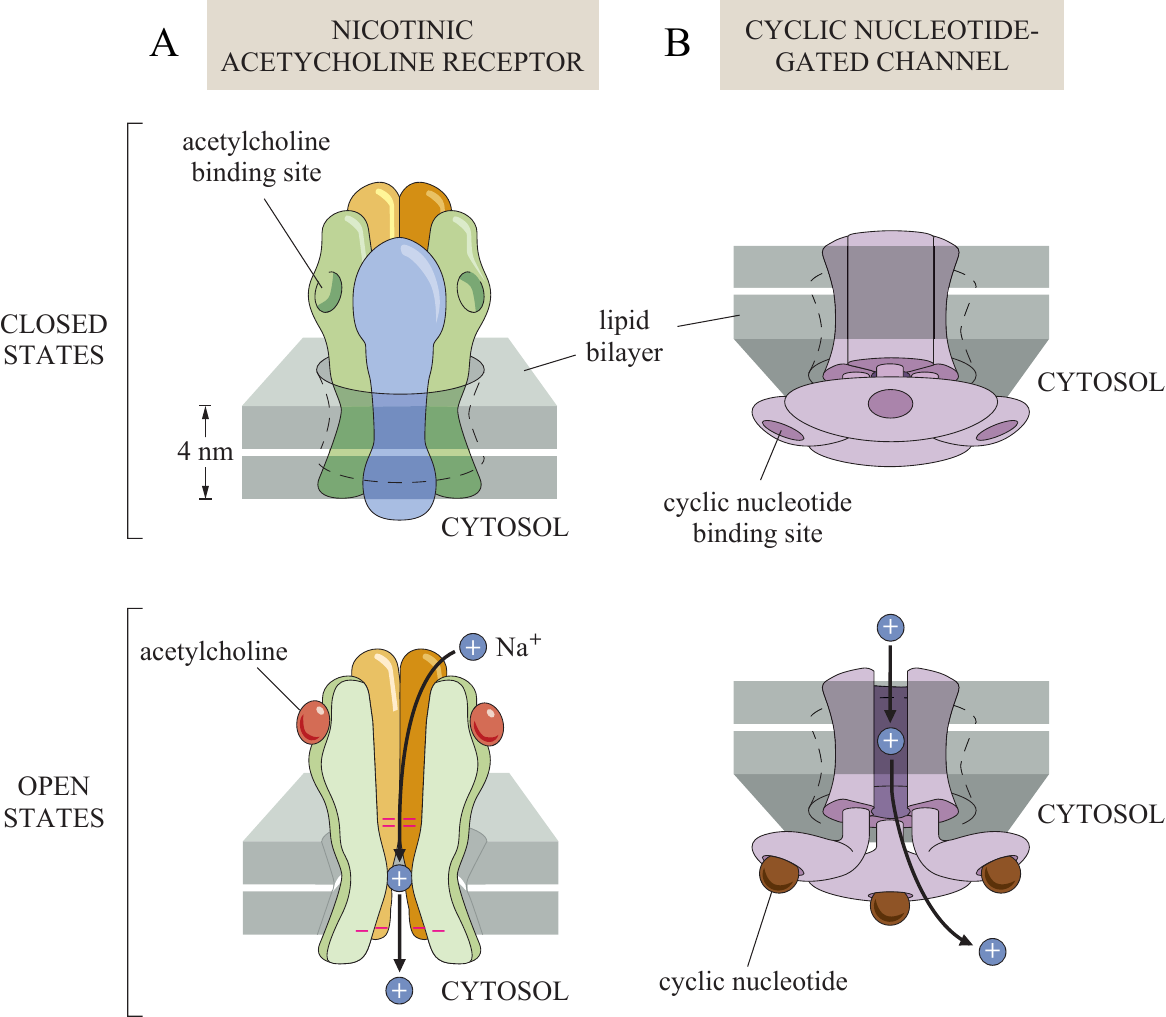}
	\caption{\captionStroke{Schematic of nAChR and CNGA2 ion channels.} \letterParen{A} The
		heteropentameric nicotinic acetylcholine receptor (nAChR) has two ligand
		binding sites for acetylcholine outside the cytosol. \letterParen{B} The
		homotetrameric cyclic nucleotide-gated (CNGA2) has four ligand binding sites,
		one on each subunit, for cAMP or cGMP located inside the cytosol. Both ion
		channels have a higher probability of being closed in the absence of ligand and
		open when bound to ligand.} \label{figReceptorHighLevelImage}
\end{figure}

The MWC model has long been used in the contexts of both nAChR and CNG ion
channels \cite{Karlin1967, Goulding1994, Changeux2005}. Although careful
analysis of these systems has revealed that some details of ligand-gated ion
channel dynamics are not captured by this model (e.g. the existence and
interaction of multiple conducting states \cite{Karpen1997a, Biskup2007}), the
MWC framework nevertheless captures many critical features of a channel's
response and offers one of the simplest settings to explore its underlying
mechanisms. For example, knowledge of both the nAChR and CNG systems' molecular
architecture and our ability to measure their complex kinetics has only recently
become sufficiently advanced to tie the effects of mutations to key biophysical
parameters. Purohit and Auerbach used combinations of mutations to infer the
nAChR gating energy, finding that unliganded nAChR channels open up for a
remarkably brief $80\,\mu\text{s}$ every 15 minutes \cite{Purohit2009}. Statistical
mechanics has been used to show how changes to the energy difference between
conformations in allosteric proteins translates to different functional behavior
(i.e. how it modifies the leakiness, dynamic range, $[EC_{50}]$ and the
effective Hill coefficient) \cite{Martins2011,Marzen2013}, and we extend this
work to find simple analytic approximations that are valid within the context of
ion channels. Using this methodology, we systematically explore the full range
of behaviors that may be induced by different types of mutations. This analysis
enables us to quantify the inherent trade-offs between key properties of ion
channel dose-response curves and potentially paves the way for future
biophysical models of evolutionary fitness in which the genotype (i.e. amino
acid sequence) of allosteric molecules is directly connected to phenotype (i.e.
properties of a channel's response).

To this end, we consider two distinct classes of mutants which tune different
sets of MWC parameters -- either the ion channel gating energy or the
ligand-channel dissociation constants. Previous work by Auerbach \textit{et al.}~demonstrated
that these two sets of physical parameters can be independently tuned within the nAChR
ion channel; pore mutations only alter the channel gating energy whereas
mutations within the ligand binding domain only affect the ligand-channel
dissociation constants \cite{Auerbach2012}. Utilizing this parameter
independence, we determine the full spectrum of nAChR phenotypes given an
arbitrary set of channel pore mutations and show why a linear increase in the
channel gating energy leads to a logarithmic shift in the nAChR dose-response
curve. Next, we consider recent data from CNGA2 ion channels with mutations in
their ligand binding pocket \cite{Wongsamitkul2016}. We hypothesize that
modifying the ligand binding domain should not alter the channel gating energy
and demonstrate how the entire class of CNGA2 mutants can be simultaneously
characterized with this constraint. This class of mutants sheds light on the
fundamental differences between homooligomeric channels comprised of a single
type of subunit and heterooligomeric channels whose distinct subunits can have
different ligand binding affinities.

By viewing mutant data through its effects on the underlying biophysical
parameters of the system, we go well beyond simply fitting individual
dose-response data, instead creating a framework with which we can explore the
full expanse of ion channel phenotypes available through mutations. Using this
methodology, we: (1) analytically compute important ion channel characteristics,
namely the leakiness, dynamic range, $[EC_{50}]$, and effective Hill
coefficient, (2) link the role of mutations with thermodynamic parameters, (3)
show how the behavior of an entire family of mutants can be predicted using only
a subset of the members of that family, (4) quantify the pleiotropic effect of
point mutations on multiple phenotypic traits and characterize the correlations
between these diverse effects, and (5) collapse the data from multiple ion
channels onto a single master curve, revealing that such mutants form a
one-parameter family. In doing so, we present a unified framework to collate the
plethora of data known about such channels.

\section*{Model}

Electrophysiological techniques can measure currents across a single cell's
membrane. The current flowing through a ligand-gated ion channel is proportional
to the average probability $p_{\text{open}}(c)$ that the channel will be open at
a ligand concentration $c$. For an ion channel with $\m$ identical ligand
binding sites (see \fref[figStatesWeightsBothIonChannels]), this probability is
given by the MWC model as
\begin{equation} \label{eqPOpenGeneral}
p_{\text{open}}(c) = \frac{\left(1 + \frac{c}{\KOWT}\right)^\m}{\left(1 + \frac{c}{\KOWT}\right)^\m + e^{-\beta \epsilon}\left(1 + \frac{c}{\KCWT}\right)^\m},
\end{equation}
where $\KOWT$ and $\KCWT$ represent the dissociation constants between the
ligand and the open and closed ion channel, respectively, $c$ denotes the
concentration of the ligand, $\epsilon$ (called the gating energy) denotes the
free energy difference between the closed and open conformations of the ion
channel in the absence of ligand, and $\beta = \frac{1}{\kB T}$ where $\kB$ is
Boltzmann's constant and $T$ is the temperature of the system. Wild type ion
channels are typically closed in the absence of ligand ($\epsilon < 0$) and open
when bound to ligand ($\KOWT < \KCWT$). \fref[figStatesWeightsBothIonChannels]
shows the possible conformations of the nAChR ($\m=2$) and CNGA2 ($\m=4$) ion
channels together with their Boltzmann weights. $p_{\text{open}}(c)$ is given by
the sum of the open state weights divided by the sum of all weights.
Note that the MWC parameters $\KOWT$, $\KCWT$, and $\epsilon$ may be
expressed as ratios of the experimentally measured rates of ligand binding and
unbinding as well as the transition rates between the open and
closed channel conformations (see Supporting Information section A.1).

\begin{adjustwidth}{-5in}{0in}
\begin{figure}[t]	
	\centering \includegraphics[scale=\globalScalePlots]{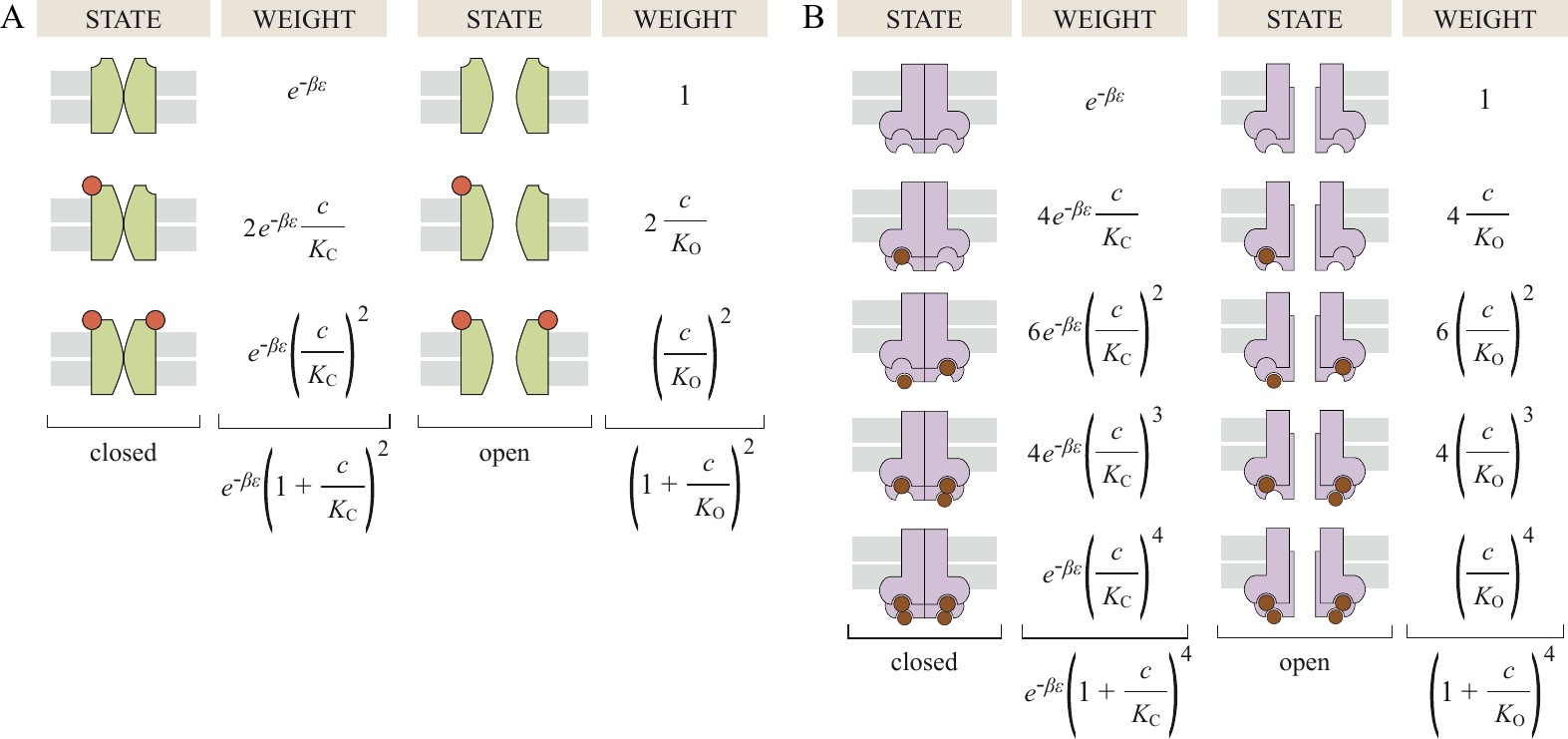}
	\caption{\captionStroke{The probability that a ligand-gated ion channel is open as given
		by the MWC model.} \letterParen{A} Microscopic states and Boltzmann weights of
		the nAChR ion channel (green) binding to acetylcholine (orange).
		\letterParen{B} Corresponding states for the CNGA2 ion channel (purple) binding to cGMP (brown).
		The behavior of these channels is determined by three physical parameters: the
		affinity between the receptor and ligand in the open ($\KOWT$) and closed
		($\KCWT$) states and the free energy difference $\epsilon$ between the closed
		and open conformations of the ion channel.}
	\label{figStatesWeightsBothIonChannels}
\end{figure}
\end{adjustwidth}

Current measurements are often reported as \textit{normalized} current, implying
that the current has been stretched vertically to run from 0 to 1, as given by
\begin{equation} \label{eqNormalizedCurrent}
\text{normalized current} = \frac{p_{\text{open}}(c)-p_{\text{open}}^{\text{min}}}{p_{\text{open}}^{\text{max}}-p_{\text{open}}^{\text{min}}}.
\end{equation}
$p_{\text{open}}(c)$ increases monotonically as a function of ligand
concentration $c$, with a minimum value in the absence of ligand given by
\begin{equation} \label{eqPMinDef}
p_{\text{open}}^{\text{min}} = p_{\text{open}}(0) = \frac{1}{1 + e^{-\beta \epsilon}},
\end{equation}
and a maximum value in the presence of saturating levels of ligand given as
\begin{equation} \label{eqPMaxDef}
p_{\text{open}}^{\text{max}} = \lim_{c\to \infty} p_{\text{open}}(c) = \frac{1}{1 + e^{-\beta \epsilon}\left(\frac{\KOWT}{\KCWT}\right)^\m}.
\end{equation}

Using the above two limits, we can investigate four important characteristics of
ion channels \cite{Martins2011, Marzen2013}. First, we examine the
\textit{leakiness} of an ion channel, or the fraction of time a channel is open
in the absence of ligand, namely, 
\begin{equation} \label{eqLeakinessDef}
\text{leakiness} = p_{\text{open}}^{\text{min}}.
\end{equation}
Next we determine the \textit{dynamic range}, or the difference between the
probability of the maximally open and maximally closed states of the ion channel, given by
\begin{equation} \label{eqDynamicRangeDef}
\text{dynamic range} = p_{\text{open}}^{\text{max}} - p_{\text{open}}^{\text{min}}.
\end{equation}
Ion channels that minimize leakiness only open upon ligand binding, and ion
channels that maximize dynamic range have greater contrast between their open
and closed states. Just like $p_{\text{open}}(c)$, leakiness and dynamic range
lie within the interval $[0,1]$. 

Two other important characteristics are measured from the normalized current.
The \textit{half maximal effective concentration} $[EC_{50}]$ denotes the
concentration of ligand at which the normalized current of the ion channel
equals \sfrac{1}{2}, namely, 
\begin{equation} \label{eqEC50Def}
p_{\text{open}}([EC_{50}]) = \frac{p_{\text{open}}^{\text{min}} + p_{\text{open}}^{\text{max}}}{2}.
\end{equation}
The \textit{effective Hill coefficient} $h$ equals twice the log-log slope of
the normalized current evaluated at $c = [EC_{50}]$,
\begin{equation} \label{eqhDef}
h = 2 \frac{d}{d \log c} \log \left( \frac{p_{\text{open}}(c)-p_{\text{open}}^{\text{min}}}{p_{\text{open}}^{\text{max}}-p_{\text{open}}^{\text{min}}} \right)_{c = [EC_{50}]},
\end{equation}
which reduces to the standard Hill coefficient for the Hill function
\cite{Hill1985}. The $[EC_{50}]$ determines how the normalized current shifts
left and right, while the effective Hill coefficient corresponds to the slope at
$[EC_{50}]$. Together, these two properties determine the approximate window of
ligand concentrations for which the normalized current transitions from 0 to 1.

In the limit $1 \ll e^{-\beta \epsilon} \ll \left(\frac{\KCWT}{\KOWT}\right)^m$,
which we show below is relevant for both the nAChR and CNGA2 ion channels, the
various functional properties of the channel described above can be approximated
to leading order as (see Supporting Information section B):
\begin{align}
\text{leakiness} &\approx e^{\beta \epsilon} \label{leakinessValue}\\
\text{dynamic range} &\approx 1 \label{dynamicRangeValue}\\
[EC_{50}] &\approx e^{-\beta \epsilon/m} \KOWT \label{eqEC50Value}\\
h &\approx m.\label{eqhValue}
\end{align}

\section*{Results}

\subsection*{nAChR Mutants can be Categorized using Free Energy} 

Muscle-type nAChR is a heteropentamer with subunit stoichiometry $\alpha_2\beta
\gamma\delta$, containing two ligand binding sites for acetylcholine at the
interface of the $\alpha$-$\delta$ and $\alpha$-$\gamma$ subunits
\cite{Changeux1998}. The five homologous subunits have M2 transmembrane domains
which move symmetrically during nAChR gating to either occlude or open the ion
channel \cite{Unwin1995}. By introducing a serine in place of the leucine at a
key residue (L251S) within the M2 domain present within each subunit, the
corresponding subunit is able to more easily transition from the closed to open
configuration, shifting the dose-response curve to the left (see
\fref[figNormalizedCurrent]\letter{A}) \cite{Labarca1995}. For example, wild
type nAChR is maximally stimulated with $100\,\mu\text{M}$ of acetylcholine,
while a mutant ion channel with one L251S mutation is more sensitive and only
requires $10\,\mu\text{M}$ to saturate its dose-response curve.

\begin{figure}[t]
		\centering \includegraphics[scale=\globalScalePlots]{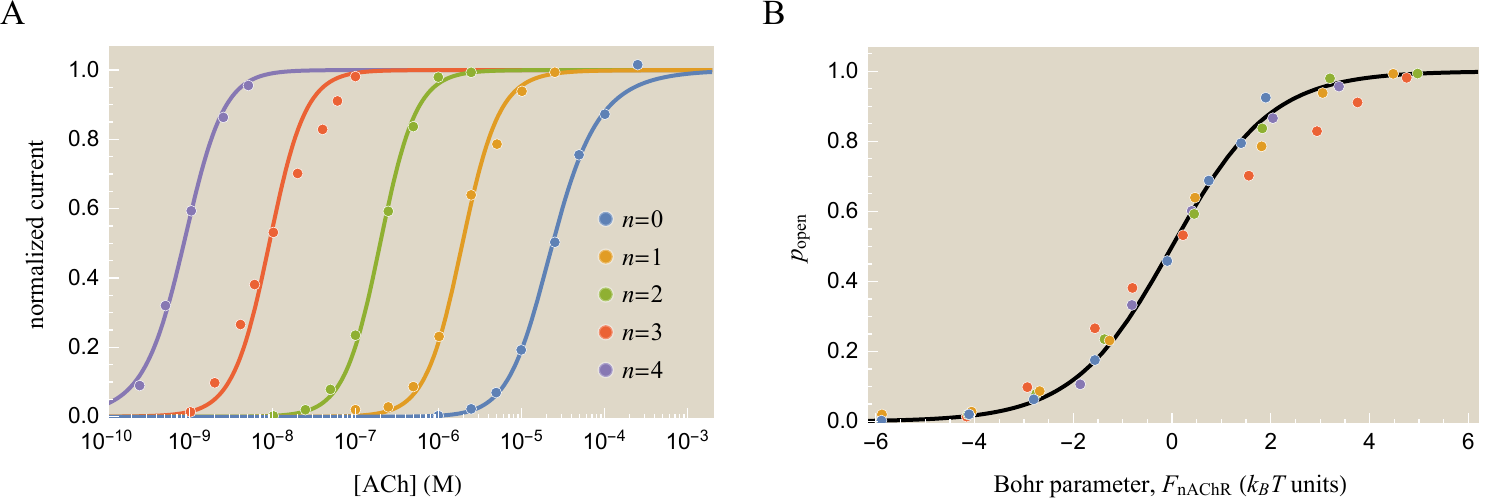}
		\caption{\captionStroke{Characterizing nicotinic acetylcholine receptors with
			$\captionSymbol{n}$ subunits carrying the L251S mutation.} 
			\letterParen{A} Normalized currents of 
			mutant nAChR ion channels at different concentrations of the agonist
			acetylcholine (ACh) \cite{Labarca1995}. The curves from right to left show a
			receptor with $n=0$ (wild type), $n=1$ ($\alpha_2\beta \gamma^* \delta$),
			$n=2$ ($\alpha_2^*\beta \gamma \delta$), $n=3$ ($\alpha_2\beta^* \gamma^*
			\delta^*$), and $n=4$ ($\alpha_2^* \beta \gamma^* \delta^*$) mutations, where
			asterisks ($*$) denote a mutated subunit. Fitting the data
			(solid lines) to \eref[eqPOpenGeneral][eqNormalizedCurrent] with $\m = 2$
			ligand binding sites determines the three MWC parameters $\KOWT = 0.1 \times
			10^{-9}\,\,\text{M}$, $\KCWT = 60 \times 10^{-6}\,\,\text{M}$, and $\beta
			\epsilon^{(n)} = \left[-4.0, -8.5, -14.6, -19.2, -23.7 \right]$ from left
			($n=4$) to right ($n=0$). With each additional mutation, the dose-response
			curve shifts to the left by roughly a decade in concentration while the
			$\epsilon$ parameter increases by roughly $5~\kB T$. \letterParen{B} The 
			probability $p_{\text{open}}(c)$ that the five ion channels are open
			can be collapsed onto the same curve using the Bohr parameter
			$F_{\text{nAChR}}(c)$ given by \eref[eqBohrParameterGeneral]. A positive Bohr
			parameter indicates that $c$ is above the $[EC_{50}]$. See Supporting Information section
			C for details on the fitting procedure.}
		\label{figNormalizedCurrent}
\end{figure}

Labarca \textit{et al.}~used L251S mutations to create ion channels with $n$
mutated subunits \cite{Labarca1995}. \fref[figNormalizedCurrent]\letter{A} shows
the resulting normalized current for several of these mutants; from right to
left the curves represent $n=0$ (wild type) to $n=4$ (an ion channel with four
of its five subunits mutated). One interesting trend in the data is that each
additional mutation shifts the normalized current to the left by approximately
one decade in concentration (see Supporting Information section
A.2). This constant shift in the dose-response curves
motivated Labarca \textit{et al.}~to postulate that mutating each subunit
increases the gating free energy $\epsilon$ by a fixed amount.

To test this idea, we analyze the nAChR data at various concentrations $c$ of
the ligand acetylcholine using the MWC model \eref[eqPOpenGeneral] with $\m = 2$
ligand binding sites. Because the L251S mutation is approximately
$4.5\,\text{nm}$ from the ligand binding domain \cite{Unwin1993}, we assume that
the ligand binding affinities $\KOWT$ and $\KCWT$ are unchanged for the wild
type and mutant ion channels, an assumption that has been repeatedly verified
by Auerbach \textit{et al.}~for nAChR pore mutations \cite{Auerbach2012}.
\fref[figNormalizedCurrent]\letter{A} shows the best-fit theoretical curves
assuming all five nAChR mutants have the same $\KOWT$ and $\KCWT$ values but
that each channel has a distinct gating energy $\epsilon^{(n)}$ (where the
superscript $n$ denotes the number of mutated subunits). These gating energies
were found to increase by roughly $5~\kB T$ per $n$, as would be expected for a
mutation that acts equivalently and independently on each subunit.

One beautiful illustration of the power of the MWC model lies in its ability to
provide a unified perspective to view data from many different ion channels. Following earlier work in the context of both chemotaxis and quorum sensing
\cite{Keymer2006,Swem2008},
we rewrite the probability that the nAChR receptor is open as
\begin{equation} \label{eqBohrParameterGeneral}
p_{\text{open}}(c) \equiv \frac{1}{1+e^{-\beta F(c)}},
\end{equation}
where this last equation defines the \textit{Bohr parameter} \cite{Phillips2015a}
\begin{equation} \label{eqBohrParameterInTermsOfMWCParameters}
F(c) = - \kB T \log \left( e^{-\beta \epsilon} \frac{\left(1 + \frac{c}{\KCWT} \right)^m}{\left( 1 + \frac{c}{\KOWT} \right)^m} \right).
\end{equation}
The Bohr parameter quantifies the trade-offs between the physical parameters of
the system (in the case of nAChR, between the entropy associated with the ligand
concentration $c$ and the gating free energy $\beta \epsilon$). When the Bohr
parameters of two ion channels are equal, both channels will elicit the same
physiological response. Using \eref[eqPOpenGeneral][eqBohrParameterGeneral] to
convert the normalized current data into the probability $p_{\text{open}}$
(see Supporting Information section A.3), we can collapse the dose-response data of
the five nAChR mutants onto a single master curve as a function of the Bohr
parameter for nAChR, $F_{\text{nAChR}}(c)$, as shown in
\fref[figNormalizedCurrent]\letter{B}. In this way, the Bohr parameter maps the
full complexity of a generic ion channel response into a single combination of
the relevant physical parameters of the system.

\subsubsection*{Full Spectrum of nAChR Gating Energy Mutants}

We next consider the entire range of nAChR phenotypes achievable by only
modifying the gating free energy $\epsilon$ of the wild type ion channel. For
instance, any combination of nAChR pore mutations would be expected to not
affect the ligand dissociation constants and thus yield an ion channel within
this class (see Supporting Information section A.4 for one such example). For
concreteness, we focus on how the $\epsilon$ parameter tunes key features of the
dose-response curves, namely the leakiness, dynamic range, $[EC_{50}]$, and
effective Hill coefficient $h$ (see
\eref[eqLeakinessDef][eqDynamicRangeDef][eqEC50Value][eqhValue]), although we
note that other important phenotypic properties such as the intrinsic noise and
capacity have also been characterized for the MWC model \cite{Martins2011}.
\fref[figCharacteristics] shows these four characteristics, with the open
squares representing the properties of the five best-fit dose-response curves
from \fref[figNormalizedCurrent]\letter{A}.

\begin{figure}[t]
	\centering \includegraphics[scale=\globalScalePlots]{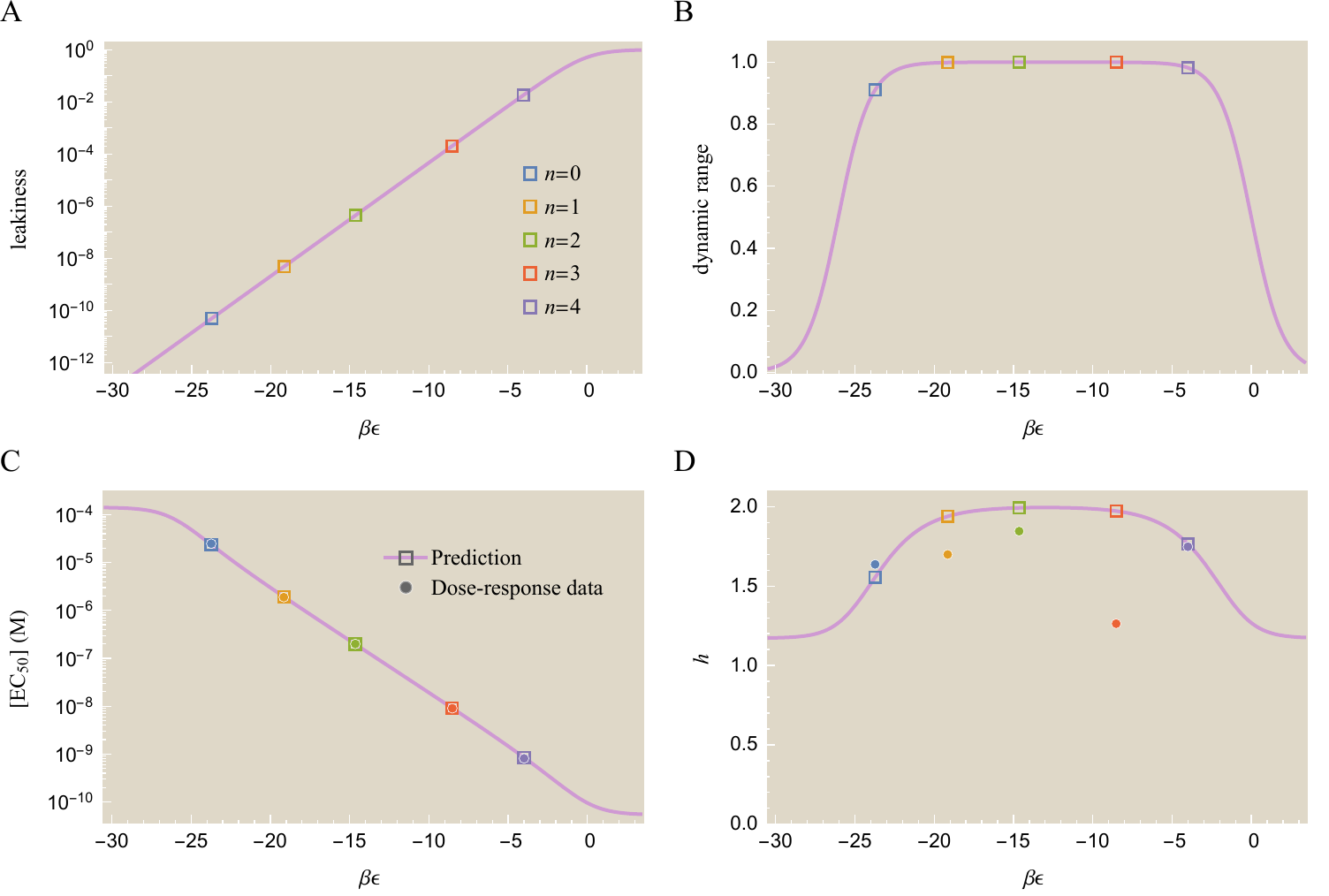}
	\caption{\captionStroke{Theoretical prediction and experimental measurements for mutant nAChR
		ion channel characteristics.} The open squares mark the $\beta \epsilon$ values
		of the five dose response curves from \fref[figNormalizedCurrent]\letter{A}.
		\letterParen{A} The leakiness given by \eref[eqLeakinessDef] increases
		exponentially with each mutation. \letterParen{B} The dynamic range from
		\eref[eqDynamicRangeDef] is nearly uniform for all mutants. \letterParen{C}
		The $[EC_{50}]$ decreases exponentially with each mutation. \letterParen{D}
		The effective Hill coefficient $h$ is predicted to remain approximately
		constant. $[EC_{50}]$ and $h$ offer a direct comparison between the best-fit
		model predictions (open squares) and the experimental measurements (solid
		circles) from \fref[figNormalizedCurrent]\letter{A}. While the $[EC_{50}]$
		matches well between theory and experiment, the effective Hill coefficient $h$
		is significantly noisier. } \label{figCharacteristics}
\end{figure}

\fref[figCharacteristics]\letter{A} implies that all of the mutants considered here have negligible leakiness; according to the MWC parameters found here, the
probability that the wild type channel ($\beta \epsilon^{(0)} = -23.7$) will be
open is less than $10^{-10}$. Experimental measurements have shown that such
spontaneous openings occur extremely infrequently in nAChR \cite{Nayak2012},
although direct measurement is difficult for such rare events. Other mutational
analysis has predicted gating energies around $\beta \epsilon^{(0)} \approx -14$
(corresponding to a leakiness of $10^{-6}$) \cite{Auerbach2012}, but we note
that such a large wild type gating energy prohibits the five mutants in
\fref[figNormalizedCurrent] from being fit as a single class of mutants with the
same $\KOWT$ and $\KCWT$ values (see Supporting Information section
C.2). If this large wild type gating energy
is correct, it may imply that the L251S mutation also affects the $\KOWT$ and
$\KCWT$ parameters, though the absence of error bars on the original data make
it hard to quantitatively assess the underlying origins of these discrepancies.

\fref[figCharacteristics]\letter{B} asserts that all of the mutant ion channels
should have full dynamic range except for the wild type channel, which has a
dynamic range of $0.91$. In comparison, the measured dynamic range of wild type
nAChR is $0.95$, close to our predicted value \cite{Auerbach2012}. Note that
only when the dynamic range approaches unity does the normalized current become
identical to $p_{\text{open}}$; for lower values, information about the
leakiness and dynamic range is lost by only measuring normalized currents.

We compare the $[EC_{50}]$ (\fref[figCharacteristics]\letter{C}) and effective
Hill coefficient $h$ (\fref[figCharacteristics]\letter{D}) with the nAChR data
by interpolating the measurements (see Supporting Information section C.3) in
order to precisely determine the midpoint and slope of the response. The
$[EC_{50}]$ predictions faithfully match the data over four orders of magnitude.
Because each additional mutation lowers the $[EC_{50}]$ by approximately one
decade, the analytic form \eref[eqEC50Value] implies that $\epsilon$ increases
by roughly $5~\kB T$ per mutation, enabling the ion channel to open more easily.
In addition to the L251S mutation considered here, another mutation (L251T) has
also been found to shift $[EC_{50}]$ by a constant logarithmic amount (see
Supporting Information section A.4) \cite{Filatov1995}. We also note that many
biological systems logarithmically tune their responses by altering the energy
difference between two allosteric states, as seen through processes such as
phosphorylation and calmodulin binding \cite{Olsman2016}. This may give rise to an
interesting interplay between physiological time scales where such processes occur and
evolutionary time scales where traits such as the $[EC_{50}]$ may be accessed
via mutations like those considered here \cite{Milo2007}.

Lastly, the Hill coefficients of the entire class of mutants all lie between 1.5
and 2.0 except for the $n=3$ mutant whose dose-response curve in
\fref[figNormalizedCurrent]\letter{A} is seen to be flatter than the MWC
prediction. We also note that if the L251S mutation moderately perturbs the
$\KOWT$ and $\KCWT$ values, it would permit fits that more finely attune to the
specific shape of each mutant's data set. That said, the dose-response curve for
the $n=3$ mutant could easily be shifted by small changes in the measured values
and hence without recourse to error bars, it is difficult to make definitive
statements about the value adopted for $h$ for this mutant.

Note that the simplified expressions
\eref[leakinessValue][dynamicRangeValue][eqEC50Value][eqhValue] for the
leakiness, dynamic range, $[EC_{50}]$, and effective Hill coefficient apply when
$1 \ll e^{-\beta \epsilon} \ll \left(\frac{\KCWT}{\KOWT}\right)^m$, which given
the values of $\KCWT$ and $\KCWT$ for the nAChR mutant class translates to $-22
\lesssim \beta \epsilon \lesssim -5$. The $n=1$, 2, and 3 mutants all fall
within this range, and hence each subsequent mutation exponentially increases
their leakiness and exponentially decreases their $[EC_{50}]$, while their
dynamic range and effective Hill coefficient remain indifferent to the L251S
mutation. The $\beta \epsilon$ parameters of the $n=0$ and $n=4$ mutants lie at
the edge of the region of validity, so higher order approximations can be used
to more precisely fit their functional characteristics (see Supporting Information section
B).

\subsection*{Heterooligomeric CNGA2 Mutants can be Categorized using an Expanded MWC Model}

The nAChR mutant class discussed above had two equivalent ligand binding sites,
and only the gating free energy $\beta \epsilon$ varied for the mutants we
considered. In this section, we use beautiful data for the olfactory CNGA2 ion
channel to explore the unique phenotypes that emerge from a heterooligomeric ion
channel whose subunits have different ligand binding strengths.

The wild type CNGA2 ion channel is made up of four identical subunits, each with
one binding site for the cyclic nucleotide ligands cAMP or cGMP
\cite{Kusch2014}. Within the MWC model, the  probability that this channel is
open is given by  \eref[eqPOpenGeneral] with $\m = 4$ ligand binding sites (see
\fref[figStatesWeightsBothIonChannels]\letter{B}). Wongsamitkul \textit{et
	al.}~constructed a mutated subunit with lower affinity for ligand and formed
tetrameric CNGA2 channels from different combinations of mutated and wild type
subunits (see \fref[figStatesWeightsCNGA2]) \cite{Wongsamitkul2016}. Since the
mutation specifically targeted the ligand binding sites, these mutant subunits
were postulated to have new ligand dissociation constants but the same free
energy difference $\beta \epsilon$.

\begin{figure}[t]
	\centering \includegraphics[scale=\globalScalePlots]{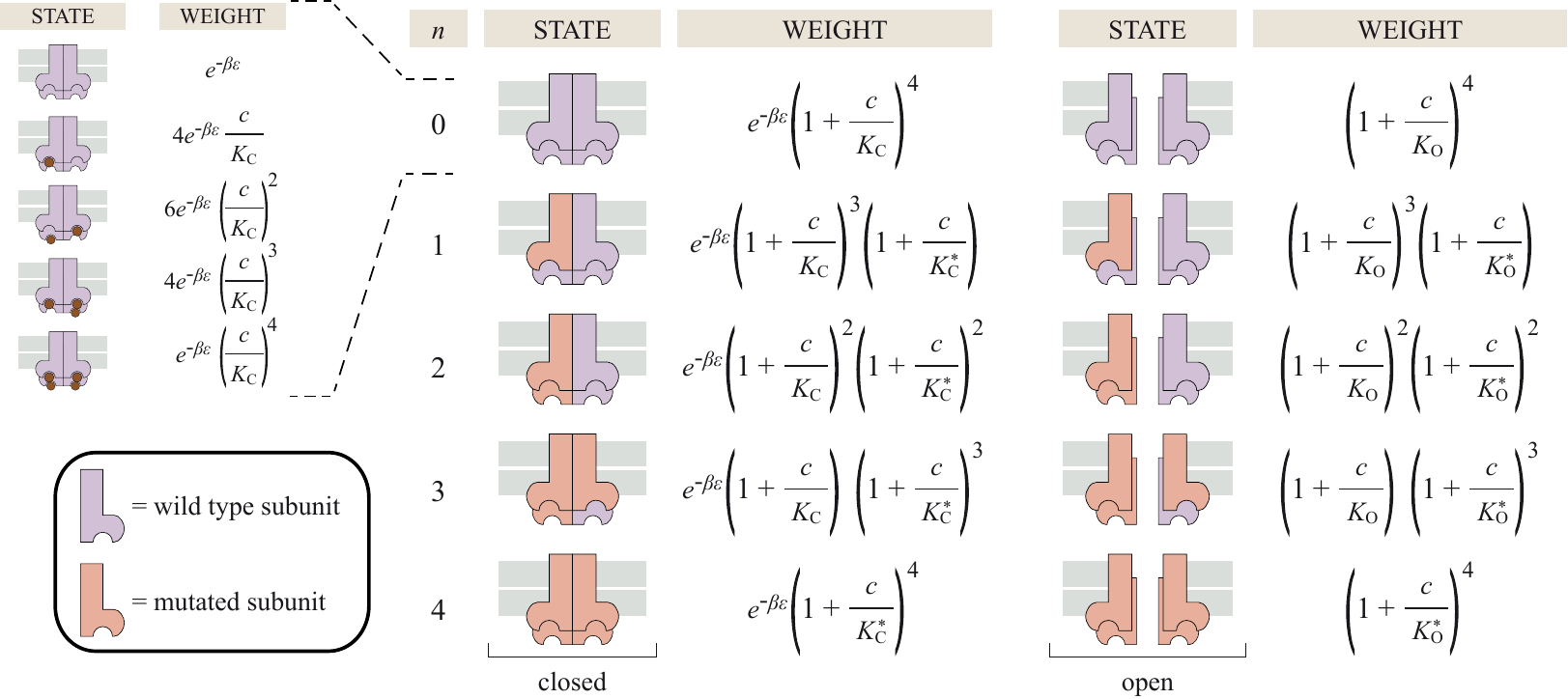}
	\caption{\captionStroke{States and weights for mutant CNGA2 ion channels.} CNGA2 mutants with $\m = 4$ subunits were
		constructed using $n$ mutated (light red) and $\m-n$ wild type subunits
		(purple). The affinity between the wild type subunits to ligand in the open and
		closed states ($\KOWT$ and $\KCWT$) is stronger than the affinity of the
		mutated subunits ($\KOMut$ and $\KCMut$). The weights shown account for all
		possible ligand configurations, with the inset explicitly showing all of the
		closed states for the wild type ($n=0$) ion channel from
		\fref[figStatesWeightsBothIonChannels]\letter{B}. The probability that a
		receptor with $n$ mutated subunits is open is given by its corresponding open
		state weight divided by the sum of open and closed weights in that same row.}
	\label{figStatesWeightsCNGA2}
\end{figure}

We can extend the MWC model to compute the probability $p_{\text{open}}$ that
these CNGA2 constructs will be open. The states and weights of an ion channel
with $\nWT$ mutated subunits (with ligand affinities $\KOMut$ and $\KCMut$) and
$\m-\nWT$ wild type subunits (with ligand affinities $\KOWT$ and $\KCWT$) is
shown in \fref[figStatesWeightsCNGA2], and its probability to be open is given
by
\begin{equation} \label{eqPActive4Mutants}
p_{\text{open}}(c) = \frac{\left(1 + \frac{c}{\KOWT}\right)^{\m-\nWT} \left(1 + \frac{c}{\KOMut}\right)^{\nWT}}{\left(1 + \frac{c}{\KOWT}\right)^{\m-\nWT} \left(1 + \frac{c}{\KOMut}\right)^{\nWT} + e^{-\beta \epsilon} \left(1 + \frac{c}{\KCWT}\right)^{\m-\nWT} \left(1 + \frac{c}{\KCMut}\right)^{\nWT}}.
\end{equation}
Measurements have confirmed that the dose-response curves of the mutant CNGA2
channels only depend on the total number of mutated subunits $n$ and not on the
positions of those subunits (for example both $n=2$ with adjacent mutant subunits
and $n=2$ with mutant subunits on opposite corners have identical dose-response
curves) \cite{Wongsamitkul2016}.

\fref[figNormalizedCurrentCNGA2]\letter{A} shows the normalized current of all
five CNGA2 constructs fit to a single set of $\KOWT$, $\KCWT$, $\KOMut$,
$\KCMut$, and $\epsilon$ parameters. Since the mutated subunits have weaker
affinity to ligand (leading to the larger dissociation constants $\KOMut>\KOWT$
and $\KCMut>\KCWT$), the $[EC_{50}]$ shifts to the right as $n$ increases. As in
the case of nAChR, we can collapse the data from this family of mutants onto a
single master curve using the Bohr parameter $F_{\text{CNGA2}}(c)$ from
\eref[eqBohrParameterGeneral][eqPActive4Mutants], as shown in
\fref[figNormalizedCurrentCNGA2]\letter{B}.

\begin{figure}[t]
		\centering \includegraphics[scale=\globalScalePlots]{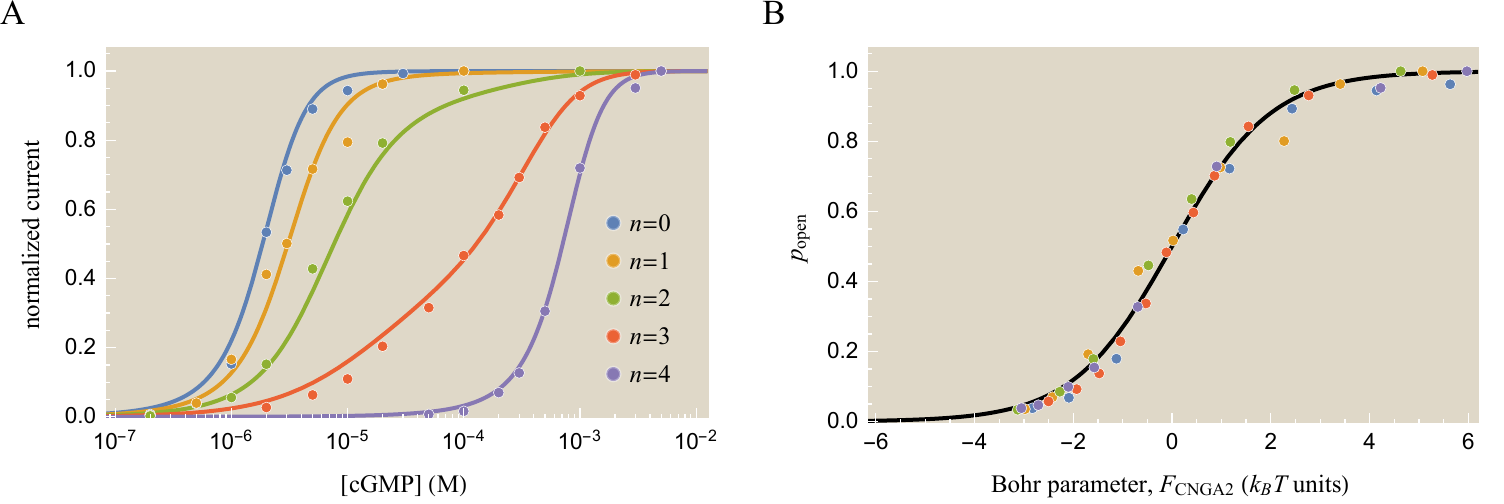}
		\caption{\captionStroke{Normalized currents for CNGA2 ion channels with a varying number $\captionSymbol{n}$
			of mutant subunits.} \letterParen{A} Dose-response curves for CNGA2 mutants comprised of $4-n$ wild type subunits and $n$ mutated subunits with weaker affinity for the ligand cGMP
			\cite{Wongsamitkul2016}. Once the free energy $\epsilon$ and the ligand
			dissociation constants of the wild type subunits ($\KOWT$ and $\KCWT$) and
			mutated subunits ($\KOMut$ and $\KCMut$) are fixed, each mutant is completely
			characterized by the number of mutated subunits $\nWT$ in
			\eref[eqPActive4Mutants]. Theoretical best-fit curves are shown using the
			parameters $\KOWT = 1.2 \times 10^{-6}\,\,\text{M}$, $\KCWT = 20 \times
			10^{-6}\,\,\text{M}$, $\KOMut = 500 \times 10^{-6}\,\,\text{M}$, $\KCMut = 140
			\times 10^{-3}\,\,\text{M}$, and $\beta \epsilon = -3.4$. \letterParen{B} Data from all five mutants collapses onto a single master curve when plotted as a function of
			the Bohr parameter given by \eref[eqBohrParameterGeneral]. See Supporting Information section C 
			for details on the fitting.}
\label{figNormalizedCurrentCNGA2}
\end{figure}

Although we analyze the CNGA2 ion channels in equilibrium, we can glimpse the
dynamic nature of the system by computing the probability of each channel
conformation. \fref[figCNGA2AllStates]\letter{A} shows the ten possible states
of the wild type ($n=0$) channel, the five open states $O_j$ and the five closed
states $C_j$ with $0 \le j \le 4$ ligands bound.
\fref[figCNGA2AllStates]\letter{B} shows how the probabilities of these states
are all significantly shifted to the right in the fully mutated ($n=4$) channel
since the mutation diminishes the channel-ligand affinity. The individual state
probabilities help determine which of the intermediary states can be ignored
when modeling. One extreme simplification that is often made is to consider the
Hill limit, where all of the states are ignored save for the closed, unbound ion
channel ($C_0$) and the open, fully bound ion channel ($O_4$). The drawbacks of
such an approximation are two-fold: (1) at intermediate ligand concentrations
($c \in [10^{-7},10^{-5}]\,\text{M}$ for $n=0$ and $c \in
[10^{-4},10^{-2}]\,\text{M}$ for $n=4$) the ion channel spends at least 10\% of
its time in the remaining states which results in fundamentally different
dynamics than what is predicted by the Hill response and (2) even in the limits
such as $c = 0$ and $c \to \infty$ where the $C_0$ and $O_4$ states dominate the
system, the Hill limit ignores the leakiness and dynamic range of the ion
channel (requiring them to exactly equal 0 and 1, respectively), thereby
glossing over these important properties of the system.

\begin{figure}[t]
	\centering \includegraphics[scale=\globalScalePlots]{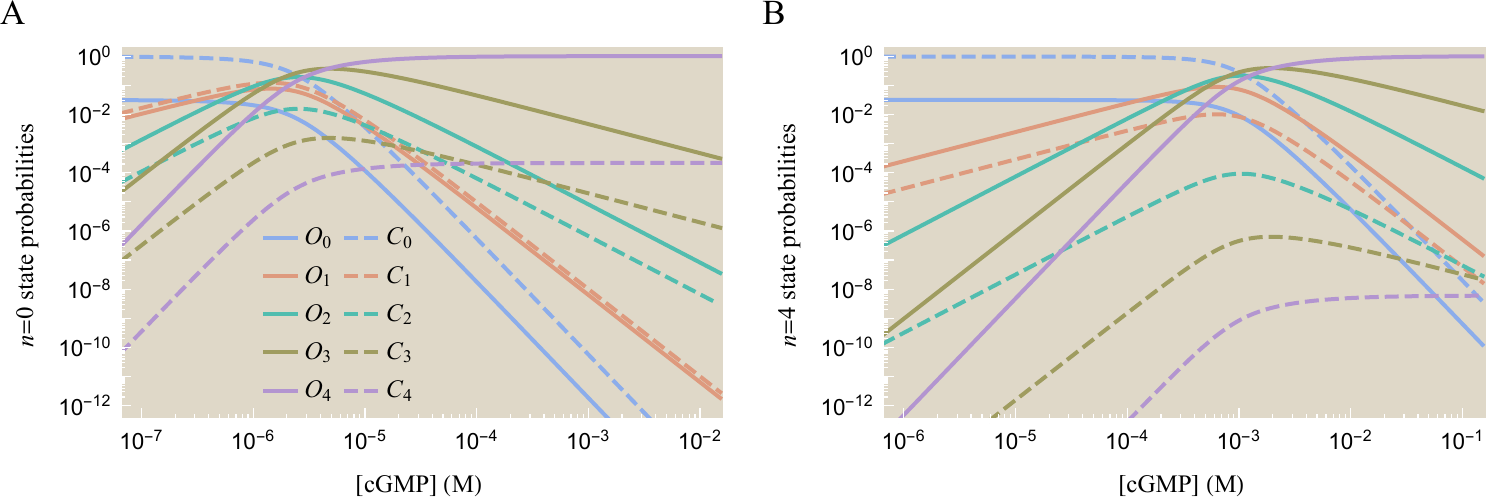}
	\caption{\captionStroke{Individual state probabilities for the wild type and mutant CNGA2
		ion channels.} \letterParen{A} The state probabilities for the wild type 
		($n=0$) ion channel. The subscripts of the open ($O_j$) and
		closed ($C_j$) states represent the number of ligands bound to the channel.
		States with partial occupancy, $1 \le j \le 3$, are most likely to occur in a
		narrow range of ligand concentrations $[\text{cGMP}] \in [10^{-7},
		10^{-5}]\,\text{M}$, outside of which either the completely empty $C_0$ or fully
		occupied $O_4$ states dominate the system. \letterParen{B} The state probabilities 
		for the $n=4$ channel. Because the mutant subunits have a
		weaker affinity to ligand ($\KOMut>\KOWT$ and $\KCMut>\KCWT$), the state
		probabilities are all shifted to the right. } \label{figCNGA2AllStates}
\end{figure}

\subsubsection*{Characterizing CNGA2 Mutants based on Subunit Composition}

We now turn to the leakiness, dynamic range, $[EC_{50}]$, and effective Hill
coefficient $h$ of a CNGA2 ion channel with $n$ mutated and $m-n$ wild
type subunits. Detailed derivations for the following results are provided in
Supporting Information section B:
\begin{align}
\text{leakiness} &= \frac{1}{1 + e^{-\beta \epsilon}} \label{eqLeakinessCNGA2} \\
\text{dynamic range} &= \frac{1}{1 + e^{-\beta \epsilon} \left(\frac{\KOWT}{\KCWT}\right)^{\m-\nWT}\left(\frac{\KOMut}{\KCMut}\right)^{\nWT}} - \frac{1}{1 + e^{-\beta \epsilon}} \label{eqDynamicRangeCNGA2}\\
[EC_{50}] &\approx \begin{cases} 
e^{-\beta \epsilon/m} \KOWT & n=0 \\
e^{-2\beta\epsilon/m} \frac{\KOWT \KOMut}{\KOWT + \KOMut} & n=\frac{m}{2} \\
e^{-\beta \epsilon/m} \KOMut & n=m
\end{cases} \label{eqEC50ValueCNGA2}\\
h &\approx \begin{cases} 
\m & \,\,\,\,\,\,\,\,\,n=0 \\
\frac{\m}{2} & \,\,\,\,\,\,\,\,\,n=\frac{m}{2} \\
\m & \,\,\,\,\,\,\,\,\,n=m.
\end{cases}\label{eqhValueCNGA2}
\end{align}
Note that we recover the original MWC model results
\eref[eqLeakinessDef][eqDynamicRangeDef][eqEC50Value][eqhValue] for the $n=0$
wild type ion channel. Similarly, the homooligomeric $\nWT=\m$ channel is also
governed by the MWC model with $\KOWT \to \KOMut$ and $\KCWT \to \KCMut$. We also
show the $[EC_{50}]$ and $h$ formulas for the $n=\frac{\m}{2}$ case to demonstrate the
fundamentally different scaling behavior that this heterooligomeric channel
exhibits with the MWC parameters.

\fref[figLeakDynEC50EffHCNGA2]\letter{A} shows that all of the CNGA2 mutants
have small leakiness, which can be understood from their small $\epsilon$ value
and \eref[eqLeakinessCNGA2]. In addition, the first term in the dynamic range
\eref[eqDynamicRangeCNGA2] is approximately 1 because the open state affinities
are always smaller than the closed state affinities by at least a factor of ten,
which is then raised to the fourth power. Thus, all of the mutants have a
large dynamic range as shown in \fref[figLeakDynEC50EffHCNGA2]\letter{B}.
Experimentally, single channel measurements confirmed that the wild type $n=0$
channel is nearly always closed in the absence of ligand; in the opposite limit
of saturating cGMP, it was found that $p_{\text{open}}^{\text{max}}=0.99$ for
both the $n=0$ and $n=\m$ ion channels (see Supporting Information section
C.2) \cite{Wongsamitkul2016}.

\begin{figure}[t!]
	\centering \includegraphics[scale=\globalScalePlots]{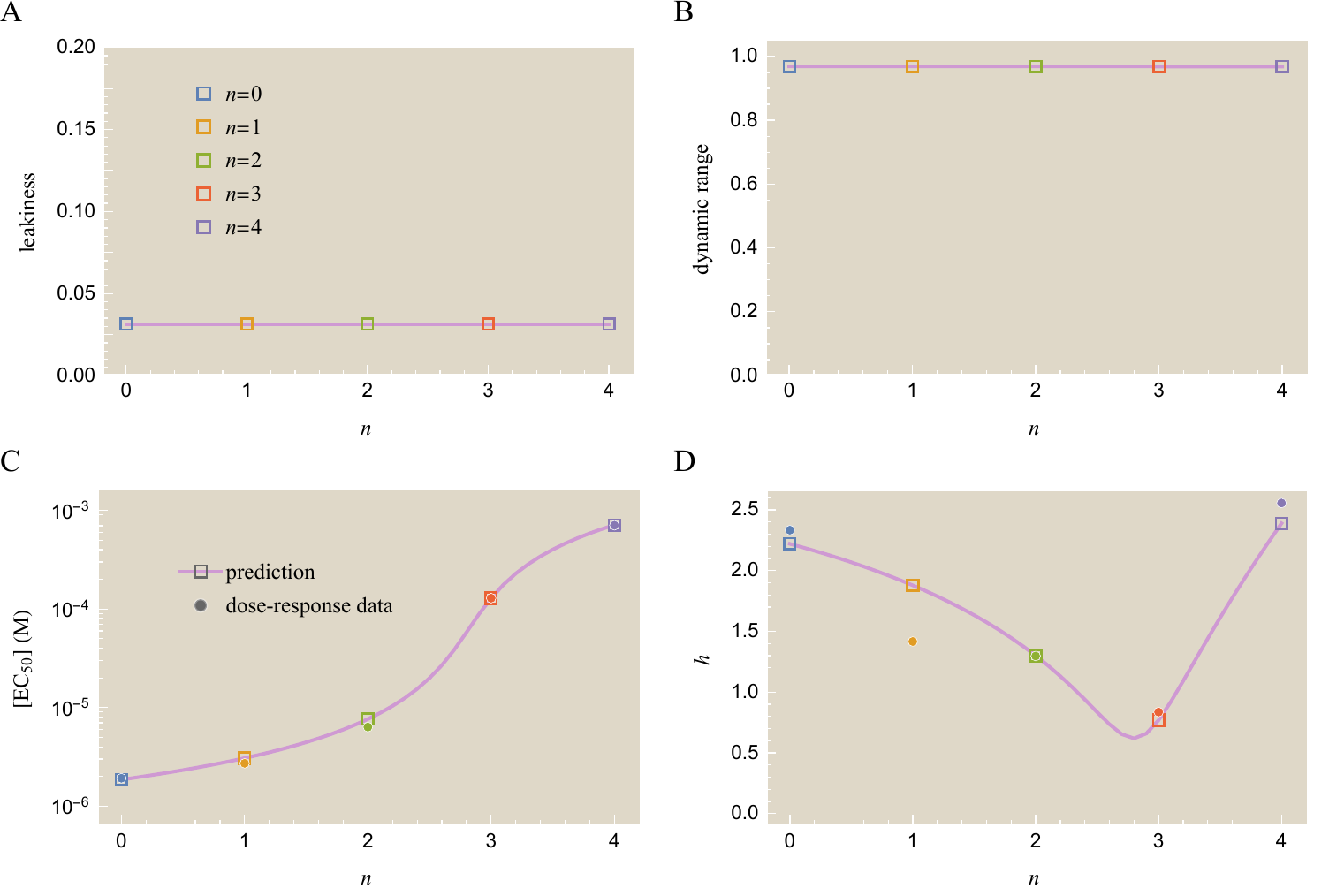}
	\caption{\captionStroke{Theoretical prediction and experimental measurements for mutant CNGA2
		ion channel characteristics.} The open squares represent the five mutant
		ion channels in \fref[figNormalizedCurrentCNGA2] with $n$ mutated subunits.
		\letterParen{A} All ion channels have small leakiness. \letterParen{B} The
		dynamic range of all channels is near the maximum possible value of unity,
		indicating that they rarely open in the absence of ligand and are always open
		in the presence of saturating ligand concentrations. \letterParen{C} The
		$[EC_{50}]$ increases non-uniformly with the number of mutant subunits. Also
		shown are the measured values (solid circles) interpolated from the data.
		\letterParen{D} The effective Hill coefficient has a valley due to the
		competing influences of the wild type subunits (which respond at $\mu$M ligand
		concentrations) and the mutant subunits (which respond at mM concentrations).
		Although the homotetrameric channels ($n=0$ and $n=4$) both have sharp
		responses, the combined effect of having both types of subunits ($n=1$, 2, and
		3) leads to a flatter response.} \label{figLeakDynEC50EffHCNGA2}
\end{figure}

The $[EC_{50}]$ and effective Hill coefficient $h$ are shown in
\fref[figLeakDynEC50EffHCNGA2]\letter{C} and \letter{D}. In contrast to the
nAChR case, where each additional mutation decreased $[EC_{50}]$, each CNGA2
mutation tends to increase $[EC_{50}]$, although not by a uniform amount. The
effective Hill coefficient has a particularly complex behavior, first decreasing
with each of the first three subunit mutations and then finally increasing back
to the wild type level for the fully mutated ion channel. To explain this
decrease in $h$ for the heterooligomeric channels, we first note that the wild
type $n=0$ channel has a sharp response about its $[EC_{50}] \approx e^{-\beta
	\epsilon/\m} \KOWT$ while the fully mutated $n=\m$ channel has a sharp response
about $[EC_{50}] \approx e^{-\beta \epsilon/\m} \KOMut$. Roughly speaking, the
response of the heterooligomeric channels with $1 \le n \le 3$ will occur
throughout the full range between $e^{-\beta \epsilon/\m} \KOWT$ and $e^{-\beta
	\epsilon/\m} \KOMut$, which causes the dose-response curves to flatten out and
leads to the smaller effective Hill coefficient. Such behavior could
influence, for example, the response of the heterooligomeric nAChR ion channel
if the two acetylcholine binding pockets diverged to have different ligand
affinities.

Although we have focused on the particular mutants created by Wongsamitkul
\textit{et al.}, it is straightforward to apply this framework to other types of
mutations. For example, in Supporting Information section B.2 we
consider the effect of modifying the $\KOWT$ and $\KCWT$ parameters of all four
CNGA2 channels simultaneously. This question is relevant for physiological CNGA2
channels where a mutation in the gene would impact all of the subunits
in the homooligomer, in contrast to the Wongsamitkul constructs
where the four subunits were stitched together within a single gene. We find
that when $\KOWT$ and $\KCWT$ vary for all subunits, the leakiness, dynamic
range, and effective Hill coefficient remain nearly fixed for all parameter
values, and that only the $[EC_{50}]$ scales linearly with $\KOWT$ as per
\eref[eqEC50Value]. In order to affect the other properties, either the gating
energy $\beta \epsilon$ or the number of subunits $m$ would need to be changed.

\subsection*{Extrapolating the Behavior of a Class of Mutants}

In this section, we explore how constant trends in both the nAChR and CNGA2 data
presented above provide an opportunity to characterize the full class of mutants
based on the dose-response curves from only a few of its members. Such trends
may well be applicable to other ion channel systems, enabling us to
theoretically probe a larger space of mutants than what is available from
current data.

First, we note that because the $[EC_{50}]$ of the five nAChR mutants fell on a
line in \fref[figCharacteristics]\letter{C}, we can predict the response of the
entire class of mutants by only considering the dose-response curves of two of
its members and extrapolating the behavior of the remaining mutants using linear
regression. Experimentally, such a characterization arises because the L251S
mutation acts nearly identically and independently across subunits to change the
gating free energy of nAChR \cite{Auerbach2012, Labarca1995, Filatov1995}. This
implies that mutating $n$ subunits would yield an ion channel with gating energy
\begin{equation} \label{eqExtrapolatingEpsilonValues}
\epsilon^{(n)} = \epsilon^{(0)} + n \Delta \epsilon,
\end{equation}
where $\epsilon^{(0)}$ is the wild type gating energy and $\Delta \epsilon$ is
the change in free energy per mutation. This functional form is identical to the
mismatch model for transcription factor binding, where each additional
mutation-induced mismatch adds a constant energy cost to binding
\cite{Berg2004}. \fref[figNormalizedCurrentv2]\letter{A} demonstrates how
fitting only the $n=0$ and $n=4$ dose-response curves (solid lines) accurately
predicts the behavior of the $n=1$, 2, and 3 mutants (dashed lines). In Supporting Information section
D, we carry out such predictions using all possible
input pairs.

\begin{figure}[t]
	\centering \includegraphics[scale=\globalScalePlots]{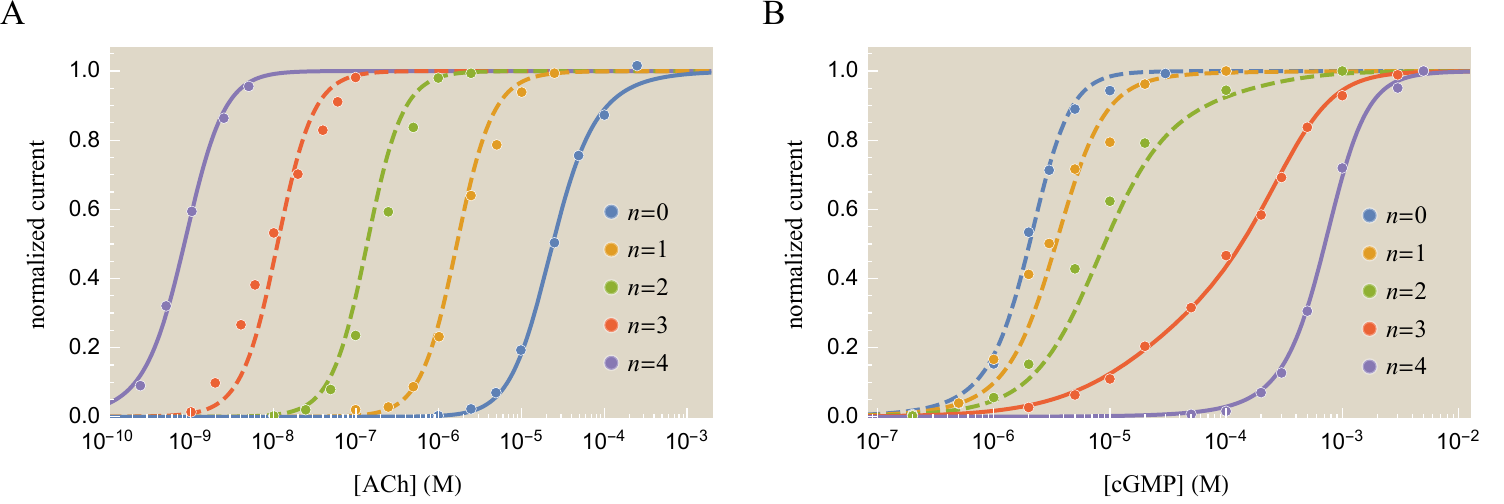}
	\caption{\captionStroke{Predicting the dose-response of a class of mutants using a subset of its members.}
		\letterParen{A} The MWC parameters of the nAChR mutants can be fixed
		using only two data sets (solid lines), which together with
		\eref[eqExtrapolatingEpsilonValues] predict the dose-response curves of the
		remaining mutants (dashed lines). \letterParen{B} For the CNGA2 channel, the properties of both the
		wild type and mutant subunits can also be fit using two data sets,
		accurately predicting the responses of the remaining three mutants. Supporting Information section
		D demonstrates the results of using alternative pairs
		of mutants to fix the thermodynamic parameters in both systems.
	} \label{figNormalizedCurrentv2}
\end{figure}

We now turn to the CNGA2 ion channel where, once the $\KOWT$, $\KCWT$, $\KOMut$,
$\KCMut$, and $\epsilon$ parameters are known, the dose-response curve of any
mutant can be predicted by varying $\nWT$ in \eref[eqPActive4Mutants].
\fref[figNormalizedCurrentv2]\letter{B} demonstrates that the wild type ion
channel ($n=0$) and the ion channel with only one mutated subunit ($n=1$) can
accurately predict the dose-response curves of the remaining mutants. Supporting Information section
D explores the resulting predictions using all possible
input pairs.

\subsection*{MWC Model allows for Degenerate Parameter Sets}

One critical aspect of extracting model parameters from data is that degenerate
sets of parameters may yield identical outputs, which suggests that there are
fundamental limits to how well a single data set can fix parameter values
\cite{Transtrum2015, Milo2007}. This phenomenon, sometimes dubbed
``sloppiness,'' may even be present in models with very few parameters such as
the MWC model considered here. \fref[fignAChRandCNGA2Sloppiness] demonstrates
the relationship between the best-fit parameters within the nAChR and CNGA2
systems. For concreteness, we focus solely on the nAChR system.

\begin{figure}[t]
	\centering \includegraphics[scale=\globalScalePlots]{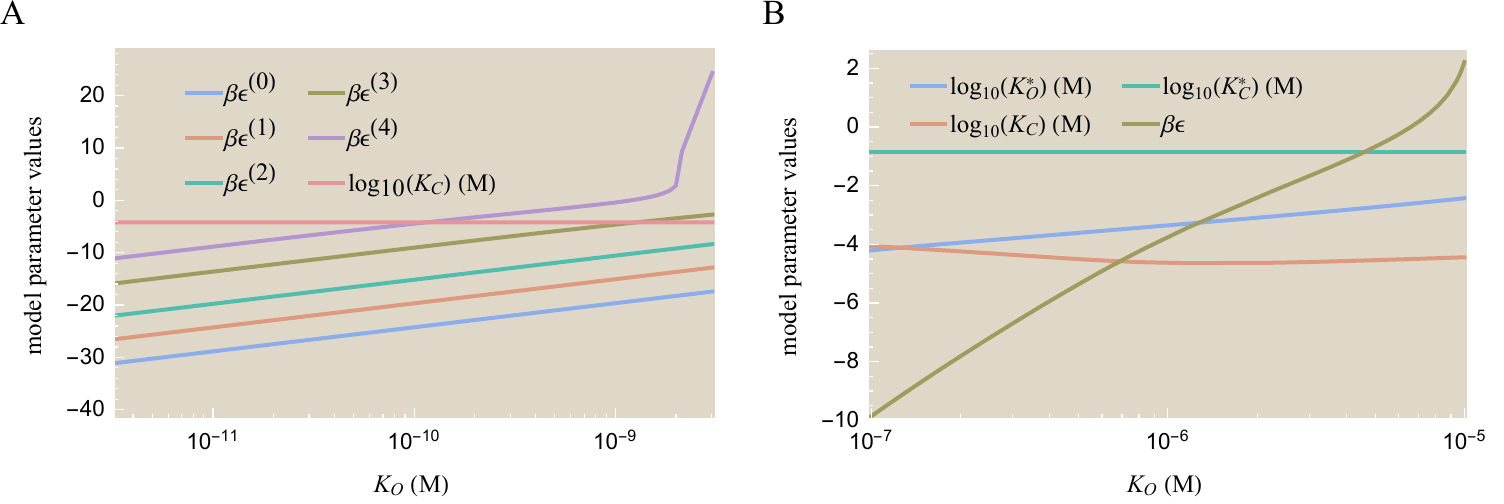}
	\caption{\captionStroke{Degenerate parameter sets for nAChR and CNGA2 model fitting.}
		Different sets of biophysical parameters can yield the same system response.
		\letterParen{A} Data for the nAChR system in \fref[figNormalizedCurrent] is
		fit by constraining $\KOWT$ to the value shown on the $x$-axis. The remaining
		parameters can compensate for this wide range of $\KOWT$ values.
		\letterParen{B} The CNGA2 system in \fref[figNormalizedCurrentCNGA2] can
		similarly be fit by constraining the $\KOWT$ value, although fit quality
		decreases markedly outside the narrow range shown. Any set of parameters shown
		for either system leads to responses with $R^2 > 0.96$.
	} \label{fignAChRandCNGA2Sloppiness}
\end{figure}

After fixing the value of $\KOWT$ (to that shown on the $x$-axis of
\fref[fignAChRandCNGA2Sloppiness]\letter{A}), the remaining parameters are
allowed to freely vary in order to best fit the nAChR data. Although every value
of $\KOWT \in \left[ 10^{-11} \,\hyphen\, 10^{-9} \right]\text{M}$ yields a
nearly identical response curve in excellent agreement with the data (with a
coefficient of determination $R^2 > 0.96$), we stress that dissociation
constants are rarely found in the range $\KOWT \ll 10^{-10}\,\,\text{M}$. In
addition, a dissociation constant above the nM range, $\KOWT \gg
10^{-9}\,\,\text{M}$, cannot fit the data well and is therefore invalidated by
the model. Thus, we may suspect the true parameter values will fall around the
interval $\KOWT \in \left[10^{-10} \,\hyphen\, 10^{-9}\right]\text{M}$ for the
nAChR system. $\KOWT$ could ultimately be fixed by measuring the leakiness
\eref[eqLeakinessDef] (and thereby fixing $\beta \epsilon$) for any of the ion
channel mutants.

Two clear patterns emerge from \fref[fignAChRandCNGA2Sloppiness]\letter{A}: (1)
The value of $\KCWT$ is approximately constant for all values of $\KOWT$ and (2)
the five free energies all vary as $\beta \epsilon^{(n)} = 2 \log \left( \KOWT
\right) + n \left( \text{constant} \right)$. This suggests that $\KCWT$ and
$e^{-\beta \epsilon/2} \KOWT$ are the fundamental parameters combinations of the
system. In Supporting Information section C.4, we discuss how these parameter
combinations arise from our model.

We end by noting that the notion of sloppiness, while detrimental to fixing the
physical parameter values of a system, nevertheless suggests that multiple
evolutionary paths may lead to optimal ion channel phenotypes, providing another
mechanism by which allostery promotes a protein's capacity to adapt
\cite{Raman2016}.

\section*{Discussion}

There is a deep tension between the great diversity of biological systems and
the search for unifying perspectives to help integrate the vast data that has
built up around this diversity. Several years ago at the Institut Pasteur, a
meeting was convened to celebrate the 50$^\text{th}$ anniversary of the allostery concept,
which was pioneered in a number of wonderful papers in the 1960s and which since
then has been applied to numerous biological settings \cite{MONOD1963,
	MONOD1965, Koshland1966, Changeux2012, Gerhart2014}. Indeed, that meeting
brought together researchers working in physiology, neuroscience, gene
regulation, cell motility, and beyond, making it very clear that allostery has
great reach as a conceptual framework for thinking about many of the key
macromolecules that drive diverse biological processes.

In this paper, we have built upon this significant previous work and explored
how the Monod-Wyman-Changeux model can serve as a unifying biophysical framework
for whole suites of ion channel mutants (see
\fref[figNormalizedCurrent][figNormalizedCurrentCNGA2]). Specifically, we used
two well-studied ligand-gated ion channels to explore the connection between
mutations, the MWC parameters, and the full spectrum of dose-response curves
which may be induced by those mutations. In addition, we have shown how earlier
insights into the nature of ``data collapse'' in the context of bacterial
chemotaxis and quorum sensing \cite{Keymer2006, Swem2008} can be translated into
the realm of ion channels. By introducing the Bohr parameter, we are able to
capture the non-linear combination of thermodynamic parameters which governs the
system's response.

For both the nAChR and CNGA2 ion channels, we showed that precise predictions of
dose-response curves can be made for an entire class of mutants by only using
data from two members of this class (\fref[figNormalizedCurrentv2]). In other
words, the information contained in a single dose-response curve goes beyond
merely providing data for that specific ion channel. Ultimately, because the
total space of all possible mutants is too enormous for any significant fraction
to be explored experimentally, we hope that a coupling of theory with experiment
will provide a step towards mapping the relation between channel function
(phenotype) and the vast space of protein mutations.

Moreover, we used the MWC model to determine analytic formulas for key
properties such as the leakiness, dynamic range, $[EC_{50}]$, and the effective
Hill coefficient, which together encapsulate much of the information in
dose-response curves. These relationships tie into the extensive knowledge about
phenotype-genotype maps \cite{Gerland2002, Berg2004, Raman2016}, enabling us to
quantify the trade-offs inherent in an ion channel's response. For example, when
modifying the ion channel gating free energy, the changes in the leakiness and
$[EC_{50}]$ are always negatively correlated (\fref[figCharacteristics]),
whereas modifying the ligand binding domain will not affect the leakiness but
may change the $[EC_{50}]$ (\fref[figLeakDynEC50EffHCNGA2] and Supporting Information section
B.2). The ability to navigate between the
genotype and phenotype of proteins is crucial in many bioengineering settings,
where site-directed mutagenesis is routinely employed to find mutant proteins
with specific characteristics (e.g. a low leakiness and large dynamic range)
\cite{Flytzanis2014, Vogt2015, Wietek2014}.

While general formulas for these phenotypic properties were elegantly derived in
earlier work \cite{Martins2011}, we have shown that such relations can be
significantly simplified in the context of ion channels where $1 \ll e^{-\beta
	\epsilon} \ll \left(\frac{\KCWT}{\KOWT}\right)^m$ (see
\eref[leakinessValue][dynamicRangeValue][eqEC50Value][eqhValue]). This
approximation is applicable for the range of parameters spanned by both the
nAChR and CNGA2 systems, and we suspect it may hold for many other ion channels.
These formulas provide a simple, intuitive framework to understand the effects
of mutations. For example, they suggest: (1) Channel pore mutations that
increase $\epsilon$ will exponentially increase the leakiness of the channel,
although the constraint $1 \ll e^{-\beta \epsilon}$ ensures that this leakiness
will still be small. Ligand domain mutations are not expected to affect
leakiness at all. (2) Channel pore mutations will exponentially decrease the
$[EC_{50}]$ with increasing $\epsilon$, although this effect is diminished for
ion channels with multiple subunits. For mutations in the ligand binding domain,
the $[EC_{50}]$ will increase linearly with the dissociation constant $\KOWT$
between the ligand and the open ion channel (see Supporting Information section
B.2). (3) Neither the dynamic range nor
the effective Hill coefficient will be significantly perturbed by either type of
mutation. (4) Transforming a homooligomeric ion channel into a heterooligomer
can generate a significantly flatter response. For example, even though the
CNGA2 channel comprised of either all wild type or all mutant subunits had a
very sharp response, a channel comprised of both subunits had a smaller
effective Hill coefficient (see \fref[figLeakDynEC50EffHCNGA2]\letter{D}).

The framework presented here could be expanded in several exciting ways. First,
it remains to be seen whether channel pore mutations and ligand binding domain
mutations are completely independent, or whether there is some cross-talk
between them. This question could be probed by creating a channel pore mutant
(whose dose-response curves would fix its new $\tilde{\epsilon}$ values), a
ligand domain mutant (whose new $\KOMut$ and $\KCMut$ values would be
characterized from its response curve), and then creating the ion channel with
both mutations. If these two mutations are independent, the response of the
double mutant can be predicted \textit{a priori} using $\tilde{\epsilon}$, $\KOMut$, and
$\KCMut$.

We also note that the MWC model discussed here does not consider several
important aspects relating to the dynamics of ion channel responses. Of
particular importance is the phenomenon of desensitization which significantly
modifies an ion channel's response in physiological settings
\cite{Edelstein1996, Plested2016}. In addition, some ion channels have multiple
open and closed conformations \cite{Karpen1997, VanSchouwen2016, Jackson1986} while other
channels exhibit slow switching between the channel states \cite{Bicknell2016}.
Exploring these additional complexities within generalizations of the MWC model
would be of great interest.

Finally, we believe that the time is ripe to construct an explicit biophysical
model of fitness to calculate the relative importance of mutation, selection,
and drift in dictating the diversity of allosteric proteins such as the ion
channels considered here. Such a model would follow in the conceptual footsteps
laid in the context of fitness of transcription factors binding
\cite{Gerland2002, Berg2004, Lynch2015}, protein folding stability
\cite{Zeldovich2008, Serohijos2014, Peleg2014}, and influenza evolution
\cite{Luksza2014}. This framework would enable us to make precise, quantitative
statements about intriguing trends; for example, nearly all nAChR pore mutations
appear to increase a channel's leakiness, suggesting that minimizing leakiness
may increase fitness \cite{Auerbach2012}. One could imagine that computing
derivatives such as $\frac{d\text{leakiness}}{d\epsilon}$, a quantity analogous
to the magnetic susceptibility in physics, would be correlated with how likely
an $\epsilon$ mutation is to be fixed. The goal of such fitness functions is to
map the complexity of the full evolutionary space (i.e. changes to a protein
amino acid sequence) onto the MWC parameters, and then the fitness function
determines how these parameters evolve in time. In this way, the complexity of
sequence and structure would fall onto the very low dimensional space governed by
$\epsilon$, $\KOWT$, and $\KCWT$.

\begin{acknowledgement}

It is with sadness that we dedicate this paper to the memory of Klaus Schulten
with whom one of us (RP) wrote his first paper in biophysics. Klaus was an
extremely open and kind man, a broad and deep thinker who will be deeply missed.
We thank Stephanie Barnes, Nathan Belliveau, Chico Camargo, Griffin Chure, Lea
Goentoro, Michael Manhart, Chris Miller, Muir Morrison, Manuel Razo-Mejia, Noah
Olsman, Allyson Sgro, and Jorge Za\~{n}udo for their sharp insights and valuable
feedback on this work. We are also grateful to Henry Lester, Klaus Benndorf, and
Vasilica Nache for helpful discussions as well as sharing their ion channel
data. All plots were made in \textit{Mathematica} using the CustomTicks package
\cite{Caprio2005}. This work was supported by La Fondation Pierre-Gilles de
Gennes, the Rosen Center at Caltech, the National Science Foundation under NSF
PHY11-25915 at the Kavli Center for Theoretical Physics, and the National
Institutes of Health through DP1 OD000217 (Director's Pioneer Award), 5R01
GM084211C, R01 GM085286, and 1R35 GM118043-01 (MIRA).

\end{acknowledgement}

\begin{suppinfo}
The following Supporting Information is available:
\begin{itemize}
  \item Details on aforementioned derivations and calculations (PDF)
  \item All of the plots in the main text can be reproduced with the supplementary \textit{Mathematica} notebook. The dose-response data for both the nAChR and CNGA2 systems may be found in this notebook (ZIP)
\end{itemize}
\end{suppinfo}
\setcounter{page}{2}
\setcounter{figure}{0}
\setcounter{table}{0}
\setcounter{equation}{0}

\appendix

\section{Additional Ion Channel Data} \label{moreData}

In this section, we explore some of the additional experimental measurements
available for the nAChR and CNGA2 systems studied above and elaborate on several
calculations mentioned in the text. In \ref{AppendixDynamics}, we analyze the
time scale required for an ion channel to reach equilibrium. In
\ref{morenAChRMutants}, we present data on additional L251S nAChR mutants. Using
these mutants, we examine the approximation made in the text that only the
\textit{total} number of mutations, and not the identity of the subunits
mutated, influences the resulting nAChR mutant behavior. In \ref{appendixPopen},
we examine $p_{\text{open}}(c)$ for the classes of ion channels considered in
the text and comment on how this probability differs from the normalized
current. In \ref{nAChRleu9Thr}, we examine data from a similar class of L251T
mutations and show that their qualitative behavior is similar to the L251S
mutants. In \ref{appendixCombiningIonChannels}, we discuss measurements of
combinations of CNGA2 ion channels.


\subsection{Dynamics Towards Equilibrium} \label{AppendixDynamics}

\begin{figure}[h]
	\centering \includegraphics[scale=\globalScalePlots]{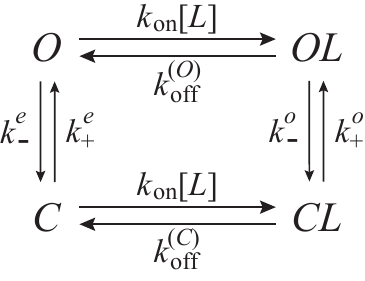} 
	\caption{\captionStroke{Rates for an ion channel with one ligand binding site.} The ion
		channel tends to transition from the closed ($C$) state to the open ($O$) state
		after binding to ligand ($L$). We assume both ion channel states have the same
		diffusion-limited on-rate $k_{\text{on}} = 10^9\,\frac{1}{\text{M} \cdot
			\text{s}}$. The remaining rates of the bound states should satisfy
		$k_{\text{off}}^{(C)} > k_{\text{off}}^{(O)}$ and $k_+^{o} > k_-^{o}$ so that
		ligand binding drives the ion channel to the open state $OL$.}
	\label{figOneLigandRates}
\end{figure}

In this section we derive an exact expression for the time constant for which an
ion channel with one ligand binding site will come to equilibrium. This analysis
can be readily extended numerically to include multiple ligand binding sites.

\fref[figOneLigandRates] shows the rates between the four possible ion channel
states: the unbound open ($O$) and closed ($C$) states as well as the bound open
($OL$) and closed ($CL$) states. We assume that there is a sufficient ligand
$[L]$ in the system so that when the ligand binds to the ion channels its
concentration does not appreciably diminish. Hence the rate equations for the
system can be written in matrix form (with bold denoting vectors and matrices)
as
\begin{equation} \label{eqTransientRateOfChange}
\frac{d\boldsymbol{E}}{dt}= \boldsymbol{K} \boldsymbol{E}
\end{equation}
where the right hand side represents the product of the transition matrix
\begin{equation} \label{eqDynamicsMatrix}
\boldsymbol{K}=\left(
\begin{array}{cccc}
-(k_+^{e} + k_{\text{on}} [L]) & k_{\text{off}}^{(C)} 				& k_-^{e} 			
& 0 \\
k_{\text{on}} [L]				 & - (k_+^{o} + k_{\text{off}}^{(C)}) 	& 0 
& k_-^{o} \\
k_+^{e} 						 & 0 									& -(k_-^{e} + k_{\text{on}} [L]) 
& k_{\text{off}}^{(O)} \\
0 								 & k_+^{o} 								& k_{\text{on}} [L] 
& -(k_-^{o} + k_{\text{off}}^{(O)}) \\
\end{array}
\right)
\end{equation}
and the vector representing the occupancy of each ion channel state
\begin{equation} \label{eqDynamicsMatrixVector}
\boldsymbol{E}=\left(
\begin{array}{c}
\left[C\right] \\
\left[CL\right] \\
\begin{array}{c}
\left[O\right] \\
\left[OL\right] \\
\end{array}
\\
\end{array}
\right).
\end{equation}

The matrix $\boldsymbol{K}$ can be decomposed as
\begin{equation}
\boldsymbol{K}=\boldsymbol{V}^{-1} \boldsymbol{\Lambda} \boldsymbol{V}
\end{equation}
where \(\boldsymbol{V}\)'s columns are the eigenvectors of \(\boldsymbol{K}\)
and \(\boldsymbol{\Lambda}\) is a diagonal matrix whose entries are the
eigenvalues of \(\boldsymbol{K}\). In general, it is known that the eigenvalues
of such a matrix $\boldsymbol{K}$ representing the dynamics of any graph such as
\fref[figOneLigandRates] has one eigenvalue that is 0 while the remaining
eigenvalues are non-zero and have negative real parts \cite{Mirzaev2013}.
(Indeed, because all of the columns of \(\boldsymbol{K}\) add up to zero,
\(\boldsymbol{K}\) is not full rank and hence one of its eigenvalues must be
zero.) Defining the vector
\begin{equation}
\boldsymbol{\tilde{E}} \equiv \boldsymbol{V} \boldsymbol{E}=\left(
\begin{array}{c}
\tilde{E}_1 \\
\tilde{E}_2 \\
\tilde{E}_3 \\
\tilde{E}_4 \\
\end{array}
\right),
\end{equation}
\eref[eqTransientRateOfChange] can be rewritten as
\begin{equation}
\frac{d\boldsymbol{\tilde{E}}}{dt}= \boldsymbol{\Lambda} \boldsymbol{\tilde{E}}.
\end{equation}
If the eigenvalues of $\Lambda$ are $\lambda _1$, $\lambda _2$, $\lambda _3$,
and 0, then \(\tilde{E}_j=c_je^{\lambda _jt}\) for \(j=1,2,3\) and
\(\tilde{E}_4=c_4\) where the \(c_j\)'s are constants determined by initial
conditions. Since the \(\tilde{E}_j\)'s are linear combinations of
\(\left[C\right],\left[CS\right],\left[O\right],\) and \(\left[OS\right]\), this
implies that the \(-\frac{1}{\lambda _1}$,$-\frac{1}{\lambda _2},\) and
\(-\frac{1}{\lambda _3}\) (or $-\frac{1}{\Re(\lambda_j)}$ if the eigenvalues are
complex) are the time scales for the normal modes of the system to come to
equilibrium, with the largest value representing the time scale $\tau$ for the
entire system to reach equilibrium,
\begin{equation} \label{eqtauExactAppendix}
\tau=\max \left(-\frac{1}{\lambda _1},-\frac{1}{\lambda _2},-\frac{1}{\lambda _3}\right).
\end{equation}

Although the eigenvalues of this matrix can be calculated in closed form, as
roots of a cubic function, the full expression is complicated. Instead, we write
the Taylor expansion of $\lambda_1$, $\lambda_2$, and $\lambda_3$ in the limit
$k_+^{o} \to \infty$, since we suspect that the transition from $CS \to OS$ is
extremely fast. In this limit, the $\lambda_j$s take the forms
\begin{align}
\lambda_1 &= -(k_{\text{off}}^{(O)} + k_{\text{on}} [L]) + O\left(\frac{1}{k_+^{o}}\right) \label{eqLambda1}\\
\lambda_2 &= -(k_{\text{off}}^{(C)} + k_-^{o} + k_+^{o}) + O\left(\frac{1}{k_+^{o}}\right) \label{eqLambda2}\\
\lambda_3 &= -(k_{\text{on}} [L] + k_-^{e} + k_+^{e}) + O\left(\frac{1}{k_+^{o}}\right). \label{eqLambda3}
\end{align}

\fref[figDynamics] shows an example of how the system attains its equilibrium
starting from a random initial condition. The exact time scale
\eref[eqtauExactAppendix] using the matrix eigenvalues leads to $\tau = 1.1
\times 10^{-3}\,\text{s}$, which is very close to the approximation using
\eref[eqLambda1][eqLambda2][eqLambda3] which yields $\tau^{(approx)} = 1.0
\times 10^{-3}\,\text{s}$. The exact time scale is shown in \fref[figDynamics]
as a dashed line, and states achieve near total equilibrium by
$t=10^{-2}\,\text{s}$. 

As a point of reference for this time scale described above for the system to
come to equilibrium, there are two other relevant times scales for an ion
channel: (1) the time scale for an ion channel to switch between the open and
closed conformations and (2) the time scale for an ion channel to stay in its
open conformation before switching to the closed conformation. The former occurs
on the microsecond scale for nAChR \cite{Unwin1995}, while the latter occurs on
the millisecond scale \cite{Labarca1995, Wongsamitkul2016}. Thus, the time to
transition between the closed and open conformations can be ignored, and the
system reaches equilibrium after only a few transitions between the open and
closed states.

\begin{figure}[h]
	\centering \includegraphics[scale=\globalScalePlots]{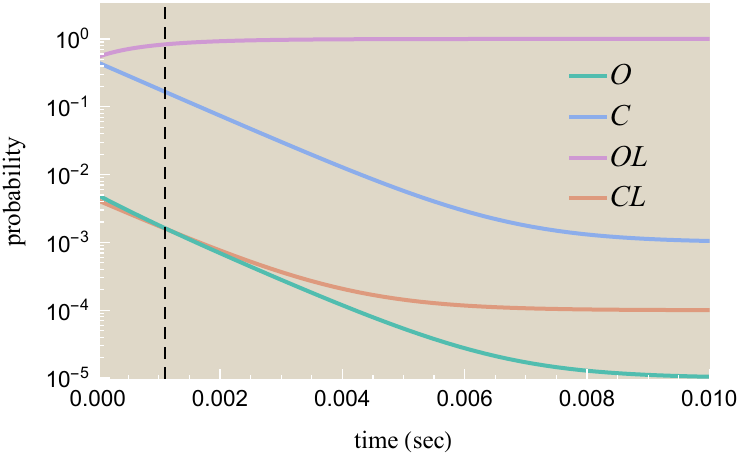}
	\caption{\captionStroke{Kinetics of a system heading towards equilibrium.} The relative
		probabilities of the four states are computed using
		\eref[eqTransientRateOfChange][eqDynamicsMatrix] and the rate constants
		$k_{\text{on}} [L] = 10^3\frac{1}{\text{s}}$, $k_{\text{off}}^{(O)} =
		10^{-2}\frac{1}{\text{s}}$, $k_{\text{off}}^{(C)} = 10^4\frac{1}{\text{s}}$,
		$k_+^{o} = 10^4\frac{1}{\text{s}}$, $k_-^{o} = 10\frac{1}{\text{s}}$, $k_+^{e}
		= 10\frac{1}{\text{s}}$, and $k_-^{e} = 10^3\frac{1}{\text{s}}$. Note that the
		rate constants must satisfy the cycle condition: the product of rates moving
		clockwise equals the product of rates going counterclockwise
		\cite{Auerbach2012}. The dashed line indicates the exact time scale
		\eref[eqtauExactAppendix] for the system to reach equilibrium. Initial
		conditions were chosen randomly as $p_O = 0.005$, $p_C = 0.45$, $p_{OL} =
		0.54$, and $p_{CL} = 0.005$.} \label{figDynamics}
\end{figure}

Lastly, we compute the fractional occupancy of the four states ion channel
states in steady state, $\frac{d\boldsymbol{E}}{dt} = \boldsymbol{0}$. We first
make the standard assumption that the system is not expending energy to drive a
cyclic flux in the system. Formally, this implies that the rate constants
satisfy the cycle condition: the product of rates moving clockwise in
\fref[figOneLigandRates] equals the product of rates going counterclockwise
\cite{Auerbach2012},
\begin{equation}
k_{\text{on}} [L] k_-^o k_{\text{off}}^{(C)} k_+^e = k_-^e k_{\text{off}}^{(O)} k_{\text{on}} [L] k_+^o.
\end{equation}
With this condition, the fractional occupancy of each state is given by
\begin{align}
[C] &= \frac{\frac{k_-^e}{k_+^e}}{\left( 1 + \frac{k_{\text{on}} [L]}{k_{\text{off}}^{(O)}} \right) + \frac{k_-^e}{k_+^e} \left( 1 + \frac{k_{\text{on}} [L]}{k_{\text{off}}^{(C)}} \right)}\\
[CL] &= \frac{\frac{k_-^e}{k_+^e} \frac{k_{\text{on}} [L]}{k_{\text{off}}^{(C)}}}{\left( 1 + \frac{k_{\text{on}} [L]}{k_{\text{off}}^{(O)}} \right) + \frac{k_-^e}{k_+^e} \left( 1 + \frac{k_{\text{on}} [L]}{k_{\text{off}}^{(C)}} \right)} \\
[O] &= \frac{1}{\left( 1 + \frac{k_{\text{on}} [L]}{k_{\text{off}}^{(O)}} \right) + \frac{k_-^e}{k_+^e} \left( 1 + \frac{k_{\text{on}} [L]}{k_{\text{off}}^{(C)}} \right)} \\
[OL] &= \frac{\frac{k_{\text{on}} [L]}{k_{\text{off}}^{(O)}}}{\left( 1 + \frac{k_{\text{on}} [L]}{k_{\text{off}}^{(O)}} \right) + \frac{k_-^e}{k_+^e} \left( 1 + \frac{k_{\text{on}} [L]}{k_{\text{off}}^{(C)}} \right)},
\end{align}
A system in steady state which satisfies the cycle condition must necessarily be
in thermodynamic equilibrium \cite{Gunawardena2012}, which implies that these
fractional occupancies must be identical to the result derived from the
Boltzmann distribution (see \fref[figStatesWeightsBothIonChannels]). And indeed,
this correspondence is made explicit if we define
\begin{align}
e^{-\beta \epsilon} &= \frac{k_-^e}{k_+^e} \\
\KOWT &= \frac{k_{\text{off}}^{(O)}}{k_{\text{on}}} \\
\KCWT &= \frac{k_{\text{off}}^{(C)}}{k_{\text{on}}}.
\end{align}
In this way, the MWC parameters can be defined through the ratios of the rate parameters of the system.


\subsection{Additional nAChR Mutants} \label{morenAChRMutants}

\begin{figure}[h!]
	\centering \includegraphics[scale=\globalScalePlots]{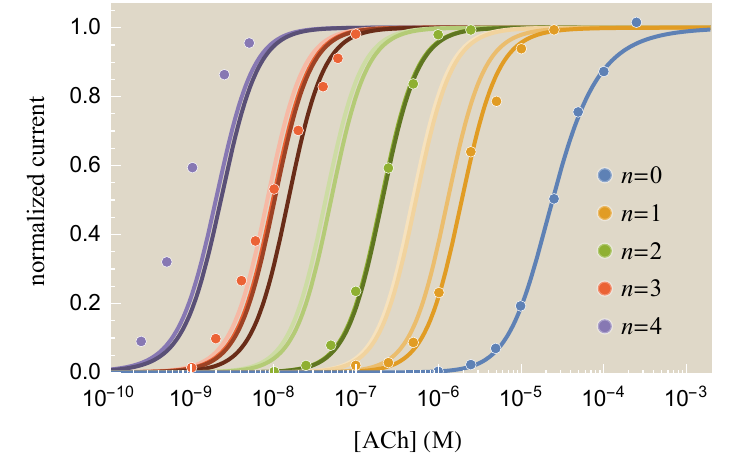}
	\caption{\captionStroke{Categorizing the full set of ion channel mutants.} Using the
		best-fit $\KOWT$ and $\KCWT$ values obtained from the five mutants in
		\fref[figNormalizedCurrent], we can use the measured $[EC_{50}]$ value for
		each mutant in \tref[tablenAChRAppendix] to determine its $\beta \epsilon$
		parameter. Thus, a single data point for each mutant enables us to predict its
		complete dose-response curve. All mutants with the same total number $n$ of
		mutations are plotted in shades of the same color, together with the complete
		dose-response curves from \fref[figNormalizedCurrent]. Note that while each
		mutant family spans a range of $[EC_{50}]$ values, the classes are distinct
		and do not overlap.
	} \label{figAllMutants}
\end{figure}

In addition to the five constructs shown in \fref[figNormalizedCurrent], namely
$n=0$ (wild type), $n=1$ ($\alpha_2\beta \gamma^* \delta$), $n=2$
($\alpha_2^*\beta \gamma \delta$), $n=3$ ($\alpha_2\beta^* \gamma^* \delta^*$),
and $n=4$ ($\alpha_2^* \beta \gamma^* \delta^*$), Labarca \textit{et
	al.}~constructed multiple other ion channel mutants listed in
\tref[tablenAChRAppendix] \cite{Labarca1995}. While complete dose-response
curves are not available for these other constructs, their $[EC_{50}]$ values
were measured. Using the $\KOWT$ and $\KCWT$ values for this entire class of
mutants given in \fref[figNormalizedCurrent], we can use the $[EC_{50}]$
measurements to fit the $\beta \epsilon$ value of each mutant, thereby providing
us with a complete description of each mutant.

In particular, we can predict the dose-response curves of each of these mutants,
as shown in \fref[figAllMutants]. We overlay the data from
\fref[figNormalizedCurrent] on top of these theoretical curves, where mutants
with the same total number $n$ of mutations are drawn as shades of the same
color. Note that there was some error in the original measurements, since the
reported $[EC_{50}]$ value for the $n=4$ ($\alpha_2^* \beta \gamma^* \delta^*$)
mutant shown in purple dots in \fref[figAllMutants] should clearly be less
than $10^{-9}\,\,\text{M}$, even though it was given as $\left( 2.0 \pm 0.6
\right) \times 10^{-9}\,\,\text{M}$ in Ref.~\citenum{Labarca1995}.

\begin{table}[h!]
	\centering \caption{\captionStroke{Dose-response relations for mouse muscle ACh receptors
			containing various numbers of mutated L251S subunits ($\captionSymbol{n}$).} Mutated
		subunits are indicated by an asterisk ($*$). Standard error of the mean for $[EC_{50}]$ was less than 10\% of the mean,
		except where given. Responses for the $\alpha_2^*\beta^* \gamma^* \delta^*$
		mutant were too small for reliable measurements. Reproduced from
		Ref.~\citenum{Labarca1995}.}
	\begin{tabular}{@{\vrule height 10.5pt depth4pt width0pt}cccc}
		\hline
		$n$ &
		subunits& 
		$[EC_{50}]$ (nM) \cr
		\hline
		\hline
		\noalign{\vskip 2pt} 
		0	& 	$\alpha_2\beta \gamma \delta$ 			& 24,010 	\cr
		1 	& 	$\alpha \alpha^*\beta \gamma \delta$ 	& 1,290 	\cr
		&	$\alpha_2\beta^* \gamma \delta$ 		& 531 			\cr
		&	$\alpha_2\beta \gamma^* \delta$ 		& 1,910 		\cr
		&	$\alpha_2\beta \gamma \delta^*$ 		& 486 			\cr
		2	&	$\alpha_2^*\beta \gamma \delta$ 		& 202 		\cr 
		&	$\alpha_2\beta^* \gamma^* \delta$ 		& 49.7 			\cr
		&	$\alpha_2\beta^* \gamma \delta^*$ 		& $208\pm69$ 	\cr
		&	$\alpha_2\beta \gamma^* \delta^*$ 		& 42.7 			\cr
		3	&	$\alpha_2^*\beta^* \gamma \delta$ 		& 10.3 		\cr 
		&	$\alpha_2^*\beta \gamma^* \delta$ 		& 15.1 			\cr
		&	$\alpha_2^*\beta \gamma \delta^*$ 		& $8.4\pm1.3$ 	\cr
		&	$\alpha_2\beta^* \gamma^* \delta^*$		& $9.8\pm1.3$ 	\cr
		4	&	$\alpha_2^*\beta^* \gamma \delta^*$ 	& 2.3 		\cr 
		&	$\alpha_2^*\beta \gamma^* \delta^*$ 	& $2.0\pm0.6$ 	\cr
		5	&	$\alpha_2^*\beta^* \gamma^* \delta^*$ 	& $<1$ 		\cr 
		\hline
	\end{tabular}
	\label{tablenAChRAppendix}
\end{table}

\fref[figAllMutants] demonstrates that not all subunit mutations cause a tenfold
decrease in $[EC_{50}]$, but rather that there is a small spread in $[EC_{50}]$
depending on precisely which subunit was mutated. This variation is not
unreasonable given that $\alpha_2\beta \gamma \delta$ nAChR is a heteropentamer.
Indeed, such subunit-dependent spreading in $[EC_{50}]$ values has also been
seen in other heteromeric ion channels \cite{Krashia2010, George2012} but is
absent within homomeric ion channels such as the CNGA2 ion channel explored in
the text \cite{Wongsamitkul2016}.

To explore this subunit-dependent shift in the dose-response curves, we now
relax the assumption that mutating any of the four nAChR subunits results in an
identical increase of roughly $5~\kB T$ to the allosteric gating energy
$\epsilon$. Instead, we allow each type of subunit to shift $\epsilon$ by a
different amount upon mutation. We begin by writing the $\epsilon$ parameter of
wild type nAChR as
\begin{equation}
\epsilon_{\alpha_2 \beta \gamma \delta} = 2\epsilon_{\alpha} + \epsilon_{\beta} + \epsilon_{\gamma} + \epsilon_{\delta},
\end{equation}
where $\epsilon_j$ denotes the gating energy contribution from subunit $j$ and
we have assumed that the five subunits independently contribute to channel
gating. Upon mutation, we define the free energy differences of each type of
subunit as
\begin{align}
\Delta \epsilon_{\alpha} &\equiv \epsilon_{\alpha^*} - \epsilon_{\alpha} \label{eqDeltaA}\\
\Delta \epsilon_{\beta} &\equiv \epsilon_{\beta^*} - \epsilon_{\beta} \label{eqDeltaB}\\
\Delta \epsilon_{\gamma} &\equiv \epsilon_{\gamma^*} - \epsilon_{\gamma} \label{eqDeltaC}\\
\Delta \epsilon_{\delta} &\equiv \epsilon_{\delta^*} - \epsilon_{\delta} \label{eqDeltaD},
\end{align}
where $\epsilon_{j^*}$ denotes the gating energy from the mutated subunit $j$.

The allosteric energy of any nAChR mutant can be found using the wild type
energy $\epsilon_{\alpha_2\beta \gamma \delta} = -23.7~\kB T$ from the main text
together with \(\Delta \epsilon_{\alpha}\), \(\Delta \epsilon_{\beta}\),
\(\Delta \epsilon_{\delta}\), and \(\Delta \epsilon_{\gamma}\). For example, the
gating energy of \(\alpha_2 \beta^* \gamma \delta\) is given by
\(\epsilon_{\alpha_2 \beta^* \gamma \delta} = \epsilon_{\alpha_2 \beta \gamma
	\delta}+\Delta \epsilon_{\beta}\) while that of \(\alpha_2^* \beta \gamma \delta^*\) is given by
\(\epsilon_{\alpha_2^* \beta \gamma \delta^*} = \epsilon_{\alpha_2 \beta \gamma
\delta} + 2 \Delta \epsilon_{\alpha} + \Delta \epsilon_{\delta}\).


Using the measured $[EC_{50}]$ values of all the mutants in Table
\ref{tablenAChRAppendix}, we can fit the four \(\Delta \epsilon_j\)'s to
determine how the different subunits increase the ion channel gating energy upon
mutation. We find the values $\Delta \epsilon_{\alpha} = 4.4~\kB T$, $\Delta
\epsilon_{\beta} = 5.3~\kB T$, $\Delta \epsilon_{\gamma} = 5.4~\kB T$, and
$\Delta \epsilon_{\delta} = 5.2~\kB T$, which show a small spread about the
value of roughly $5~\kB T$ found in the text by assuming that all four $\Delta
\epsilon_{j}$'s are identical. To show the goodness of fit, we can compare the
$[EC_{50}]$ values from this model to the experimental measurements in Table
\ref{tablenAChRAppendix}, as shown in
\fref[figIndividualSubunitEpsilons].

\begin{figure}[h!]
	\centering \includegraphics[scale=\globalScalePlots]{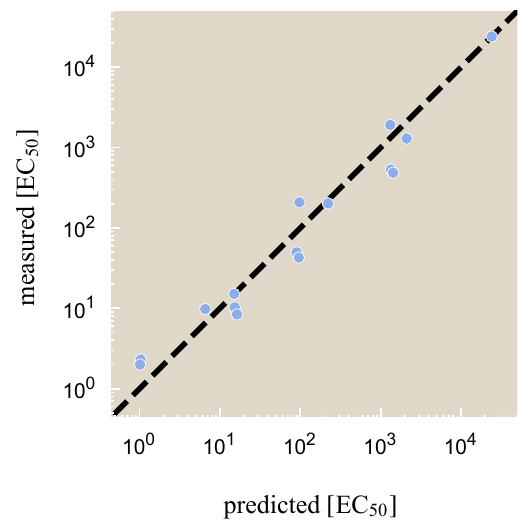}
	\caption{\captionStroke{Mutating different nAChR subunits changes the gating energy $\captionSymbol{\epsilon}$
		by different amounts.} Using a linear model where each subunit independently
	contributes to channel gating, we fit all of the $[EC_{50}]$ values in Table
	\ref{tablenAChRAppendix} to compute the increase of the gating energy
	$\epsilon$ when each subunit of $\alpha_2 \beta \gamma \delta$ nAChR is mutated
	(see \eref[eqDeltaA][eqDeltaB][eqDeltaC][eqDeltaD]). Upon mutation, a subunit
	of type $j$ increases the gating energy by $\Delta \epsilon_{j}$, where $\Delta
	\epsilon_{\alpha} = 4.4~\kB T$, $\Delta \epsilon_{\beta} = 5.3~\kB T$, $\Delta
	\epsilon_{\gamma} = 5.4~\kB T$, and $\Delta \epsilon_{\delta} = 5.2~\kB T$.
	For each mutant in Table \ref{tablenAChRAppendix}, the $[EC_{50}]$ from the model can be compared to the corresponding experimental measurement, with the black dashed line denoting the line of equality $y=x$.} \label{figIndividualSubunitEpsilons}
\end{figure}

\subsection{$\boldsymbol{p_{\text{open}}(c)}$ Curves} \label{appendixPopen}

Although the dose-response curves we analyze for nAChR were all presented using
normalized current, the underlying physical process - namely, the opening and
closing of the ion channel - is not required to go from 0 to 1.
\fref[figPopennAChR] shows the normalized dose-response curves from
\fref[figNormalizedCurrent]\letter{A} together with the average probability that
each ion channel mutant will be open, $p_{\text{open}}(c)$. Note that these
$p_{\text{open}}(c)$ curves have exactly the same shape as the normalized
current curves but are compressed in the vertical direction to have the
leakiness and dynamic range specified by \fref[figCharacteristics]\letter{A} and
\letter{B}.

From the viewpoint of these $p_{\text{open}}(c)$ curves, various nuances of this
ion channel class stand out more starkly. For example, the four mutant channels
have $p_{\text{open}}^{\text{max}} \approx 1$, noticeably larger than the
$p_{\text{open}}^{\text{max}} \approx 0.9$ value of the wild type channel. In
addition, the $n=4$ mutant is the only ion channel with non-negligible
leakiness, and \fref[figCharacteristics]\letter{A} suggests that an $n=5$ mutant
with all five subunits carrying the L251S mutation would have an even larger
leakiness value greater than \sfrac{1}{2}. In other words, the $n=5$ ion
channel is open more than half the time even in the absence of ligand, which
could potentially cripple or kill the cell. This may explain why Labarca \textit{et
	al.}~made the $n=5$ strain but were unable to measure its properties
\cite{Labarca1995}.

\begin{figure}[h]
	\centering \includegraphics[scale=\globalScalePlots]{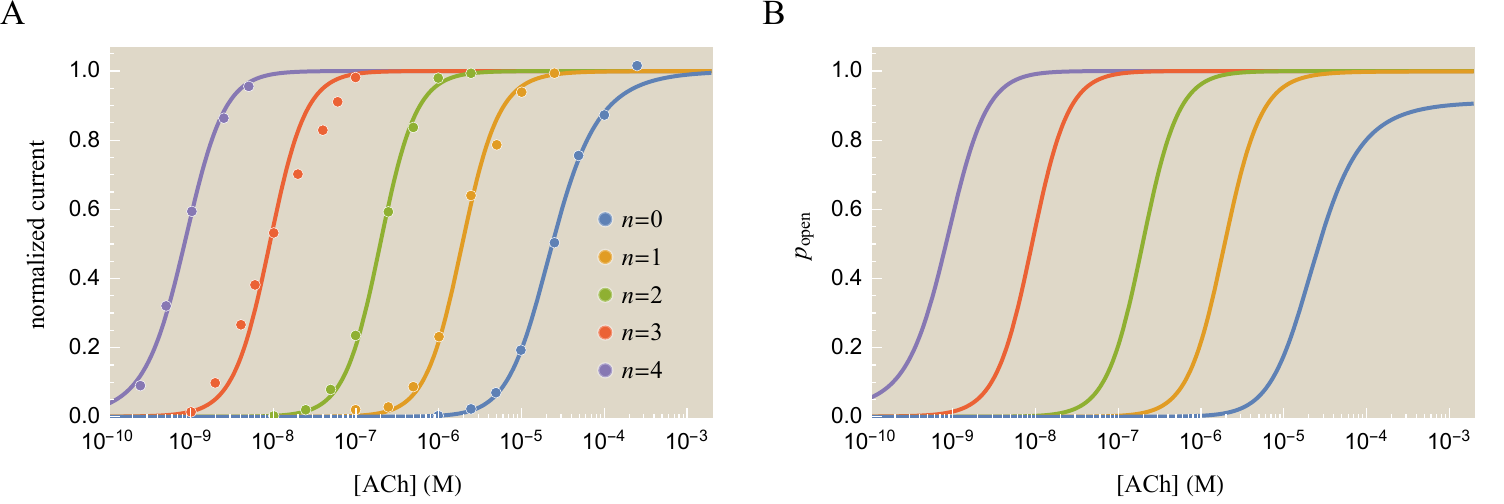}
	\caption{\captionStroke{Probability that an nAChR mutant will be open.} \letterParen{A}
		Normalized current curves of the five nAChR mutants from
		\fref[figNormalizedCurrent]\letter{A}. \letterParen{B} The probability that
		each ion channel will be open is given by \eref[eqPOpenGeneral]. Note that the wild
		type ion channel has a smaller dynamic range and the $n=4$ mutant has a
		noticeably larger leakiness than the other mutants.} \label{figPopennAChR}
\end{figure}

\fref[figPopenCNGA2] repeats this same analysis for the CNGA2 dose-response
curves from \fref[figNormalizedCurrentCNGA2]\letter{A}. In this case, all of the
ion channel mutants have uniformly small values of $p_{\text{open}}^{\text{min}}
\approx 0.03$ and uniformly large $p_{\text{open}}^{\text{max}} \approx 1$, as
indicated by \fref[figLeakDynEC50EffHCNGA2]\letter{A} and \letter{B}. Therefore,
the $p_{\text{open}}(c)$ curves look very similar to the normalized currents.

\begin{figure}[h]
	\centering \includegraphics[scale=\globalScalePlots]{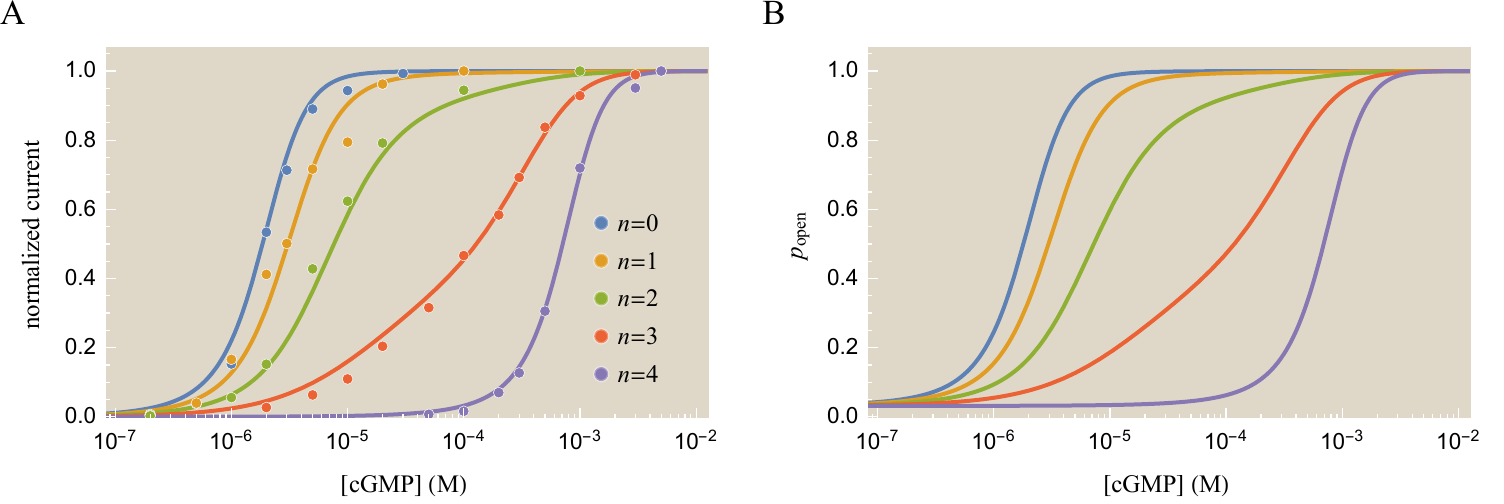}
	\caption{\captionStroke{Probability that a CNGA2 mutant will be open.} \letterParen{A}
		CNGA2 dose-response curves from \fref[figNormalizedCurrentCNGA2]\letter{A}.
		\letterParen{B} The probability that each ion channel will be open is given by
		\eref[eqPOpenGeneral]. Since all of the channels have small leakiness ($\approx 0.03$) and large
		dynamic range, the $p_{\text{open}}(c)$ curves are nearly identical to the
		normalized current curves.} \label{figPopenCNGA2}
\end{figure}

\subsection{nAChR L251T Mutation} \label{nAChRleu9Thr}

In this section we consider a separate nAChR data from the one considered in the
main paper. Filatov and White constructed nAChR ion channel mutants closely
related to those of Labarca \textit{et al.} but employing a L251T mutation
\cite{Filatov1995}. They measured the $[EC_{50}]$ of multiple such constructs
with the L251T mutation on different subsets of nAChR subunits, with the results
shown in \fref[figLeu9ThrnAChRMutants]\letter{A} as a function of the total
number of mutated subunits $n$.

As in the case of the L251S mutations from Labarca (see \fref[figAllMutants] and
Table~\ref{tablenAChRAppendix}), there was some variation in $[EC_{50}]$ between
different mutants with the same total number of mutations $n$, but the entire
class of mutants is well approximated as having $[EC_{50}]$ exponentially
decrease with each additional mutation. Utilizing our analytical formula for the
$[EC_{50}]$ of nAChR, \eref[eqEC50Value], and assuming that each mutation
changes $\epsilon$ by a fixed amount $\Delta \epsilon$, the shift in $[EC_{50}]$
due to $n$ mutations is given by
\begin{equation} \label{eqEC50Leu9Thr}
[EC_{50}] = e^{-\beta (\epsilon^{(0)} + n \Delta \epsilon)/2} \KOWT.
\end{equation}
We can fit the logarithm of the $[EC_{50}]$ values in
\fref[figLeu9ThrnAChRMutants]\letter{A} to a linear function going through the
wild type ($n=0$) data point to obtain $\Delta \epsilon = 3.64~\kB T$ from the
slope of this line. This value is comparable to that found for the L251S
mutation (where $\Delta \epsilon = 5~\kB T$).

\begin{figure}[t]
	\centering \includegraphics[scale=\globalScalePlots]{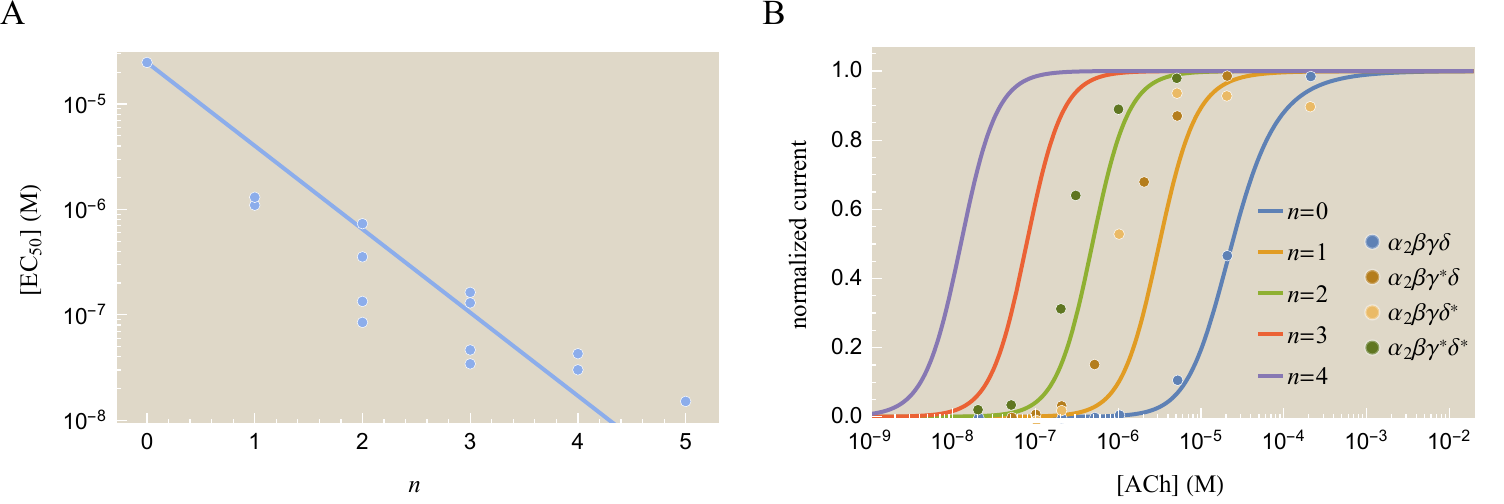}
	\caption{\captionStroke{Effects of L251T mutations on nAChR.} \letterParen{A} $[EC_{50}]$
		values for another class of L251T mutations introduced at different
		combinations of subunits \cite{Filatov1995}. This data set is separate from the
		L251S mutation considered in the main text. The $[EC_{50}]$ mainly depends on
		the total number of mutations, $[EC_{50}] \propto e^{-1.82 n}$, although there
		is slight variation depending upon which subunits are mutated. From
		\eref[eqEC50Leu9Thr], we find that each mutation imparts $\Delta \epsilon =
		3.64~\kB T$. \letterParen{B} Once the MWC parameters have been fixed from the
		$[EC_{50}]$ measurements, we can predict the full dose-response curves for the
		entire class of L251T nAChR mutants. Overlaid on these theoretical prediction
		are four experimentally measured response curves for the wild type
		($\alpha_2\beta \gamma \delta$), two $n=1$ single mutants ($\alpha_2\beta
		\gamma^* \delta$ and $\alpha_2\beta \gamma \delta^*$), and the $n=2$ double
		mutant ($\alpha_2\beta \gamma^* \delta^*$). We expect the predicted
		dose-response curves to match the data on average for the entire class of
		mutants, but Part \letter{A} shows that the $[EC_{50}]$ of the $n=1$ and $n=2$
		mutants will be overestimated while that of the $n=4$ and $n=5$ mutants will be
		underestimated. Asterisks ($*$) in the legend denote L251T mutations.}
	\label{figLeu9ThrnAChRMutants}
\end{figure}

With the gating energy now fully determined for any number of mutations $n$, and
using the $\KOWT$ and $\KCWT$ parameters from \fref[figNormalizedCurrent], we
now have a complete theoretical model of the L251T nAChR mutant class. For
example, we can plot the predicted dose-response curves for all such mutants.
\fref[figLeu9ThrnAChRMutants]\letter{B} shows these predictions together with
experimentally measured responses from the wild type channel and three mutant
constructs. The dose-response predictions should match the data on average for
the entire class of mutants, although individual channel responses may be
slightly off. For example, \fref[figLeu9ThrnAChRMutants]\letter{A} indicates
that the $[EC_{50}]$ of the $n=1$ and $n=2$ mutants will be lower than predicted
while that of the $n=4$ and $n=5$ mutants (whose dose-response data was not
provided) will be higher than predicted.

\subsection{Combining Multiple Ion Channels} \label{appendixCombiningIonChannels}

In this section, we consider the dose-response curve for the case in which the
cell harbors both wild type and mutant ion channels. Given $n_1$ ion channels
whose dose-response curves are governed by $p_{1,\text{open}}(c)$ and $n_2$ ion
channels with a different response $p_{2,\text{open}}(c)$, the current produced
by the combination of these two ion channels is given by
\begin{equation}
\text{current} \propto n_1 p_{1,\text{open}}(c) + n_2 p_{2,\text{open}}(c).
\end{equation}
Experimental measurements are computed on a relative scale so that the data runs
from 0 to 1. Analytically, this amounts to subtracting the leakiness and
dividing by the dynamic range,
\begin{equation} \label{eqIonChannelCombinationCurrent}
(\text{normalized current})_\text{tot} = \frac{n_1 p_{1,\text{open}}(c) + n_2 p_{2,\text{open}}(c) - n_1 p_{1,\text{open}}^{\text{min}} - n_2 p_{2,\text{open}}^{\text{min}}}{n_1 p_{1,\text{open}}^{\text{max}} + n_2 p_{2,\text{open}}^{\text{max}} - n_1 p_{1,\text{open}}^{\text{min}} - n_2 p_{2,\text{open}}^{\text{min}}}.
\end{equation}
Wongsamitkul \textit{et al.}~constructed cells expressing both the $n=0$ wild
type ion channels and the $n=4$ fully mutated ion channels in a ratio of 1:1
(i.e. $n_1 = n_2$) as shown in \fref[figNormalizedCurrentCNGA2DoubleChannel]
\cite{Wongsamitkul2016}. 

Recall from \fref[figLeakDynEC50EffHCNGA2] that these ion channels have very
small leakiness ($p_{1,\text{open}}^{\text{min}} \approx
p_{2,\text{open}}^{\text{min}} \approx 0$) and nearly full dynamic range
($p_{1,\text{open}}^{\text{max}} \approx p_{2,\text{open}}^{\text{max}} \approx
1$). This implies that $p_{1,\text{open}}(c) \approx (\text{normalized
	current})_1$ and $p_{2,\text{open}}(c) \approx (\text{normalized current})_2$,
so that the total normalized current due to the combination of ion channels is given
by
\begin{equation} \label{eqIonChannelCombinationCurrent2}
(\text{normalized current})_\text{tot} = \frac{(\text{normalized current})_{1} + (\text{normalized current})_{2}}{2}.
\end{equation}
\fref[figNormalizedCurrentCNGA2DoubleChannel] shows that this simple
prediction compares well to the measured data.

\begin{figure}[h]
	\centering \includegraphics[scale=\globalScalePlots]{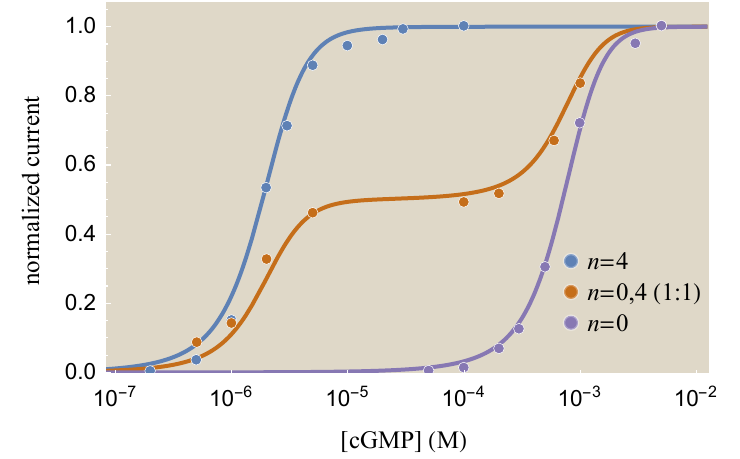}
	\caption{\captionStroke{Normalized currents for combinations of CNGA2 ion channels.}
		Channel currents of cells producing equal amounts of wild type $n=0$ and the
		$n=4$ mutant ion channels. As shown in \eref[eqIonChannelCombinationCurrent2],
		the resulting dose-response curve equals the average of the $n=0$ and $n=4$
		individual response curves.} \label{figNormalizedCurrentCNGA2DoubleChannel}
\end{figure}

\pagebreak
\section{Computing nAChR and CNGA2 Characteristics} \label{AppendixnPropertiesOfBothIonChannels}

In \ref{AppendixCharacteristicsMWCModel}, we derive
\eref[leakinessValue][dynamicRangeValue][eqEC50Value][eqhValue], the
approximations for the leakiness, dynamic range, $[EC_{50}]$, and the effective
Hill coefficient $h$ for the general MWC model \eref[eqPOpenGeneral]. We begin
by Taylor expanding the well known exact expressions from
Ref.~\citenum{Martins2011} in the limit $1 \ll e^{-\beta \epsilon} \ll
\left(\frac{\KCWT}{\KOWT}\right)^m$, which we found to be appropriate for both
the nAChR and CNGA2 ion channels, and find the lowest order approximations.

Following that, in \ref{AppendixLigandDissociationConstants} we consider how
mutations in the ligand dissociation constants $\KOWT$ and $\KCWT$ affect these
four properties. We show that ion channel dose-response curves are robust to
changes in $\KOWT$ and $\KCWT$ aside from left-right shifts dictated by
$[EC_{50}] = e^{-\beta \epsilon/m} \KOWT$. This discussion complements the nAChR
section of the text where we considered mutations of the $\beta \epsilon$
parameter.

Lastly, in \ref{AppendixHeterooligomericCNGA2Channel} we determine how ion
channels comprised of a mix of wild type subunits (with ligand dissociation
constants $\KOWT$ and $\KCWT$) and mutant subunits (with dissociation constants
$\KOMut$ and $\KCMut$) influences the four properties. Specifically, we focus on
the analytically tractable case where half of the subunits are wild type and the
other half are mutated (see \eref[eqEC50ValueCNGA2][eqhValueCNGA2] in the text).

\subsection{Characteristics of the MWC Model} \label{AppendixCharacteristicsMWCModel}

Using $1 \ll e^{-\beta \epsilon}$, the leakiness \eref[eqLeakinessDef] can be expanded as
\begin{equation} \label{eqAppendixLeakinessApproximation}
\text{leakiness} = \frac{1}{1 + e^{-\beta \epsilon}} \approx e^{\beta \epsilon}.
\end{equation}
Therefore, ion channels have a very small leakiness which scales exponentially
with $\beta \epsilon$. \fref[figAllFourCharacteristicsAppendix]\letter{A} shows
that this is a good approximation across the entire range of parameters within
the class of nAChR mutants, $-24 \le \beta \epsilon \le -4$.

The dynamic range \eref[eqDynamicRangeDef] can be similarly expanded to obtain
\begin{equation} \label{eqAppendixDynamicRangeApproximation}
\text{dynamic range} = \frac{1}{1 + e^{-\beta \epsilon}\left(\frac{\KOWT}{\KCWT}\right)^\m} - \frac{1}{1 + e^{-\beta \epsilon}} \approx 1 - e^{-\beta \epsilon}\left(\frac{\KOWT}{\KCWT}\right)^\m - e^{\beta \epsilon}.
\end{equation}
Keeping only the lowest order term yields the approximation
\eref[dynamicRangeValue] that ion channels have full dynamic range.
\fref[figAllFourCharacteristicsAppendix]\letter{B} shows that keeping the
first order terms in \eref[eqAppendixDynamicRangeApproximation] also captures
the behavior of the wild type channel ($\beta \epsilon^{(0)} = -23.7$) and the
$n=4$ mutant ($\beta \epsilon^{(4)} = -4.0$).

\begin{figure}[t]
	\centering \includegraphics[scale=\globalScalePlots]{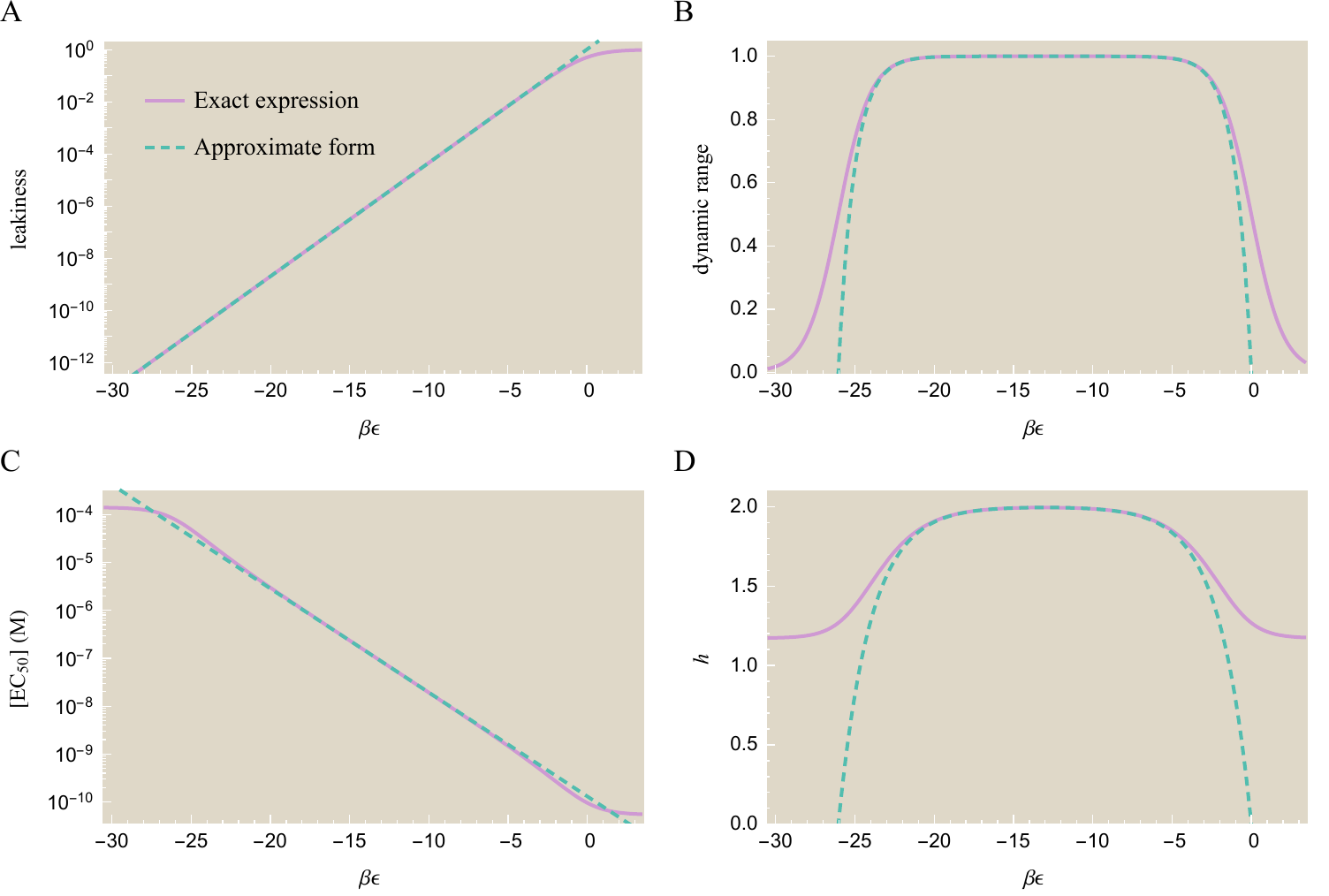}
	\caption{\captionStroke{Exact and approximate expressions for nAChR characteristics.} The approximations \eref[eqAppendixLeakinessApproximation][eqAppendixDynamicRangeApproximation][eqAppendixEffectiveHillApproximation] (dashed, teal) are valid in the limit $1 \ll e^{-\beta \epsilon} \ll
		\left(\frac{\KCWT}{\KOWT}\right)^m$ where they closely match the exact expressions (purple). \letterParen{A} Leakiness can be approximated as an exponentially increasing function of $\beta \epsilon$. \letterParen{B} To lowest order, the dynamic range of an ion channel should approach unity, with deviations only at very large and very small $\beta \epsilon$ values. \letterParen{C} The $[EC_{50}]$ is an exponentially decreasing function of $\beta \epsilon$. \letterParen{D} The effective Hill coefficient is roughly constant for all mutants, but as with the dynamic range it decreases for very large and very small $\beta \epsilon$ values.}
	\label{figAllFourCharacteristicsAppendix}
\end{figure}

We next turn to the $[EC_{50}]$ \eref[eqEC50Def], whose exact analytic formula
is given by \cite{Marzen2013}
\begin{equation} \label{eqAppendixEC50ExactAnalytic}
[EC_{50}] = \KOWT \frac{1 - \lambda^{\frac{1}{m}}}{\lambda^{\frac{1}{m}} - \frac{\KOWT}{\KCWT}}
\end{equation}
where
\begin{equation} \label{eqAppendixMWCEC50Lambda}
\lambda = \frac{2 - \left( p_{\text{open}}^{\text{min}} + p_{\text{open}}^{\text{max}} \right)}{e^{-\beta\epsilon} \left( p_{\text{open}}^{\text{min}} + p_{\text{open}}^{\text{max}} \right)}.
\end{equation}
The limit $1 \ll e^{-\beta \epsilon} \ll \left(\frac{\KCWT}{\KOWT}\right)^m$
suggests that we Taylor expand this formula to lowest order about $e^{-\beta
	\epsilon}\left(\frac{\KOWT}{\KCWT}\right)^\m \approx 0$ and $e^{-\beta \epsilon}
\approx \infty$, which yields
\begin{align} \label{eqAppendixEC50ApproximationDerivation}
[EC_{50}] &\approx \KOWT \frac{\frac{\KCWT}{\KOWT} \left( \left( 1 - \frac{1}{2 + e^{-\beta \epsilon}} \right)^{1/m} \right)}{\frac{\KCWT}{\KOWT} \left( \frac{1}{2 + e^{-\beta \epsilon}} \right)^{1/m} - 1} \nonumber\\
&\approx \KOWT \frac{\frac{\KCWT}{\KOWT} \left( 1 - e^{\beta \epsilon/m} \right)}{\frac{\KCWT}{\KOWT} e^{\beta \epsilon/m} - 1} \nonumber\\
&\approx \KOWT \frac{\frac{\KCWT}{\KOWT}}{\frac{\KCWT}{\KOWT} e^{\beta \epsilon/m}} \nonumber\\
&= \KOWT e^{-\beta \epsilon/m}.
\end{align}
Thus, the $[EC_{50}]$ decreases exponentially with $\epsilon$, although this
effect is diminished with the number of ligand binding sites $m$. The precise
relationship $[EC_{50}] \propto e^{-\beta \epsilon/2}$ for the nAChR data is
shown in \fref[figAllFourCharacteristicsAppendix]\letter{C}.

Finally, we turn to the effective Hill coefficient, whose exact analytic form is given by \cite{Marzen2013}
\begin{equation}
h = \frac{m [EC_{50}] \left( \KCWT - \KOWT \right) \left( p_{\text{open}}^{\text{min}} + p_{\text{open}}^{\text{max}} \right) \left( 2 - p_{\text{open}}^{\text{min}} - p_{\text{open}}^{\text{max}} \right)}{\left( p_{\text{open}}^{\text{min}} - p_{\text{open}}^{\text{max}} \right) \left( [EC_{50}] + \KOWT \right) \left( [EC_{50}] + \KCWT \right)},
\end{equation}
where we have used $p_{\text{open}}^{\text{min}}$ and
$p_{\text{open}}^{\text{max}}$ from \eref[eqPMinDef][eqPMaxDef] as well as the
$[EC_{50}]$ formula \eref[eqAppendixEC50ExactAnalytic]. Again, we make a Taylor
series of this expression about $e^{-\beta
	\epsilon}\left(\frac{\KOWT}{\KCWT}\right)^\m \approx 0$ and $e^{-\beta \epsilon}
\approx \infty$ to obtain the lowest order approximation, which is given by
\begin{align}
h &\approx m \frac{\frac{\KCWT}{\KOWT} + 1}{\frac{\KCWT}{\KOWT} - 1} - m \frac{\left( \frac{1}{2 + e^{-\beta \epsilon}} \right)^{-1/m}}{\frac{\KCWT}{\KOWT} - 1} - m \frac{\frac{\KCWT}{\KOWT} \left( \frac{1}{2 + e^{-\beta \epsilon}} \right)^{1/m}}{\frac{\KCWT}{\KOWT} - 1} - 2 m e^{\beta \epsilon}  \frac{\frac{\KCWT}{\KOWT} \left( \frac{1}{2 + e^{-\beta \epsilon}} \right)^{1/m}}{\frac{\KCWT}{\KOWT} - 1} \nonumber\\
&\approx m - m \frac{e^{-\beta \epsilon/m}}{\frac{\KCWT}{\KOWT}} - m \frac{\frac{\KCWT}{\KOWT} e^{\beta \epsilon/m}}{\frac{\KCWT}{\KOWT}} - 2 m e^{\beta \epsilon}  \frac{\frac{\KCWT}{\KOWT} e^{\beta \epsilon/m}}{\frac{\KCWT}{\KOWT}} \nonumber\\
&\approx m - m \frac{\KOWT}{\KCWT}e^{-\beta \epsilon/m} - m e^{\beta \epsilon/m} \label{eqAppendixEffectiveHillApproximation}.
\end{align}
Note that in the second step, we used the stronger constraint that
$\frac{\KCWT}{\KOWT} \gg 1$, although it is still reasonably satisfied for both
the nAChR ($\frac{\KCWT}{\KOWT} = 6 \times 10^5$) and CNGA2
($\frac{\KCWT}{\KOWT} = 17$) ion channels considered in the text. By keeping the
lowest order term, we recoup \eref[eqhValue] that all ion channels have the same
sharp response, and that this sharpness increases linearly with the number of
ligand binding sites. \fref[figAllFourCharacteristicsAppendix]\letter{D} shows
that by keeping the first order terms in
\eref[eqAppendixEffectiveHillApproximation], the shallower responses of the wild
type ($\beta \epsilon^{(0)} = -23.7$) and $n=4$ mutant ($\beta
\epsilon^{(4)} = -4.0$) can also be well approximated.

\subsection{Mutations Affecting the Ligand-Channel Dissociation Constants} \label{AppendixLigandDissociationConstants}

In this section, we discuss how the leakiness, dynamic range, $[EC_{50}]$, and
effective Hill coefficient $h$ vary when the channel-ligand dissociation
constants $\KOWT$ and $\KCWT$ are perturbed, as can be accomplished by mutating
the ligand binding domain. Recall that ligand-gated ion channels are typically
closed in the absence of ligand ($\epsilon < 0$) and open when bound to ligand
($\KOWT < \KCWT$).

\fref[figSILigandBindingFourProperties] shows the four ion channel properties
using the parameters of the wild type CNGA2 ion channel ($\beta \epsilon = -3.4$
and $m=4$ ligand binding sites) and letting the ratio $\frac{\KOWT}{\KCWT}$ of
dissociation constants vary. All four graphs demonstrate that the ion channel's
traits are nearly insensitive to changes in the dissociation constants provided
that $\KOWT$ does not approach $\KCWT$. In the limit $\KOWT \to \KCWT$, the
ligand no longer drives the ion channel to open, causing the dynamic range to
shrink to zero. As such, the behavior of the $[EC_{50}]$ and $h$ in this limit
should be considered as artifacts from taking this limit (since neither trait is
well defined when the dynamic range shrinks to zero). For reference, the wild
type CNGA2 channel has $\frac{\KOWT}{\KCWT} = 0.06$.

Note that the $y$-axis in \fref[figSILigandBindingFourProperties]\letter{C}
plots $\frac{[EC_{50}]}{\KOWT}$, so if both $\KOWT$ and $\KCWT$ are reduced by a
constant factor, then the $[EC_{50}]$ will also be reduced by this same factor.
In the limit $\frac{\KOWT}{\KCWT} \to 0$, $[EC_{50}] = \KOWT \left( 2 +
e^{-\beta \epsilon} \right)^{1/\m} - 1 \approx 1.4 \KOWT$ for the wild type CNGA2
channel (see \eref[eqAppendixEC50ExactAnalytic]).

\begin{figure}[h!]
	\centering \includegraphics[scale=\globalScalePlots]{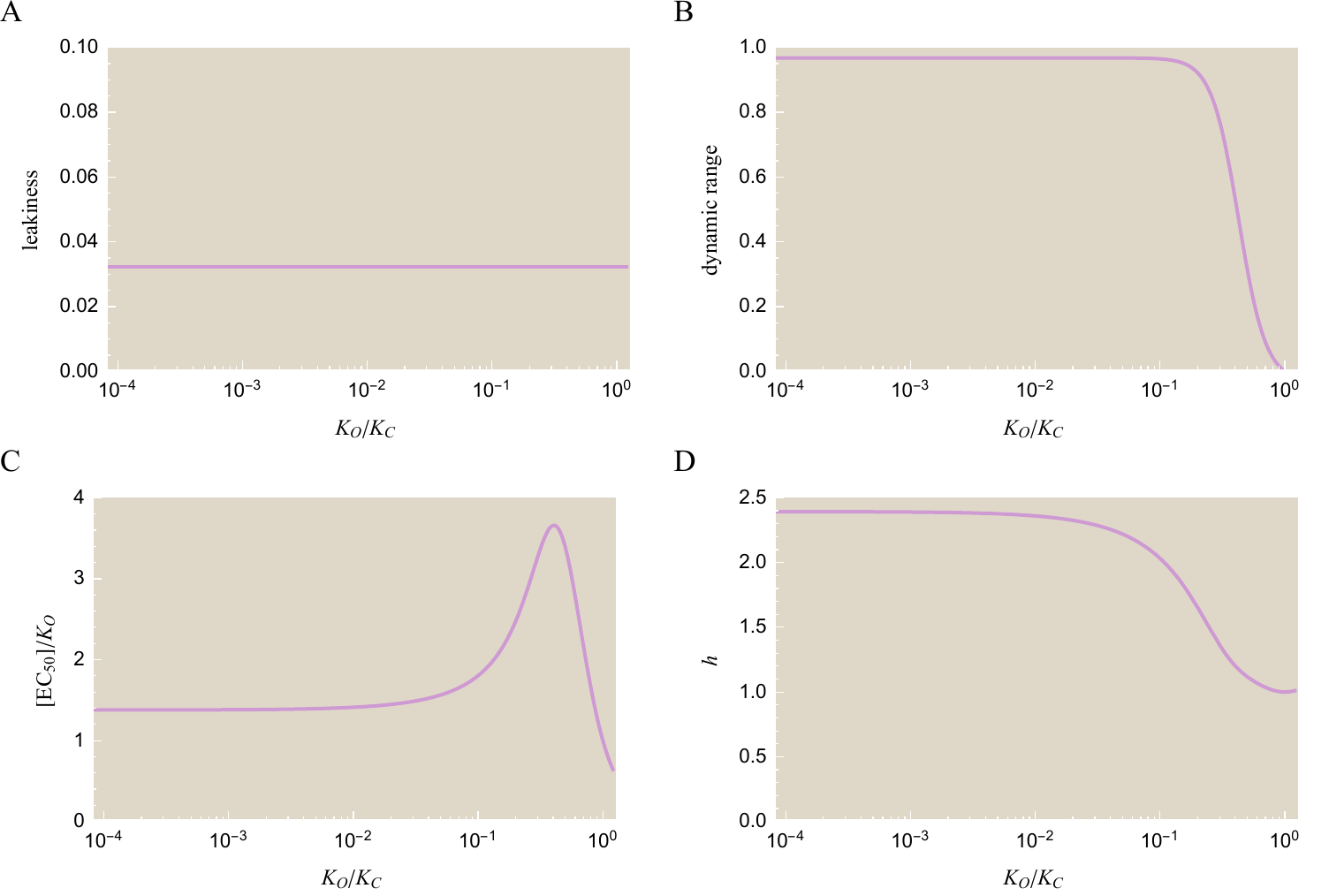}
	\caption{\captionStroke{CNGA2 properties are robust to changes in the ligand dissociation
			constants.} \letterParen{A} The leakiness does not depend on either
		dissociation constant. \letterParen{B} The dynamic range is near unity for set
		of dissociation constants where $\frac{\KOWT}{\KCWT} \le 0.1$, as was found for
		both the nAChR and CNGA2 systems. For larger ratios of the dissociation
		constants, the ligand no longer drives the ion channel to open. \letterParen{C}
		When $\frac{\KOWT}{\KCWT} \le 0.1$, $[EC_{50}] \approx 1.4 \KOWT$ is
		proportional to $\KOWT$ but robust to the ratio of dissociation constants.
		\letterParen{D} The effective Hill coefficient is also robust to changes in the
		dissociation constants, with $h \approx 2.4$ when $\frac{\KOWT}{\KCWT} \le
		0.1$.} \label{figSILigandBindingFourProperties}
\end{figure}

\subsection{The Heterooligomeric CNGA2 Channel} \label{AppendixHeterooligomericCNGA2Channel}

We now consider an ion channel with $m$ subunits, each of which could either be
a wild type subunit (with dissociation constants $\KOWT$ and $\KCWT$) or a
mutated subunit (with dissociation constants $\KOMut$ and $\KCMut$). Each
subunit contains a single ligand binding site. We compute the leakiness,
dynamic range, $[EC_{50}]$, and the effective Hill coefficient $h$ of an ion
channel composed of $n$ mutated subunits and $m-n$ wild type subunits. In the
text, we analyzed the specific case of the CNGA2 ion channel with $m=4$
subunits.

We begin by taking the limits of $p_{\text{open}}(c)$, \eref[eqPActive4Mutants],
in the absence of ligand, which is given by
\begin{equation} \label{eqSIpOpenMinCNGA2}
p_{\text{open}}^{\text{min}} = \frac{1}{1 + e^{-\beta \epsilon}},
\end{equation}
and in the presence of saturating levels of ligand, which is given as
\begin{equation} \label{eqPMaxCNGA2}
p_{\text{open}}^{\text{max}} = \frac{1}{1 + e^{-\beta \epsilon} \left(\frac{\KOWT}{\KCWT}\right)^{m-n}\left(\frac{\KOMut}{\KCMut}\right)^{n}}.
\end{equation}
Throughout this analysis, we assume $\KOWT < \KCWT$ and $\KOMut < \KCMut$ so
that ligand binding makes both the wild type and mutant subunits more likely to
open.

\begin{figure}[t!]
	\centering \includegraphics[scale=\globalScalePlots]{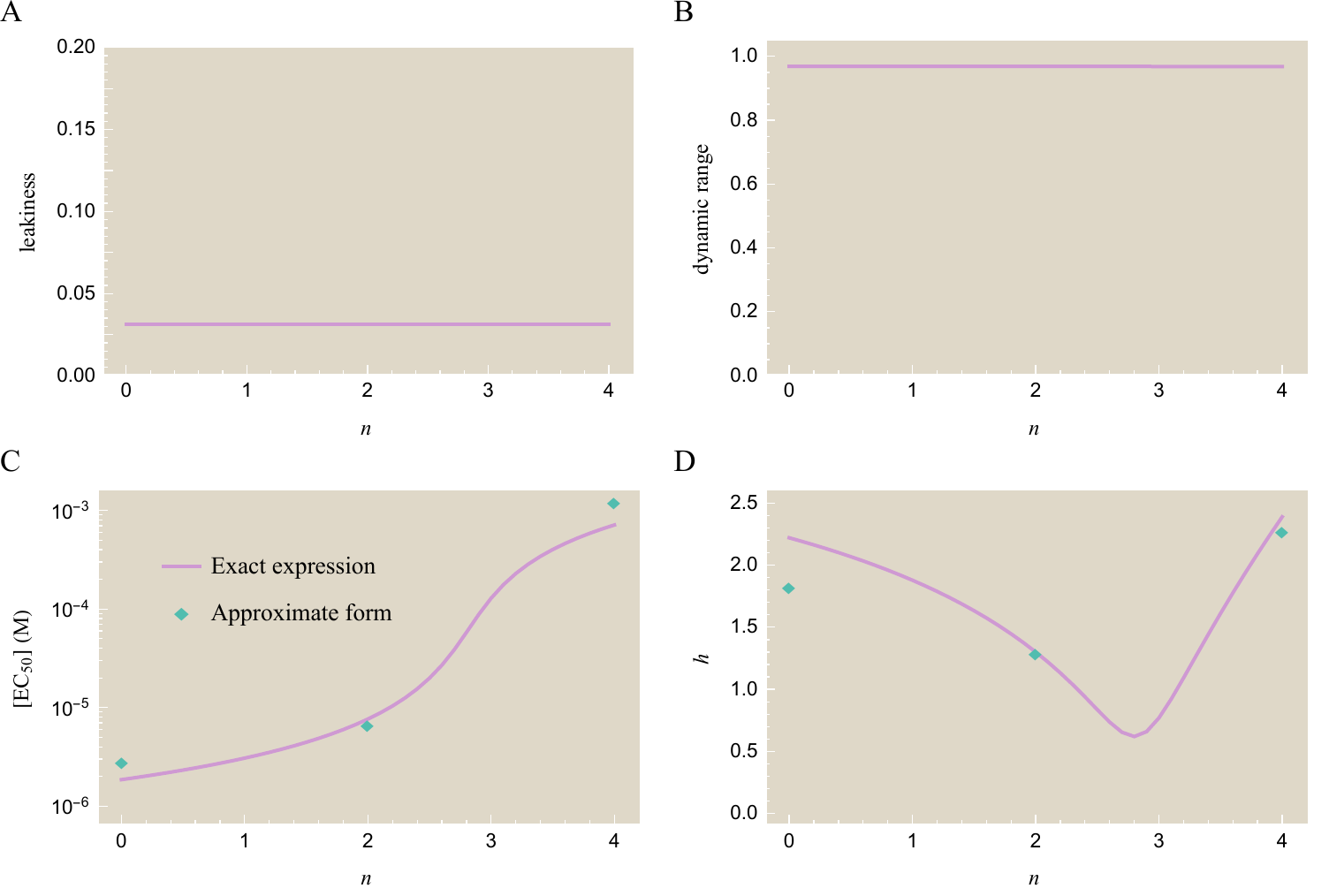}
	\caption{\captionStroke{Effects of mixing two types of subunits in the CNGA2 ion
			channel.} The CNGA2 ion channel is composed of $\m=4$ subunits, each of which
		has one ligand binding site. $\nWT$ of these subunits are mutated so as to have 
		weaker ligand binding affinity. \letterParen{A} The leakiness of
		the CNGA2 ion channel \eref[eqAppendixLeakinessCNGA2] is uniformly small.
		\letterParen{B} All of the mutants have nearly full dynamic range
		\eref[eqCNGADynamicRange] because the open channel dissociations constants
		($\KOWT$ and $\KOMut$) are significantly larger than the closed channel
		dissociation constants ($\KCWT$ and $\KCMut$). \letterParen{C} The exact
		expression (solid, purple) for the $[EC_{50}]$ is shown along with
		approximations for the $n=0$, 2, and 4 ion channels (teal diamonds) from
		\eref[eqAppendixEC50ApproximationDerivation][eqAppendixEC]. Because the mutated
		subunits bind poorly to ligand, the $[EC_{50}]$ increases with $n$.
		\letterParen{D} The effective Hill coefficient
		\eref[eqAppendixEffectiveHillApproximation][eqAppendixHill] can be approximated in the same manner
		as the $[EC_{50}]$. Although the homooligomeric $n=0$ and $n=4$ channels have
		sharp responses, the effect of combining both types of subunits ($n=1$, 2, and
		3) leads to a flatter response.} \label{figSIFourPropertiesCNG}
\end{figure}

The two limits of $p_{\text{open}}(c)$ above allow us to directly compute the leakiness
and dynamic range of the ion channels. The former is given by
\begin{equation} \label{eqAppendixLeakinessCNGA2}
\text{leakiness} = \frac{1}{1 + e^{-\beta \epsilon}},
\end{equation}
which has an identical form to the leakiness of the MWC model
\eref[eqAppendixLeakinessApproximation]. As shown in
\fref[figSIFourPropertiesCNG]\letter{A}, the leakiness does not depend
explicitly on the number of mutated subunits $n$. We next turn to the formula for
the dynamic range,
\begin{equation} \label{eqCNGADynamicRange}
\text{dynamic range} = \frac{1}{1 + e^{-\beta \epsilon} \left(\frac{\KOWT}{\KCWT}\right)^{m-n}\left(\frac{\KOMut}{\KCMut}\right)^{n}} - \frac{1}{1 + e^{-\beta \epsilon}}.
\end{equation}
The first term in the dynamic range is approximately 1 because the open state
affinities are always smaller than the closed state affinities by at least a
factor of ten, and these factors are collectively raised to the $m^{\text{th}}$
power. Since these ion channels also exhibit small leakiness, they each have a
large dynamic range as shown in \fref[figSIFourPropertiesCNG]\letter{B}.

We next consider approximations for the $[EC_{50}]$ and effective Hill
coefficient $h$. The wild type CNGA2 channel ($n=0$) will necessarily follow the
formulas derived above for the MWC model
(\eref[eqAppendixEC50ApproximationDerivation][eqAppendixEffectiveHillApproximation]). Similarly, the homooligomeric CNGA2 ion channel comprised of all mutated subunits ($n=m$) will have the same formulas as the wild type channel but with $\KOWT \to \KOMut$ and $\KCWT \to \KCMut$.

To gain a sense of how the $[EC_{50}]$ and $h$ vary for channels comprised of a
mix of wild type and mutant subunits ($1 \le n \le m - 1$), we analyze the $n=m/2$
case (implicitly assuming that $m$ is even) where half the subunits are wild
type and the other half are mutated. We begin with the $[EC_{50}]$ formula
which by definition is given by
\begin{equation} \label{eqAppendixCNGEC50}
\frac{\left[\left(1 + \frac{c}{\KOWT}\right)\left(1 + \frac{c}{\KOMut}\right)\right]^{\m/2}}{\left[\left(1 + \frac{c}{\KOWT}\right)\left(1 + \frac{c}{\KOMut}\right)\right]^{\m/2} + e^{-\beta \epsilon} \left[\left(1 + \frac{c}{\KCWT}\right)\left(1 + \frac{c}{\KCMut}\right)\right]^{\m/2}} = \frac{1}{2} \left( p_{\text{open}}^{\text{min}} + p_{\text{open}}^{\text{max}} \right).
\end{equation}
Rearranging the terms, we find
\begin{equation}
\lambda \left[\left(1 + \frac{c}{\KOWT}\right)\left(1 + \frac{c}{\KOMut}\right)\right]^{\m/2} = \left[\left(1 + \frac{c}{\KCWT}\right)\left(1 + \frac{c}{\KCMut}\right)\right]^{\m/2},
\end{equation}
where we have introduced the same (positive) constant $\lambda$ in
\eref[eqAppendixMWCEC50Lambda] from the $[EC_{50}]$ of the standard MWC model,
\begin{equation}
\lambda = \frac{2 - \left( p_{\text{open}}^{\text{min}} + p_{\text{open}}^{\text{max}} \right)}{e^{-\beta\epsilon} \left( p_{\text{open}}^{\text{min}} + p_{\text{open}}^{\text{max}} \right)}.
\end{equation}
Upon raising both sides of \eref[eqAppendixCNGEC50] to the $\frac{2}{m}$ power,
we find the quadratic equation
\begin{equation}
A c^2 + B c + C = 0
\end{equation}
where
\begin{align}
A &= \frac{\lambda^{2/m}}{\KOWT \KOMut} - \frac{1}{\KCWT \KCMut} \\
B &= \frac{\lambda^{2/m}}{\KOWT} + \frac{\lambda^{2/m}}{\KOMut} - \frac{1}{\KCWT} - \frac{1}{\KCMut} \\
C &= \lambda^{2/m} - 1,
\end{align}
which has the exact solution
\begin{equation}
[EC_{50}] = \frac{-B + \sqrt{B^2 - 4 A C}}{2 A}.
\end{equation}
To simplify this expression, we note that $\left| 4 A C \right|$ is smaller than
$B^2$ by more than a factor of 10 for the CNGA2 parameter values, so that the
square root can be approximated as $\sqrt{B^2 - 4 A C} \approx B - \frac{2 A
	C}{B}$, and hence the $[EC_{50}]$ becomes
\begin{equation}
[EC_{50}] \approx -\frac{C}{B} = \frac{1 - \lambda^{2/m}}{\frac{\lambda^{2/m}}{\KOWT} + \frac{\lambda^{2/m}}{\KOMut} - \frac{1}{\KCWT} - \frac{1}{\KCMut}}.
\end{equation}
To further simplify this result, we utilize the relationships $1 \ll e^{-\beta
	\epsilon} \ll \left(\frac{\KCWT}{\KOWT}\right)^{m/2}$ and $1 \ll e^{-\beta
	\epsilon} \ll \left(\frac{\KCMut}{\KOMut}\right)^{m/2}$ which hold for the CNGA2
parameters. In this limit, $p_{\text{open}}^{\text{min}} \approx 0$,
$p_{\text{open}}^{\text{max}} \approx 1$, and $\lambda \approx e^{\beta
	\epsilon} \ll 1$, so that the formula for the $[EC_{50}]$ becomes
\begin{align} \label{eqAppendixEC}
[EC_{50}] &\approx \frac{1}{\frac{e^{2\beta\epsilon/m}}{\KOWT} + \frac{e^{2\beta\epsilon/m}}{\KOMut} - \frac{1}{\KCWT} - \frac{1}{\KCMut}} \nonumber\\
&\approx \frac{1}{\frac{e^{2\beta\epsilon/m}}{\KOWT} + \frac{e^{2\beta\epsilon/m}}{\KOMut}} \nonumber\\
&= e^{-2\beta\epsilon/m} \frac{\KOWT \KOMut}{\KOWT + \KOMut}.
\end{align}
Since the mutated CNGA2 subunits have significantly weaker binding affinity
($\KOWT \ll \KOMut$), $[EC_{50}] \approx e^{-2\beta\epsilon/m} \KOWT$ where the
exponent is twice as large as the homooligomeric case
\eref[eqAppendixEC50ApproximationDerivation].
\fref[figSIFourPropertiesCNG]\letter{C} shows how the $[EC_{50}]$ gradually
increases as more of mutant subunits are introduced into the ion channel, with
the approximation for the $\nWT=\frac{m}{2}$ mutant given by \eref[eqAppendixEC]
while the $\nWT=0$ and $\nWT=\m$ mutants are described by
\eref[eqAppendixEC50ApproximationDerivation].

We next turn to the effective Hill coefficient $h$. To greatly simplify the
computation, we ignore all of the dissociation constants ($\KCWT = 20 \times
10^{-6}\,\,\text{M}$, $\KOMut = 500 \times 10^{-6}\,\,\text{M}$, and $\KCMut =
140 \times 10^{-3}\,\,\text{M}$) greater than the $[EC_{50}] \approx 6 \times
10^{-6}\,\,\text{M}$, since they all enter $p_{\text{open}}(c)$ as $\left( 1 +
\frac{c}{K_j} \right)^{m/2}$ with $m=4$ for CNGA2. Thus, the probability of the
channel opening becomes
\begin{equation} \label{eqCNGAApproximation}
p_{\text{open}}(c) \approx \frac{\left(1 + \frac{c}{\KOWT}\right)^{\m/2}}{\left(1 + \frac{c}{\KOWT}\right)^{\m/2} + e^{-\beta \epsilon}}
\end{equation}
where only the effect of the smallest dissociation constant ($\KOWT = 1.2 \times
10^{-6}\,\,\text{M}$) is considered. Noting that $\KOWT \ll \KOMut$, the
effective Hill coefficient is given by
\begin{align} \label{eqAppendixHill}
h &\approx \frac{\m}{\left( 1 + e^{2\beta\epsilon/\m} \right) \left( 1 + \left( 1 + e^{2\beta\epsilon/\m} \right)^{\m/2} \right)} \nonumber\\
&\approx \frac{\m}{2} - \frac{\m}{2} \left( \frac{m}{4} + 1 \right) e^{2\beta\epsilon/\m},
\end{align}
where in the second step we used the Taylor expansion about
$e^{2\beta\epsilon/\m} \ll 1$. \fref[figSIFourPropertiesCNG]\letter{D} shows the
effective Hill coefficient together with the approximations for the
homooligomeric ($\nWT=0$ and $\nWT = 4$) ion channel
\eref[eqAppendixEffectiveHillApproximation] and the $\nWT = 2$ channel given by
\eref[eqAppendixHill]. As discussed in the text, the effective Hill coefficient
exhibits a surprising decrease for ion channels comprised of a mix of both types
of subunits ($1 \le n \le 3$). Qualitatively, this comes about because the
subunits of the $n=0$ wild type channel become sensitive to ligand at
concentrations approaching $[EC_{50}]^{(n=0)} \approx e^{-\beta \epsilon/m}
\KOWT$ while the mutant subunits respond at the much larger concentrations
$[EC_{50}]^{(n=4)} \approx e^{-\beta \epsilon/m} \KOMut$. Channels containing
both subunits consequently have a much flatter response over the range between
$e^{-\beta \epsilon/\m} \KOWT$ and $e^{-\beta \epsilon/\m} \KOMut$.

\pagebreak
\section{Data Fitting} \label{dataFitting}

In this section, we discuss the fitting procedure used on the nAChR and CNGA2
data sets. All fitting was done using nonlinear regression (NonlinearModelFit in
\textit{Mathematica}). A wide array of initial conditions were considered (for
example, dissociation constants were sampled in the range $K_D \in
[10^{-12}\,\,\text{M}, 10^0\,\,\text{M}]$ and allosteric energies were sampled
across $\beta \epsilon \in [-30, 5]$), and the best-fit parameters were chosen
from the fit with the largest coefficient of determination $R^2$. Because all
dissociation constants are necessarily positive, we employed the standard trick
of fitting the logarithms of dissociation constants, which improves both the fit
stability and accuracy.

In \ref{AppendixFittingProcedure}, we give more details on the fitting
procedure. In \ref{AppendixLiteratureValuesComparison}, we focus on the related
point of the sensitivity of the MWC model parameter values. We compare
experimentally measured values from the literature and analyze them in the
context of the nAChR and CNGA2 data sets to determine how much flexibility the
MWC model has in its ability to capture observed trends. In
\ref{AppendixNonUniformKOKC}, we discuss how values such as the $[EC_{50}]$ and
effective Hill coefficient can be extracted from experimental measurements.

Lastly, in \ref{AppendixSloppiness} we discuss a critical issue for extracting
best-fit parameters from data, namely, that multiple sets of parameters may
exist which characterize the data nearly identically. This attribute of models,
sometimes dubbed ``sloppiness,'' can even exist in the MWC model we have
developed for ion channels where there are only three parameters
\cite{Transtrum2015}, and it has also been found in an MWC model of hemoglobin
\cite{Milo2007}. We examine our results in light of this degeneracy.

\subsection{Fitting Procedure} \label{AppendixFittingProcedure}

The fit parameters from \fref[figNormalizedCurrent] are shown in
\tref[tableFit1]. If the data is fit to the MWC model
(\eref[eqPOpenGeneral][eqNormalizedCurrent]) with no constraints, then all of
the degenerate parameter sets in \fref[fignAChRandCNGA2Sloppiness]\letter{A}
would yield equally good fits. For example, any set of degenerate parameters
with $\KOWT \le 10^{-10}\,\,\text{M}$ have coefficient of determination $R^2 =
0.995 \pm 0.0002$. In other words, it is impossible to distinguish the actual
set of parameter values for nAChR without further information. As highlighted in
the main text, one method for lifting this degeneracy is to independently
measure one model parameter, which could then be used to fix the remaining
parameters. For example, measuring the leakiness of one of the nAChR mutants
would fix its corresponding $\beta \epsilon^{(n)}$ parameter, resolving the degeneracy
in \fref[fignAChRandCNGA2Sloppiness]\letter{A}. The leakiness of the $n=3$ and
$n=4$ mutants is significantly larger than that of wild type nAChR, and hence
should be possible to directly measure experimentally. 

Next, we examine how much sloppiness would remain in the system if an
experimental measurement fixed one of the $\beta \epsilon$ parameters. To do
this, we arbitrarily choose $\beta \epsilon^{(4)} = -4.0$, and we then fit the
remaining parameters with this constraint. With the degeneracy now removed from
the model, \tref[tableFit1] presents the mean parameter values and the error
based on confidence intervals. Note that the remaining MWC parameters are all
tightly constrained about their best-fit values, so that there is very little
sloppiness left in the system after one parameter value is determined.

\begin{table}[t]
	\centering \caption{\captionStroke{Best-fit parameters for the nAChR mutants given the
			constraint $\captionSymbol{\beta \epsilon^{(4)}=-4.0}$.} With this single
		parameter fixed, the remaining parameters have small uncertainties. $R^2$
		represents the coefficient of determination.}
	\begin{adjustwidth}{-0.5in}{0in}
		\begin{tabular}{@{\vrule height 10.5pt depth4pt width0pt}cccccccc}
			\hline
			$\KOWT$ (M)& 
			$\KCWT$ (M) & 
			$\beta \epsilon^{(0)}$ & 
			$\beta \epsilon^{(1)}$ & 
			$\beta \epsilon^{(2)}$ & 
			$\beta \epsilon^{(3)}$ & 
			$\beta \epsilon^{(4)}$ &
			$R^2$ \cr
			\hline
			\hline
			\noalign{\vskip 2pt} 
			$(0.13 \pm 0.16) \times 10^{-9}$ &
			$(11 \pm 7) \times 10^{-6}$ & 
			$-23.7 \pm 5.9$ & 
			$-19.2 \pm 2.5$ &
			$-14.6 \pm 2.5$ &
			$-8.5 \pm 2.3$ &
			$-4.0$ &
			$0.995$ \cr
			\hline
		\end{tabular}
		\label{tableFit1}
	\end{adjustwidth}
\end{table}

A similar fit procedure was used for the CNGA2 data set in
\fref[figNormalizedCurrentCNGA2]. Here, we arbitrarily fixed the parameter
$\KCMut = 140 \times 10^{-3}\,\,\text{M}$ and fit the remaining parameters, with the
results shown in \tref[tableFit2]. Again, we find that with one parameter fixed,
the remaining parameters are tightly constrained.

\begin{table}[t]
		\centering
		\caption{\captionStroke{Best-fit parameters for the CNGA2 mutants with the constraint $\captionSymbol{\KCMut = 140 \times 10^{-3}\,\,\text{M}}$.} As in the case of nAChR, we find that once one of the parameters has a fixed value, the degeneracy within the model is lifted and the remaining parameters have small uncertainties.}
		\begin{tabular}{@{\vrule height 10.5pt depth4pt width0pt}ccccccc}
			\hline
			$\KOWT$ (M)& 
			$\KCWT$ (M) & 
			$\KOMut$ (M)& 
			$\KCMut$ (M) & 
			$\beta \epsilon$ &
			$R^2$ \cr
			\hline
			\hline
			\noalign{\vskip 2pt} 
			$(1.2 \pm 0.1) \times 10^{-6}$ & 
			$(20 \pm 3) \times 10^{-6}$ & 
			$(500 \pm 100) \times 10^{-6}$ & 
			$140 \times 10^{-3}$ &
			$-3.4 \pm 0.2$ &
			$0.997$\cr
			\hline
		\end{tabular}
		\label{tableFit2}
\end{table}

\subsection{Comparison with Parameters Values from the Literature} \label{AppendixLiteratureValuesComparison}

In this section, we explore the degree of consistency between multiple independent
measurements of the thermodynamic parameters in both the nAChR and CNGA2
systems.

We begin with the nAChR ion channel, whose allosteric gating parameter
$\beta \epsilon$ has been exceedingly difficult to measure, since channel openings in
the absence of ligand occur extremely infrequently. Instead of direct
measurement, several groups measured the leakiness of nAChR channels with
multiple pore mutations. The wild type channel parameter $\beta  \epsilon^{(0)} \approx
-14.2$ was then extrapolated by assuming that all of these pore mutations
only affect the $\epsilon$ parameter and have energetically independent effects
(i.e. if two mutations change $\epsilon$ by $\Delta\epsilon_1$ and
$\Delta\epsilon_2$, respectively, then a channel with both mutations would
change $\epsilon$ by $\Delta\epsilon_1 + \Delta\epsilon_2$) \cite{Nayak2012,
	Jackson1986}. Subsequently, dose-response curves were used to determine the
values of the remaining thermodynamic parameters, namely, $\KOWT = 25 \times
10^{-9}\,\,\text{M}$ and $\KCWT = 150 \times 10^{-6}\,\,\text{M}$
\cite{Auerbach2012}.

We first attempted to use these literature values directly to specify $\KOWT$
and $\KCWT$ for the entire class of nAChR mutants. However, using these values
the $n=4$ nAChR mutant cannot be well characterized for any value of $\beta
\epsilon^{(4)}$ ($R^2 < 0.5$). Thus, we next examined the sensitivity of the
measured $\beta \epsilon^{(0)} = -14.2$ parameter to see how well the full nAChR
data set could be fit if this value was slightly altered.
\fref[figSInAChR3]\letter{A} demonstrates that if $\epsilon^{(0)}$ is lowered by
$4~\kB T$, the family of nAChR mutants can once again be well characterized
($R^2 > 0.99$) by a single parameter set. In fact, as seen in
\fref[figSInAChR3]\letter{B}, even a decrease of $2~\kB T$ in $\epsilon^{(0)}$
provides a reasonable fit ($R^2 = 0.98$) for the class of nAChR mutants. We note
that $2~\kB T$, roughly the energy of a hydrogen bond, is a very small energy,
and this discrepancy may represent a source of error in the assumptions used to
determine the $\beta \epsilon^{(0)} = -14.2$ value (e.g. that the effects of
multiple channel pore mutations are additive and independent).

\begin{figure}[h]
		\centering \includegraphics[scale=\globalScalePlots]{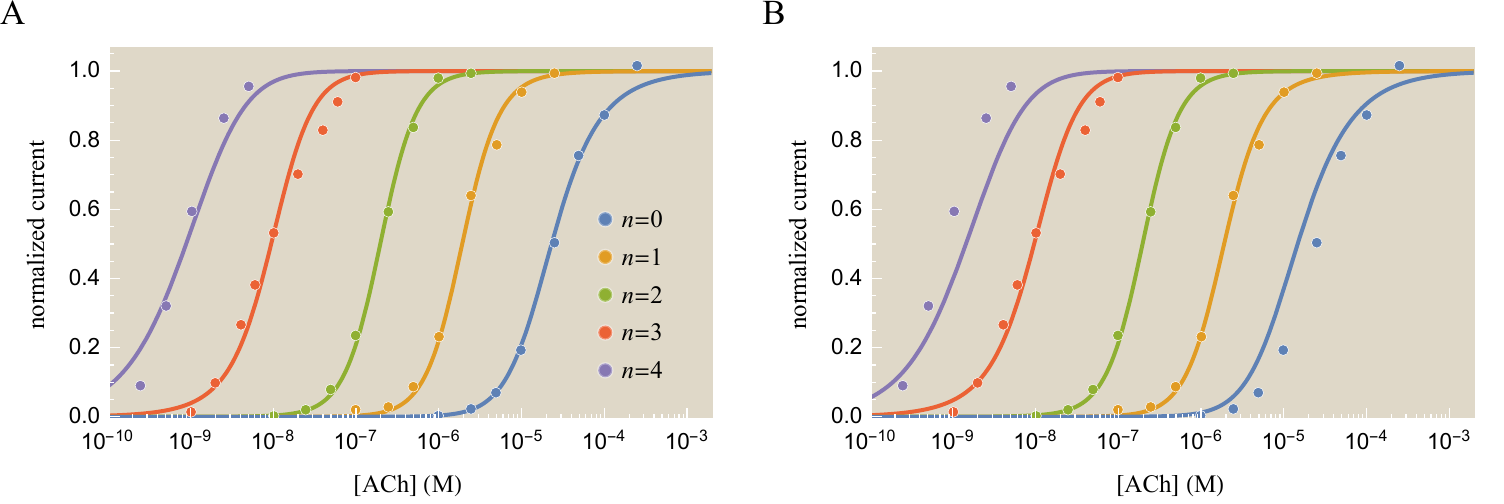}
		\caption{\captionStroke{nAChR fits can be resolved by slightly perturbing the measured
				$\captionSymbol{\beta \epsilon}$ value.} \letterParen{A} If the experimentally
			measured value of $\beta \epsilon^{(0)} = -14.2$ for wild type nAChR is
			decreased to $\beta \epsilon^{(0)} = -18$, we can characterize all of the
			nAChR mutants ($R^2 > 0.99$) using a single set of parameters given in
			\tref[tableFit3]. \letterParen{B} Even the very modest change to $\beta
			\epsilon^{(0)} = -16$ enables us to fit most of the data set well ($R^2 =
			0.98$).} \label{figSInAChR3}
\end{figure}

We now turn to how the MWC model compares to known literature values
for the CNGA2 ion channel. In their paper, Wongsamitkul \textit{et al.}~reported
single channel measurements for the ratio of the open to closed state for the wild type ($n=0$)
channel, finding
\begin{equation} \label{eqCNGA2constraint1}
\frac{[O]}{[C]} = 1.7 \times 10^{-5} = e^{\beta \epsilon}
\end{equation}
or equivalently $\beta \epsilon = -11$ \cite{Wongsamitkul2016}. However, other sources have
reported values as high as $\beta \epsilon = -6$ for this same ion channel
\cite{Nache2005, Tibbs1997}.

We find that the full spectrum of CNGA2 ion channel mutants can be fit to a
single set of thermodynamic parameters ($\KOWT$, $\KCWT$, $\KOMut$, $\KCMut$,
and $\beta \epsilon$) when $\beta \epsilon = -6$, as shown in
\fref[figSICNGA1]\letter{A} (with $R^2 > 0.99$). Alternatively, using the value
$\beta \epsilon = -9$ halfway between the experimental measurements yields
markedly worse fits (with $R^2 = 0.97$), as shown in
\fref[figSICNGA1]\letter{B}. 

\begin{figure}[h]
	\centering \includegraphics[scale=\globalScalePlots]{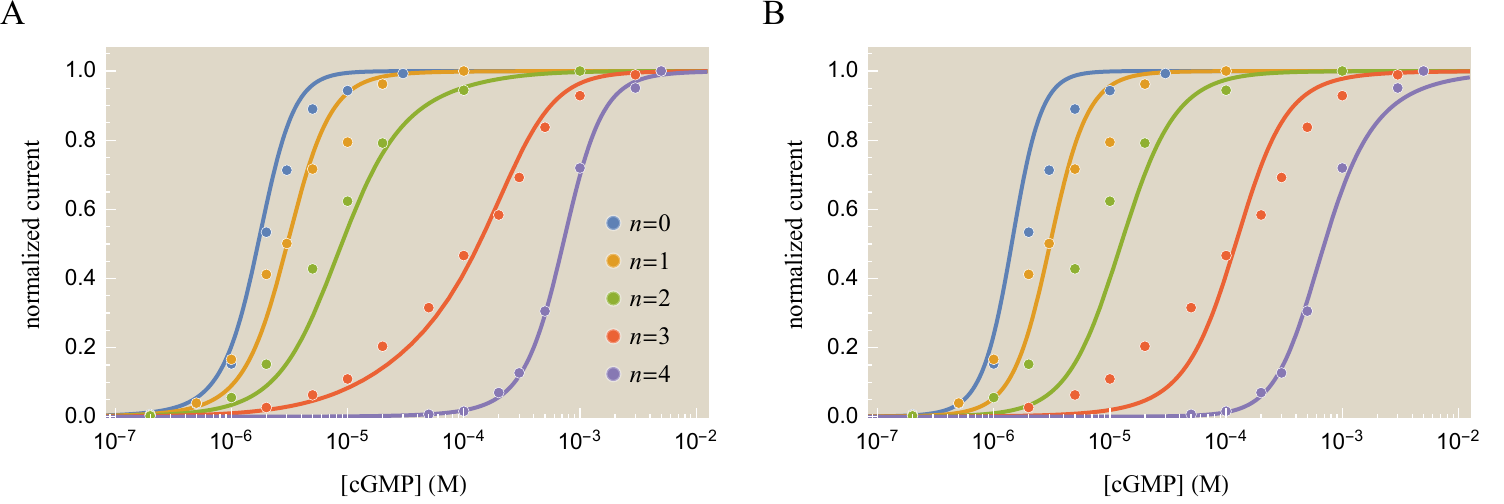}
	\caption{\captionStroke{CNGA2 fits can also be resolved with slight changes to the
			measured $\captionSymbol{\beta \epsilon}$ value.} \letterParen{A} Increasing the
		experimentally measured value of $\beta \epsilon = -11$ to $\beta \epsilon =
		-5$ permits us to recoup a single set of parameters ($R^2 > 0.99$) given by
		\tref[tableFit4] for the entire class of mutants. \letterParen{B} A more
		modest increase from $\beta \epsilon = -11$ to $\beta \epsilon = -8$ yields a
		poorer fit ($R^2 = 0.97$).} \label{figSICNGA1}
\end{figure}

\subsection{Fitting the nAChR Mutants with Non-Uniform $\boldsymbol{\KOWT}$ and $\boldsymbol{\KCWT}$} \label{AppendixNonUniformKOKC}

In order to extract the $[EC_{50}]$ and effective Hill coefficient $h$ of an
nAChR ion channel from experimental measurements
(\fref[figNormalizedCurrent]\letter{A}), the individual data points must be
connected in order to precisely infer where the normalized current reaches
\sfrac{1}{2}. This interpolation may be done in multiple ways, including
connecting the data points with straight line segments or fitting the data to a
sigmoidal function. The resulting $[EC_{50}]$ and $h$ values can then be
compared to the predictions in \fref[figCharacteristics] which were made while
constraining all of the mutants to have the same $\KOWT$ and $\KCWT$
dissociation constants (which may have resulted in worse predictions for the
characteristics of these mutants).

We chose to interpolate the nAChR data sets by fitting each mutant's data
individually to the sigmoidal MWC response
\eref[eqPOpenGeneral][eqNormalizedCurrent]. \fref[figSIIndividualnAChRFits]
shows how each data set is well fit by an individual MWC response, and that
the behavior around the midpoint of each curve when normalized current equals to
\sfrac{1}{2} is well aligned to the data, which gives us confidence that the
corresponding $[EC_{50}]$ and $h$ values that are extracted from these curves
(see \fref[figCharacteristics]\letter{C} and \letter{D}) will be precise. Note
that the resulting best-fit parameters are not meaningful, as the sole purpose
of this plot is to interpolate between the given data points in order to extract
the best possible $[EC_{50}]$ and $h$ values.

\begin{figure}[t]
	\centering \includegraphics[scale=\globalScalePlots]{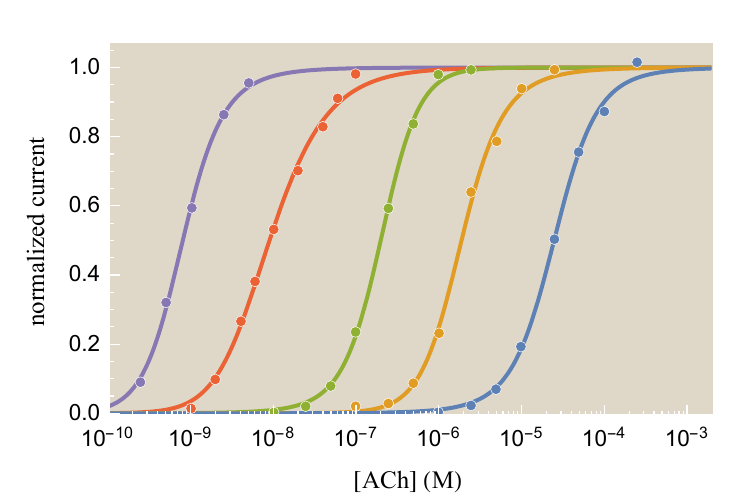}
	\caption{\captionStroke{Extracting $\captionSymbol{[EC_{50}]}$ and $\captionSymbol{h}$ from the
			nAChR data.} The individual nAChR data sets can be fit to the MWC model in
		order to interpolate between the data points and extract the best possible
		$[EC_{50}]$ and $h$ values. Note that each fit is very smooth around the
		midpoint where normalized current equals \sfrac{1}{2}, which is the key region
		where both $[EC_{50}]$ and $h$ are computed.} \label{figSIIndividualnAChRFits}
\end{figure}

\subsection{Degeneracy or ``Sloppiness'' within a Model} \label{AppendixSloppiness}

In this section, we examine the sloppiness inherent when fitting the nAChR and
CNGA2 data (see \fref[fignAChRandCNGA2Sloppiness]). This sloppiness
demonstrates the possible trade-offs between best-fit parameter values that
result in the nearly identical normalized current curves. Because of the
startlingly simple relations between the parameters found in the nAChR case, we
focus solely on that data set.

As seen in \fref[fignAChRandCNGA2Sloppiness]\letter{A} of the main text, the best-fit $\beta
\epsilon^{(n)}$ parameters all depend upon the $\KOWT$ parameter through the
simple relationship $\beta \epsilon^{(n)} = 2 \log \left( \KOWT \right) +
\text{constant}$, while the best-fit value of $\KCWT$ remains fixed for all
$\KOWT$ values. \fref[figSInAChR1] shows how these relationships can be pushed
to biologically unrealistic energy scales and still yield extremely similar
curves.

\begin{figure}[h]
	\centering \includegraphics[scale=\globalScalePlots]{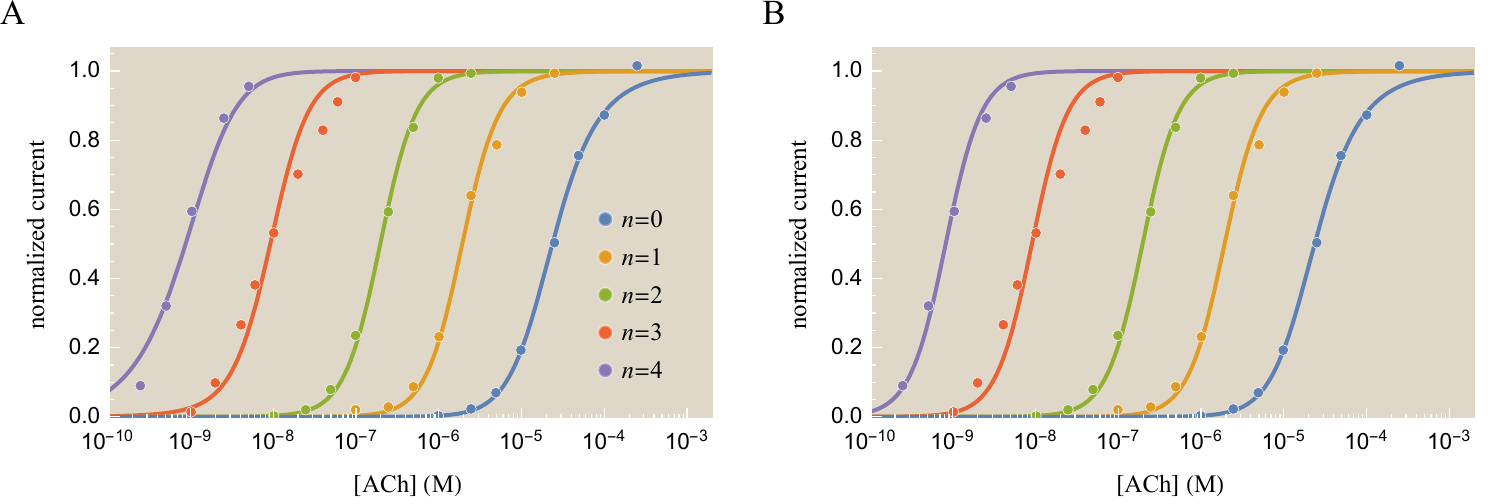}
	\caption{\captionStroke{Different sets of parameters can yield nearly identical nAChR
			normalized current curves.} \letterParen{A} The nAChR data can be fit by
		fixing the wild type energy to either be $\beta \epsilon^{(0)} = -20$ and
		letting the remaining MWC parameters fit optimally. \letterParen{B} The data
		can also be fit using $\beta \epsilon^{(0)} = -100$, which leads to equally
		good fit quality but results in energies that are far to small to be
		biologically feasible. The full set of parameters in both cases is shown in
		\tref[tableFit3], and the fit parameters obey the relationship $e^{-\beta
			\epsilon/2} \KOWT = \text{constant}$ as seen in
		\fref[fignAChRandCNGA2Sloppiness]\letter{A}.} \label{figSInAChR1}
\end{figure}

\begin{table}[t]
	\centering \caption{\captionStroke{Degenerate best-fit parameter sets for the nAChR
			mutants.} Although the curves in \fref[figSInAChR1]\letter{A} and \letter{B}
		look extremely similar and have the same goodness of fit as characterized by
		$R^2$, the latter has unreasonably small $\KOWT$ and $\beta \epsilon^{(n)}$
		values. This demonstrates the trade-offs between best-fit parameters embodied
		in \fref[fignAChRandCNGA2Sloppiness]\letter{A}. \fref[figSInAChR3]\letter{A} and \letter{B}
		demonstrate the deviations in the parameter values if the wild type energy
		$\beta \epsilon^{(0)}$ was slightly increased, as is suggested by some
		literature values.}
	\begin{tabular}{@{\vrule height 10.5pt depth4pt width0pt}ccccccccc}
		\hline
		&
		$\KOWT$ (M) & 
		$\KCWT$ (M) & 
		$\beta \epsilon^{(0)}$ & 
		$\beta \epsilon^{(1)}$ & 
		$\beta \epsilon^{(2)}$ & 
		$\beta \epsilon^{(3)}$ & 
		$\beta \epsilon^{(4)}$ &
		$R^2$ \cr
		\hline
		\hline
		\noalign{\vskip 2pt} 
		\fref[figSInAChR1]\letter{A} &
		$0.8 \times 10^{-9}$ & 
		$60 \times 10^{-6}$ & 
		$-20.0$ & 
		$-15.4$ &
		$-10.9$ &
		$-5.0$ &
		$-0.8$ &
		$0.995$ \cr
		\fref[figSInAChR1]\letter{B} &
		$4 \times10^{-27}$ & 
		$60 \times 10^{-6}$ & 
		$-100.0$ & 
		$-95.4$ &
		$-90.9$ &
		$-84.8$ &
		$-80.0$ &
		$0.995$ \cr
		\fref[figSInAChR3]\letter{A} &
		$2.1 \times 10^{-9}$ & 
		$45 \times 10^{-6}$ & 
		$-18.0$ & 
		$-13.5$ &
		$-9.1$ &
		$-3.3$ &
		$2.0$ &
		$0.995$ \cr
		\fref[figSInAChR3]\letter{B} &
		$3.5 \times 10^{-9}$ & 
		$15 \times 10^{-6}$ & 
		$-16.0$ & 
		$-12.5$ &
		$-8.2$ &
		$-2.6$ &
		$2.0$ &
		$0.986$ \cr
		\hline
	\end{tabular}
	\label{tableFit3}
\end{table}

\begin{table}[t]
	\centering \caption{\captionStroke{Dependence of best-fit CNGA2 parameters with
			$\captionSymbol{\beta \epsilon}$.} Unlike in the nAChR system, the parameters in
		the CNGA2 system are much more tightly constrained (see
		\fref[fignAChRandCNGA2Sloppiness]\letter{B}). Even lowering $\beta \epsilon$ from -5 to -8
		results in a poorer match between the theoretical predictions and the data.
		Literature values suggest that $\beta \epsilon$ may lie between -5 and -11.}
	\begin{tabular}{@{\vrule height 10.5pt depth4pt width0pt}cccccccc}
		\hline
		&
		$\KOWT$ (M) & 
		$\KCWT$ (M) & 
		$\KOMut$ (M) & 
		$\KCMut$ (M) & 
		$\beta \epsilon$ &
		$R^2$ \cr
		\hline
		\hline
		\noalign{\vskip 2pt} 
		\fref[figSICNGA1]\letter{A} &
		$0.6 \times 10^{-6}$ & 
		$40 \times 10^{-6}$ & 
		$290 \times 10^{-6}$ & 
		$200 \times 10^{-3}$ &
		$-5.0$ &
		$0.993$ \cr
		\fref[figSICNGA1]\letter{B} &
		$0.2 \times 10^{-6}$ & 
		$60 \times 10^{-6}$ & 
		$40 \times 10^{-6}$ & 
		$800 \times 10^{-6}$ &
		$-8.0$ &
		$0.976$ \cr
		\hline
	\end{tabular}
	\label{tableFit4}
\end{table}

This suggests that $\KCWT$ and $e^{-\beta \epsilon/2} \KOWT$ are the fundamental
parameters combinations of the system. One means to confirm this result is to
follow the framework in Ref.~\citenum{Transtrum2015}. We first rewrite the
normalized current in terms of the log parameters
\begin{align}
\beta \epsilon_O &= \log \left( \frac{\KOWT}{1\,\text{M}} \right) \\
\beta \epsilon_C &= \log \left( \frac{\KCWT}{1\,\text{M}} \right).
\end{align}
Next, we compute the Hessian matrix of normalized current whose components are
given by
\begin{equation}
\boldsymbol{H}_{i,j} = \frac{\partial \text{normalized current}}{\partial x_i \partial x_j},
\end{equation}
which is a function of the logarithmic parameters
\begin{equation} \label{eqHessianParameters}
\boldsymbol{x}=\left(
\begin{array}{c}
\beta \epsilon \\
\beta \epsilon_O \\
\beta \epsilon_C
\end{array}
\right).
\end{equation}
We can evaluate the eigenvalues of the Hessian using the wild type MWC
parameters ($\beta \epsilon = -23.7$, $\beta \epsilon_O = \log \left( 0.1 \times
10^{-9} \right)$, and $\beta \epsilon_C = \log \left( 60 \times 10^{-6}
\right)$) throughout the range $c \in [10^{-6},3 \times 10^{-4}]\,\text{M}$ of
ligand concentrations at which the wild type channel's dose-response curve was
measured in \fref[figNormalizedCurrent]\letter{A}. For example, at $c=
10^{-6}\,\,\text{M}$, we find that the Hessian has the three eigenvalues $\{2
\times 10^{-2}, 9 \times 10^{-4}, 2 \times 10^{-7}\}$. This last eigenvalue has
a corresponding eigenvector in the direction $\left( 2, 1, 0 \right)$. Because
this last eigenvalue is significantly smaller than the other two, it indicates a
direction in parameter space that the system can be perturbed without
significantly modifying its response, which leads to sloppiness. Note that in
the eigenvalue direction $\left( 2, 1, 0 \right)$, $e^{-\beta \epsilon/2} \KOWT$
remains constant; by considering this parameter combination we effectively
remove the sloppiness in the $\beta \epsilon$ and $\KOWT$ parameters.

As yet another way to show how the parameter combinations $\KCWT$ and $e^{-\beta
	\epsilon/2} \KOWT$ may arise in our model, we rewrite $p_{\text{open}}(c)$ from
\eref[eqPOpenGeneral] as
\begin{equation} \label{eqSIexactpOpen}
p_{\text{open}}(c) = \frac{\left(e^{\beta \epsilon/2} + e^{\beta \epsilon/2} \frac{c}{\KOWT}\right)^2}{\left(e^{\beta \epsilon/2} + e^{\beta \epsilon/2} \frac{c}{\KOWT}\right)^2 + \left(1 + \frac{c}{\KCWT}\right)^2}.
\end{equation}
The denominator can be approximated as $\left(e^{\beta \epsilon/2} + e^{\beta
	\epsilon/2} \frac{c}{\KOWT}\right)^2 + \left(1 + \frac{c}{\KCWT}\right)^2
\approx \left(e^{\beta \epsilon/2} \frac{c}{\KOWT}\right)^2 + \left(1 +
\frac{c}{\KCWT}\right)^2$ as can be seen by considering the two possible regimes
of the concentration: (1) when $c \gg \KOWT$, $e^{\beta \epsilon/2} + e^{\beta
	\epsilon/2} \frac{c}{\KOWT} \approx e^{\beta \epsilon/2} \frac{c}{\KOWT}$
while (2) when $c \ll \KOWT$ the left term $\left(e^{\beta \epsilon/2} +
e^{\beta \epsilon/2} \frac{c}{\KOWT}\right)^2$ will be negligible
compared to the right term $\left(1 + \frac{c}{\KCWT}\right)^2$ so that removing
$e^{\beta \epsilon/2}$ from the left term will not noticeably affect the denominator.
Thus, the formula for $p_{\text{open}}(c)$ takes
the form
\begin{align} \label{eqSIPActiveSloppiless}
p_{\text{open}}^{\text{approx}}(c) &\approx \frac{\left(e^{\beta \epsilon/2}\frac{c}{\KOWT}\right)^2}{\left(e^{\beta \epsilon/2}\frac{c}{\KOWT}\right)^2 + \left(1 + \frac{c}{\KCWT}\right)^2} \nonumber \\
&\equiv \frac{\left(\frac{c}{\KOeff}\right)^2}{\left(\frac{c}{\KOeff}\right)^2 + \left(1 + \frac{c}{\KCWT}\right)^2}.
\end{align}
where we have introduced the effective dissociation constant $\KOeff = \KOWT
e^{-\beta\epsilon/2}$. Thus, we once again find that only the parameter
combinations $\KCWT$ and $\KOeff$ will affect the nAChR response, whereas
changing $\KOWT$ and $e^{-\beta\epsilon}$ while keeping $\KOeff$ fixed will
result in sloppiness.

\fref[figSInAChR2]\letter{A} demonstrates that $p_{\text{open}}(c)$ and
$p_{\text{open}}^{\text{approx}}(c)$ lie on top of each other for the wild type
nAChR values in \tref[tableFit1]. To show that this similarity was not mere
happenstance, we fit the wild type data by constraining the gating energy $\beta
\epsilon^{(0)}$ to the values shown on the $x$-axis of
\fref[figSInAChR2]\letter{B} and letting the two other MWC parameters, $\KOWT$
and $\KCWT$, find their optimal values. The resulting differences between
$p_{\text{open}}(c)$ and $p_{\text{open}}^{\text{approx}}(c)$ will be very small
for all ligand concentrations provided $\beta \epsilon^{(0)}$ is substantially
negative, which is true for the entire class of nAChR mutants in
\tref[tableFit1].

\begin{figure}[h]
	\centering \includegraphics[scale=\globalScalePlots]{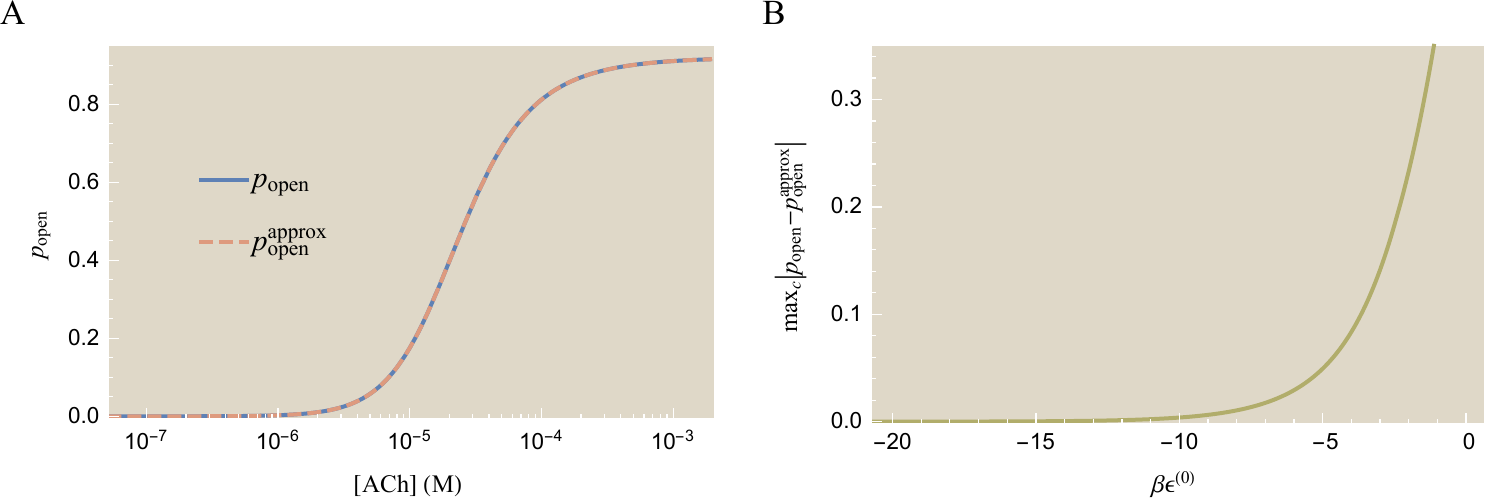}
	\caption{\captionStroke{Exact versus approximate expressions of
			$\captionSymbol{p_{\text{open}}}$.} \letterParen{A} A plot of the exact and
		approximate forms of $p_{\text{open}}$ (see
		\eref[eqSIexactpOpen][eqSIPActiveSloppiless]) using the wild type parameters
		($\KOWT$, $\KCWT$, and $\beta \epsilon^{(0)}$) from \tref[tableFit1].
		\letterParen{B} To check the accuracy of this approximation across many
		degenerate best-fit parameters, we fixed the wild type energy $\beta
		\epsilon^{(0)}$ to the value shown on the $x$-axis and fit the remaining MWC
		parameters ($\KOWT$, $\KCWT$) to the wild type data in
		\fref[figNormalizedCurrent]\letter{A}. The maximum difference between
		$p_{\text{open}}$ and $p_{\text{open}}^{\text{approx}}$ for any concentration
		$c$ is extremely small when $\beta \epsilon^{(0)} \lesssim -5$, indicating that
		the approximate form strongly resembles the exact expression.}
	\label{figSInAChR2}
\end{figure}

\pagebreak
\section{Predicting the behavior of mutants using the MWC model} \label{appendixPredictions}

This section is intended to explore two related questions. First, experiments on
nAChR ion channels with single point mutations in different subunits hint at the
possibility that each mutation incurs the same energetic cost, as described by
the MWC $\epsilon^{(n)}$ parameter (see \eref[eqExtrapolatingEpsilonValues]). In
\ref{appendixPredictionsnAChR}, we explore how well this hypothesis accords with
the data and the predictive power that it grants the MWC model of nAChR. In
\ref{appendixPredictionsCNGA2}, we turn to the CNGA2 ion channels where the
opposite hypothesis holds true, namely, that the gating energy $\epsilon$ is
unaltered by subunit mutations while the remaining MWC parameters are impacted
by these mutations. We again examine how a subset of the CNGA2 mutant data
captures the behavior of the entire class of mutants.

\subsection{nAChR} \label{appendixPredictionsnAChR}

The wild type nAChR ion channel is characterized by the three MWC parameters
$\KOWT$, $\KCWT$, and $\beta \epsilon^{(0)}$ (see
\eref[eqPOpenGeneral][eqNormalizedCurrent]), all three of which can be fit from
the wild type data set ($n=0$ in \fref[figNormalizedCurrent]\letter{A}). We
further postulate that the L251S mutations will only change the allosteric
energy $\beta \epsilon^{(0)}$, leaving the ligand binding affinities $\KOWT$ and
$\KCWT$ unchanged.

The nAChR data suggests that each L251S mutation increases the gating
equilibrium by $\Delta \epsilon$ per mutation, so that $\beta \epsilon^{(n)} =
\beta \epsilon^{(0)} + n \Delta \epsilon$. We aim to find to what extent this
hypothesis holds true. Specifically, we note that after the wild type data set
fixes $\KOWT$, $\KCWT$, and $\beta \epsilon^{(0)}$, using one additional data
set can fix $\Delta \epsilon$, enabling us to extrapolate the $\beta
\epsilon^{(n)}$ values for the remaining mutants. For example, in
\fref[figNormalizedCurrentv2]\letter{A} of the main text, we used $\beta
\epsilon^{(0)} = -23.7~\kB T$ and $\beta \epsilon^{(4)} = -4.0~\kB T$ to
determine $\Delta \epsilon = -4.9~\kB T$, from which we determined $\beta
\epsilon^{(1)} = -18.8~\kB T$, $\beta \epsilon^{(2)} = -13.9~\kB T$, and $\beta
\epsilon^{(3)} = -8.9~\kB T$. The resulting predictions characterized the data
sets for the $n=1,2,3$ nAChR mutants remarkably well ($R^2 = 0.985$).

Note that this same procedure could work with any two nAChR data sets. For
example, we could use the $n=1$ mutant's data to determine the MWC parameters
$\KOWT$, $\KCWT$, and $\beta \epsilon^{(1)}$, and then use the $n=2$ data set to
determine $\Delta \epsilon$ and extract the remaining $\beta \epsilon^{(n)}$
values. \fref[fignAChRFullPredictions] demonstrates the resulting predictions
when all ten possible input pairs are used to predict the remaining three mutant
dose-response curves. The corresponding best-fit parameters are given in
\tref[tableFit5]. In each case, the two input curves used to extract the MWC
parameters are shown as solid curves, while the three predicted responses are
shown as dashed lines.

Most of the predictions do an especially good job of predicting the behavior of
the intermediary $n=1$, 2, and 3 mutants, while predictions for the two outer
data sets $n=0$ (wild type) and $n=4$ are likely to be worse. This follows the
general rule that interpolation - predicting values inside the domain of the
training set - is more reliable than extrapolation. This suggests that when
trying to make predictions for a similar family of mutants, it is most
beneficial to acquire data for the extreme cases (i.e. the $n=0$ and $n=4$ data
sets). In terms of the overall fit performance on the three unknown data sets in
each of the ten plots, four of the fits have $R^2 > 0.9$ while four others have $0.9 > R^2 >
0.8$. This fit performance is improved if three or four input data sets are used
to predict the remaining dose-response curves, as shown in the supplementary
\textit{Mathematica} notebook.

Interestingly enough, when these predictions fail (most notably in
\fref[fignAChRFullPredictions]\letter{J}), it occurs because the fitting
captures the local details of (and noise in) the input data sets, which throws
off the extrapolation to the remaining ion channel mutant. This concept is
reminiscent of over-fitting in computer science. Indeed, it suggests that
contrary to our intuition, using a more generalized model which has more degrees
of freedom and is able to capture the tiny nuances of each individual data set
even more precisely would do worse at predicting the global behavior of this
class of mutants. In other words, having a coarse-grained model of the system
with fewer parameters may provide a better opportunity to correctly predict
protein behavior.

\begin{figure}[h!]
	\begin{adjustwidth}{-0.94in}{0in}
	\centering \includegraphics[scale=\globalScalePlots]{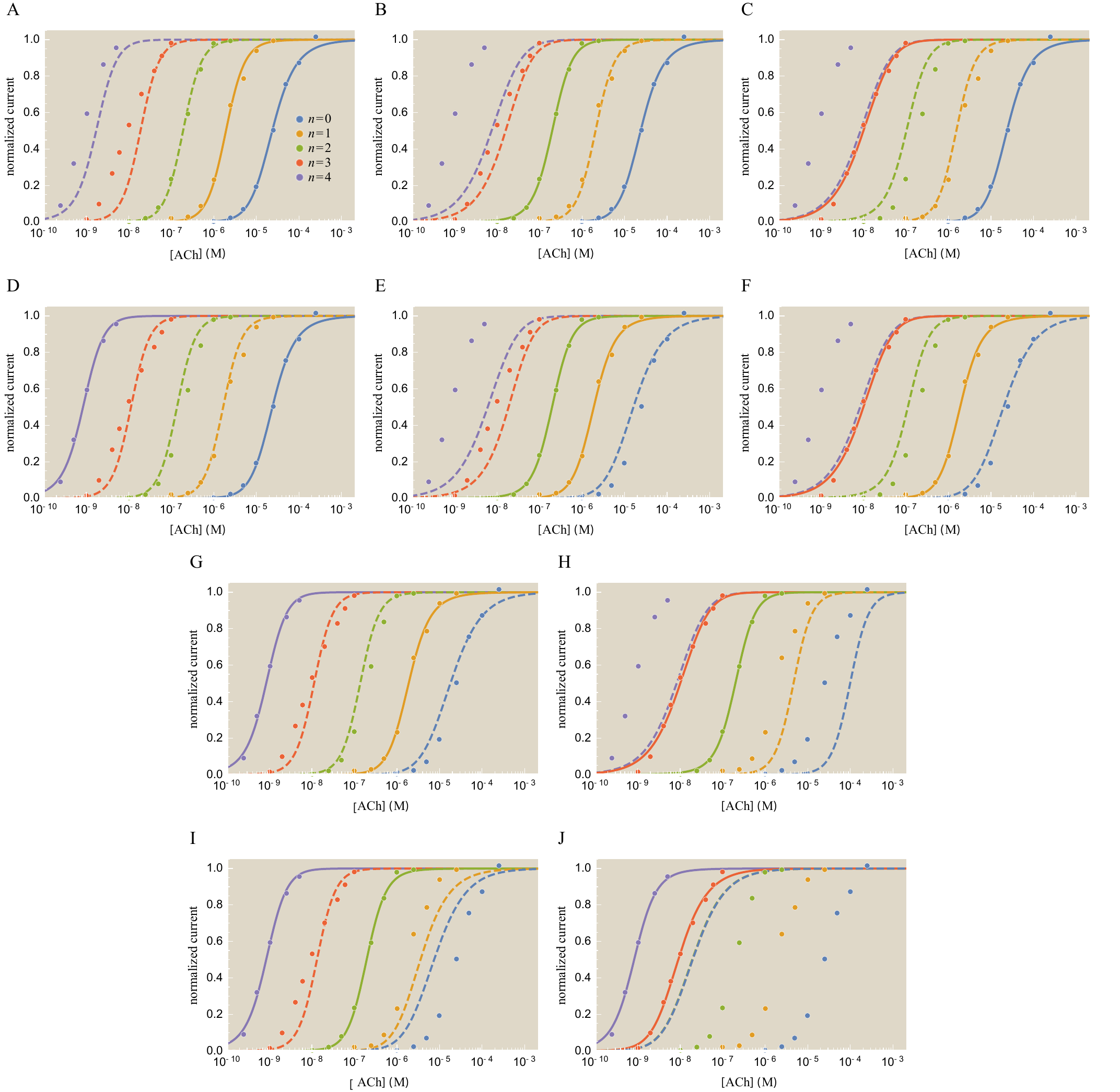} 
	\end{adjustwidth}
	\caption{\captionStroke{Predicting nAChR mutants using different training sets.} The MWC
		parameters for the entire class of nAChR mutants can be fit from two data sets
		(solid lines). Using these parameters, the dose-response curves of the
		remaining three mutants can be predicted (dashed lines) without any further
		fitting. The best-fit parameters are listed in \tref[tableFit5].}
	\label{fignAChRFullPredictions}
\end{figure}

\begin{table}
		\centering \caption{\captionStroke{nAChR parameter predictions from two input data
		sets.} Data sets from the two plain text $\beta \epsilon^{(n)}$ columns (shown
		as solid lines in their corresponding figures) were used to determine the
		$\KOWT$ and $\KCWT$ dissociation constants for the entire class of mutants and to linearly
		extrapolate the energies (bold text) of the remaining mutants using
		\eref[eqExtrapolatingEpsilonValues]. $R^2$ indicates the goodness of fit for
		the three predicted curves (shown as dashed lines in the corresponding
		figures).}
		\begin{tabular}{@{\vrule height 10.5pt depth4pt width0pt}ccccccccc}
			\hline
			&
			$\KOWT$ (M) & 
			$\KCWT$ (M) & 
			$\beta \epsilon^{(0)}$ & 
			$\beta \epsilon^{(1)}$ & 
			$\beta \epsilon^{(2)}$ & 
			$\beta \epsilon^{(3)}$ & 
			$\beta \epsilon^{(4)}$ &
			$R^2$ \cr
			\hline
			\hline
			\noalign{\vskip 2pt} 
			\fref[fignAChRFullPredictions]\letter{A} &
			$0.3 \times 10^{-9}$ & 
			$60 \times 10^{-6}$ & 
			$-22.4$ & 
			$-17.8$ &
			$\textbf{-13.2}$ &
			$\textbf{-8.6}$ &
			$\textbf{-4.1}$ &
			$0.950$ \cr
			\fref[fignAChRFullPredictions]\letter{B} &
			$20 \times 10^{-9}$ & 
			$80 \times 10^{-6}$ & 
			$-14.0$ & 
			$\textbf{-9.5}$ &
			$-5.0$ &
			$\textbf{-0.5}$ &
			$\textbf{4.0}$ &
			$0.868$ \cr
			\fref[fignAChRFullPredictions]\letter{C} &
			$20 \times 10^{-9}$ & 
			$80 \times 10^{-6}$ & 
			$-13.7$ & 
			$\textbf{-8.6}$ &
			$\textbf{-3.5}$ &
			$1.6$ &
			$\textbf{6.7}$ &
			$0.839$ \cr
			\fref[fignAChRFullPredictions]\letter{D} &
			$0.1 \times 10^{-9}$ & 
			$80 \times 10^{-6}$ & 
			$-23.8$ & 
			$\textbf{-18.8}$ &
			$\textbf{-13.9}$ &
			$\textbf{-8.9}$ &
			$-4.0$ &
			$0.985$ \cr
			\fref[fignAChRFullPredictions]\letter{E} &
			$10 \times 10^{-9}$ & 
			$10 \times 10^{-6}$ & 
			$\textbf{-13.7}$ & 
			$-9.5$ &
			$-5.4$ &
			$\textbf{-1.2}$ &
			$\textbf{2.9}$ &
			$0.867$ \cr
			\fref[fignAChRFullPredictions]\letter{F} &
			$20 \times 10^{-9}$ & 
			$10 \times 10^{-6}$ & 
			$\textbf{-13.9}$ & 
			$-8.8$ &
			$\textbf{-3.7}$ &
			$1.4$ &
			$\textbf{6.6}$ &
			$0.839$ \cr
			\fref[fignAChRFullPredictions]\letter{G} &
			$0.1 \times 10^{-9}$ & 
			$10 \times 10^{-6}$ & 
			$\textbf{-23.8}$ & 
			$-18.8$ &
			$\textbf{-13.9}$ &
			$\textbf{-8.9}$ &
			$-4.0$ &
			$0.983$\cr
			\fref[fignAChRFullPredictions]\letter{H} &
			$20 \times 10^{-9}$ & 
			$200 \times 10^{-3}$ & 
			$\textbf{-17.0}$ & 
			$\textbf{-10.9}$ &
			$-4.8$ &
			$1.4$ &
			$\textbf{7.5}$ &
			$0.729$ \cr
			\fref[fignAChRFullPredictions]\letter{I} &
			$0.1 \times 10^{-9}$ & 
			$3 \times 10^{-6}$ & 
			$\textbf{-25.0}$ & 
			$\textbf{-19.7}$ &
			$-14.5$ &
			$\textbf{-9.2}$ &
			$-4.0$ &
			$0.930$ \cr
			\fref[fignAChRFullPredictions]\letter{J} &
			$0.1 \times 10^{-9}$ & 
			$10 \times 10^{-9}$ & 
			$\textbf{-20.7}$ & 
			$\textbf{-16.6}$ &
			$\textbf{-12.5}$ &
			$-8.4$ &
			$-4.3$ &
			$0.063$\cr
			\hline
		\end{tabular}
		\label{tableFit5}
\end{table}

\subsection{CNGA2} \label{appendixPredictionsCNGA2}

The wild type CNGA2 ion channel has 4 identical subunits with ligand
affinity $\KOWT$ in the open state and $\KCWT$ in the closed state. The free
energy difference between the closed and open states is given by $\epsilon$. A
mutation was introduced in the ligand binding site of any subunit, which results
in new dissociation constants $\KOMut$ in the open state and $\KCMut$ in the
closed state, but which will leave the free energy difference $\epsilon$
unchanged. Once all of the MWC parameters are known, a CNGA2 mutant with $n$
mutated subunit and $4-n$ wild type subunits is fully described using
\eref[eqPActive4Mutants] with $m=4$.

One conceptually simple route to resolving the MWC parameters is to first fix
the wild type parameters $\KOWT$, $\KCWT$, and $\epsilon$ using the wild type
data set ($n=0$) and then fix the two mutant dissociation constants $\KOMut$ and
$\KCMut$ from the $n=4$ data set. From these parameters, the intermediate
mutants $n=1$, 2, and 3 would all follow from \eref[eqPActive4Mutants]. Yet, as
in the case of nAChR, any two data sets could be used to fix the parameter
values. In fact, in this system all five thermodynamic parameters ($\KOWT$,
$\KCWT$, $\KOMut$, $\KCMut$, and $\epsilon$) could be fit using a
\textit{single} data set from one of the $n=1$, 2, or 3 mutants, since the
dose-response curve \eref[eqPActive4Mutants] for such a mutant would contain all
five parameters. Such fits are shown in the supplementary \textit{Mathematica}
notebook, but as expected the fits that utilize two input data sets do a much
better job at predicting the remaining mutants.

\fref[figCNGAFullPredictions] shows the predictions (dashed lines) generated by
fitting the MWC parameters to all possible input pairs (solid lines), with the
best-fit parameters given in \tref[tableFit6]. As was found for the nAChR ion
channels, the worst predictions resulted from data sets that are very close
together (for example, when both input parameters came from $n=2$, $n=3$, or
$n=4$), which results in poor extrapolations for the remaining mutant data sets.
Surprisingly, the prediction based on the $n=0$ and $n=4$ data set, which could
be expected to be one of the best fits, was also poor. That said, the majority of
the predictions were quite accurate ($R^2 > 0.96$), once again demonstrating the
power of the simple statistical mechanical model we have employed.

\begin{figure}[t!]
	\begin{adjustwidth}{-0.93in}{0in}
		\centering \includegraphics[scale=\globalScalePlots]{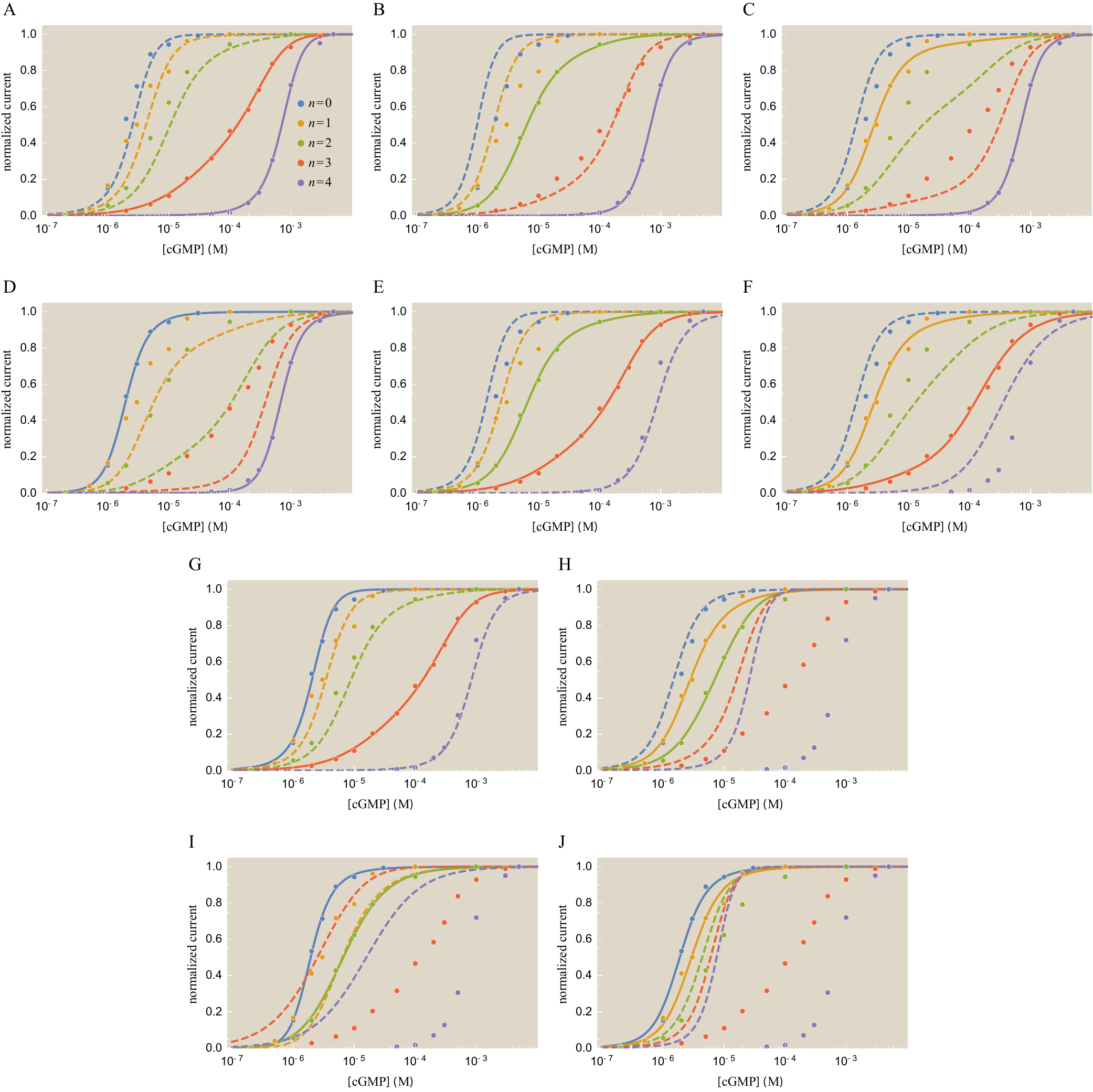} 
	\end{adjustwidth}
	\caption{\captionStroke{Predicting CNGA2 mutants using different training sets.} As was found for nAChR, two data sets (solid lines) are sufficient to extract the MWC parameters for the whole class of CNGA2 mutants, which can then be used to extrapolate the responses of the remaining mutants (dashed lines). The best-fit parameters are listed in \tref[tableFit6].}
	\label{figCNGAFullPredictions}
\end{figure}

\begin{table}
		\centering \caption{\captionStroke{CNGA2 parameter predictions from two input data
				sets.} Two data sets
			(shown as solid lines in the corresponding figures) were used to determine
			the thermodynamic parameters for the entire class of mutants.
			$R^2$ indicates the goodness of fit for the three predicted curves (shown as dashed lines in the corresponding figures).}
		\begin{tabular}{@{\vrule height 10.5pt depth4pt width0pt}cccccccc}
			\hline
			&
			$\KOWT$ (M) & 
			$\KCWT$ (M) & 
			$\KOMut$ (M) & 
			$\KCMut$ (M) & 
			$\beta \epsilon$ &
			$R^2$ \cr
			\hline
			\hline
			\noalign{\vskip 2pt} 
			\fref[figCNGAFullPredictions]\letter{A} &
			$1.5 \times 10^{-6}$ & 
			$35 \times 10^{-6}$ & 
			$470 \times 10^{-6}$ & 
			$70 \times 10^{-3}$ &
			$-3.6$ &
			$0.983$ \cr
			\fref[figCNGAFullPredictions]\letter{B} &
			$0.3 \times 10^{-6}$ & 
			$15 \times 10^{-6}$ & 
			$180 \times 10^{-6}$ & 
			$3 \times 10^{-3}$ &
			$-5.5$ &
			$0.962$ \cr
			\fref[figCNGAFullPredictions]\letter{C} &
			$0.5 \times 10^{-6}$ & 
			$6 \times 10^{-6}$ & 
			$260 \times 10^{-6}$ & 
			$5 \times 10^{-3}$ &
			$-4.6$ &
			$0.962$ \cr
			\fref[figCNGAFullPredictions]\letter{D} &
			$0.3 \times 10^{-6}$ & 
			$5 \times 10^{-6}$ & 
			$120 \times 10^{-6}$ & 
			$2 \times 10^{-3}$ &
			$-6.5$ &
			$0.857$ \cr
			\fref[figCNGAFullPredictions]\letter{E} &
			$0.6 \times 10^{-6}$ & 
			$20 \times 10^{-6}$ & 
			$290 \times 10^{-6}$ & 
			$2 \times 10^{-3}$ &
			$-4.3$ &
			$0.982$ \cr
			\fref[figCNGAFullPredictions]\letter{F} &
			$0.5 \times 10^{-6}$ & 
			$5 \times 10^{-6}$ & 
			$60 \times 10^{-6}$ & 
			$170 \times 10^{-6}$ &
			$-4.6$ &
			$0.978$ \cr
			\fref[figCNGAFullPredictions]\letter{G} &
			$1 \times 10^{-6}$ & 
			$30 \times 10^{-6}$ & 
			$370 \times 10^{-6}$ & 
			$4 \times 10^{-3}$ &
			$-3.9$ &
			$0.990$ \cr
			\fref[figCNGAFullPredictions]\letter{H} &
			$0.5 \times 10^{-6}$ & 
			$4 \times 10^{-6}$ & 
			$15 \times 10^{-6}$ & 
			$140 \times 10^{-3}$ &
			$-4.1$ &
			$0.883$ \cr
			\fref[figCNGAFullPredictions]\letter{I} &
			$0.1 \times 10^{-6}$ & 
			$3 \times 10^{-6}$ & 
			$20 \times 10^{-6}$ & 
			$20 \times 10^{-6}$ &
			$-10.2$ &
			$0.640$ \cr
			\fref[figCNGAFullPredictions]\letter{J} &
			$0.5 \times 10^{-6}$ & 
			$4 \times 10^{-6}$ & 
			$3 \times 10^{-6}$ & 
			$140 \times 10^{-3}$ &
			$-4.6$ &
			$0.713$ \cr
			\hline
		\end{tabular}
		\label{tableFit6}
\end{table}

\setcounter{page}{20}

\providecommand{\latin}[1]{#1}
\providecommand*\mcitethebibliography{\thebibliography}
\csname @ifundefined\endcsname{endmcitethebibliography}
{\let\endmcitethebibliography\endthebibliography}{}

\setcounter{page}{31}

\providecommand{\latin}[1]{#1}
\providecommand*\mcitethebibliography{\thebibliography}
\csname @ifundefined\endcsname{endmcitethebibliography}
{\let\endmcitethebibliography\endthebibliography}{}


\end{document}